\documentclass[12pt]{article}
\usepackage{epsfig}
\usepackage{colortbl}
\usepackage{ulem}

\usepackage{tikz}
\usetikzlibrary{shapes,arrows}

\usepackage{mathrsfs,bbm}
\usepackage{newtxtext,newtxmath}
\usepackage[american]{babel}

\newcommand{\x}{\mathbf{x}}

\newcommand{\bxi}{\boldsymbol{\xi}}

\newcommand{\bof}{\mathbf{f}}

\newcommand{\f}{f}

\newcommand{\ve}{\varepsilon}
\newcommand{\vf}{\varphi}

\newcommand{\bnu}{\boldsymbol{\nu}}

\newcommand{\bvf}{\boldsymbol{\varphi}}
\newcommand{\bx}{{\boldsymbol{x}}}

\newcommand{\bu}{{\boldsymbol{u}}}
\newcommand{\bv}{{\boldsymbol{v}}}
\newcommand{\bw}{{\boldsymbol{w}}}

\newcommand{\bbi}{\boldsymbol{i}}
\newcommand{\bbj}{\boldsymbol{j}}

\def\va{\raise 2pt\hbox{,}}

\def\cA{{\cal A}}

\newcommand{\p}{\partial}

\title{A multi-scale model of virus pandemic: Heterogeneous interactive entities in a globally connected world}

\author{
        Nicola Bellomo \\
                University of Granada, Departamento de Matem\'atica Aplicada,  Spain\\
        IMATI CNR, Pavia, Italy, and Politecnico of Torino, Italy \\
            \and
        Richard Bingham\\
        Departments of Mathematics and Biology 
        \\ York Cross-disciplinary Centre for Systems
Analysis \\
        University of York, U.K.\\
                	\and
        	Mark A.J.~Chaplain \\
        	Mathematical Institute, School of Mathematics and Statistics \\
        	University of St Andrews, Scotland, UK \\
        	        	\and
        	Giovanni Dosi \\
        	Institute of Economics\\ EMbeDS, Scuola Superiore Sant'Anna, Pisa, Italy \\
        	        	\and
        	Guido Forni \\
Accademia Nazionale dei Lincei, Roma, Italy
        	\and
        	Damian A.~Knopoff \\
Centro de Investigaci\'on y Estudios de Matem\'atica (CONICET) \\
Famaf (UNC), C\'ordoba, Argentina
        	\and
        	John Lowengrub \\
Department of Mathematics, University California Irvine, USA
        	\and
        	Reidun Twarock \\
        	Departments of Mathematics and Biology 
        \\ York Cross-disciplinary Centre for Systems
Analysis \\
        University of York, U.K.\\
                	\and
         	Maria Enrica Virgillito\\
Institute of Economics\\ EMbeDS, Scuola Superiore Sant'Anna, Pisa, Italy 
        	}

\begin{document}

\maketitle

\begin{abstract}
This paper is devoted to the multidisciplinary modelling of a pandemic initiated by an aggressive virus, specifically the so-called \textit{SARS--CoV--2 Severe Acute Respiratory Syndrome, corona virus n.2}.  The study is developed within a multiscale framework accounting for the interaction of different spatial scales, from the small scale of the virus itself and cells, to the large scale of individuals and further up to the collective behaviour of populations. An interdisciplinary vision is developed thanks to the contributions of epidemiologists, immunologists and economists as well as those of mathematical modellers.  The first part of the contents is devoted to understanding the complex features of the system and to the design of a modelling rationale. The modelling approach is treated in the second part of the paper by showing both how the virus propagates into infected individuals, successfully and not successfully recovered, and also the spatial patterns, which are subsequently studied by kinetic and lattice models. The third part reports the contribution of research in the fields of virology, epidemiology, immune competition, and economy focused also on social behaviours. Finally, a critical analysis is proposed  looking ahead to research perspectives.\end{abstract}



\section{Motivations and plan of the essay}\label{sec:1}

The onset of the \textit{SARS--CoV--2} virus responsible for the initial \textit{COVID-19} outbreak and the subsequent pandemic, has brought to almost all countries across the globe huge problems affecting health, safety/security, economics, and practically all expressions of collective behaviour in our societies. Data are constantly updated and presented at various dedicated websites~\cite{[COR],[GC20],[JHU],[SCIRE]}. A significant percentage of societies and governments believed this to be a so-called {\it black swan}~\cite{[TAL07]} event for our society, including a number of scientists. However, this event is definitely not a black swan, although it has already had a great impact all over the world. Indeed, this event should have been predictable (and indeed was predicted by a few~\cite{[Avishai2020],[Gate20]}) but many of our societies appear to be unprepared to tackle this problem.

If we look at COVID-19 from the view point of applied mathematicians, some preliminary remarks concerning how best to approach modelling such a pandemic, which include an interdisciplinary vision and the role of mathematical models in science and our society, are as follows:

\begin{itemize}

\item  The modelling approach should go far beyond deterministic population dynamics, since individual reactions to the infection and pandemic events are heterogeneously distributed throughout the population. Spatial dynamics and interactions are an important feature to be included in the modelling approach, since the dynamics are generated by nonlocal interactions and transportation devices.

\vskip.2cm \item The modelling ought to be developed within a multiscale vision, as the dynamics of individuals depend on the dynamics at smaller scales inside each individual by the competition between virus particles and the immune system. It is clear that applied mathematicians cannot tackle the modelling problem by a stand-alone approach -- an interdisciplinary vision is necessary through mutually enriching and beneficial interactions with scientists in other fields including virology, epidemiology, immunology and biology in general.

\vskip.2cm \item  The approach adopted and described in this paper looks firstly for a model local-in-space accounting for the infection dynamics and, subsequently for the competition inside each individual, between the proliferating virus and the immune-system specific to the individual. Subsequently the approach focuses on collective behaviour. Finally, the mathematical description of spatial propagation is studied, since we are aware that this is not a problem of diffusion, but that of directed propagation with finite speed (which can and should be related to endogenous communication networks where individuals move).

\vskip.2cm \item The scope of such a research project should not be confined only to ``biological and medical sciences'', but also be addressed to wider aspects of and other communities in our society. This requirement is motivated not only by the strong impact that a virus pandemic can have on a society, but also to the need that science should look forward with predictive aims. On their own, mathematical models do not and cannot solve the problems of biology, medicine and economics.  However, once refined and informed by empirical data, they can produce insightful provisional simulations which can even uncover dynamics which were not previously observed (cf. emergent behaviour). Hence mathematical models can and should also be viewed as a tool to generate dialogue and wider communication between the hard and applied sciences. This dialogue can in turn lead to a perspective on and insight into possible future events.

\vskip.2cm \item Since the modelling approach aims and intends to capture the complex features of living systems, this effort often requires new mathematical methods and techniques, even new mathematical theories. This requirement was already posed by Hartwell~\cite{[HART99]},  accounting for multiscale problems as had already appeared in the celebrated essay by Schr\"odinger~\cite{[SCH1944]}, but the general premise and vision goes back over a century to Hilbert~\cite{[DH1900],[DH1901],[DH1902]}. Indeed, the words of Hilbert opening his lecture ``\textit{Mathematische Probleme}'' delivered before the International Congress of Mathematicians in Paris, 1900, are worth recalling here since they are prophetically apposite: 

\vskip.2cm
\begin{quote}
\textit{Who of us would not be glad to lift the veil behind which the future lies hidden; to cast a glance at the next advances of our science and at the secrets of its development during future centuries? What particular goals will there be toward which the leading mathematical spirits of coming generations will strive? What new methods and new facts in the wide and rich field of mathematical thought will the new centuries disclose?}
\end{quote}

\end{itemize}

As noted above, the contribution of mathematics to model complex natural phenomena can be effective only if sufficiently well integrated with the work, knowledge and insight of scientists active in virology, epidemiology, immunology and biology. Once integrated in this fruitful interdisciplinary manner, the best mathematical models (e.g. through predictive computational simulations) are capable of presenting a broad vision of a range of possible system outcomes which can be used towards both therapeutic actions and, in the case of a pandemic, confinement strategies to control the dynamics of an infection. The recent events related to COVID-19 have shown that the pandemic has exerted a huge impact on societies concerning both their social behaviour and economic strategies. This paper accounts for some specific, but highly relevant, features of the pandemic and aims to construct a framework which can hopefully contribute to developing future strategies by which governments can fight to reduce and possibly control the impact of such an event. Bearing in mind all of the above, the contents of our paper, which can be viewed in three parts, can be presented as follows.

\vskip.1cm Part 1 is devoted to understanding the complex features of the system under consideration whose biological dynamics are described in Section 2. A first step towards the design of a modelling rationale is proposed in the same section to contribute to defining the general mathematical framework deemed to provide the conceptual basis for the derivation of the models.

\vskip.1cm Part 2 focuses on modelling topics. Firstly the case of spatial homogeneity is studied in Section 3, where a model is derived accounting for the heterogeneity of the immune defence of individuals and describing how the virus propagates into infected individuals.  Simulations enlighten the main features of the dynamics as well as the role of possible therapeutic actions. Subsequently, the more realistic case of the spatial development of the disease is developed in Section 4, accounting for recent studies on the modelling of crowd dynamics~\cite{[ALBI19],[ABGR20]}. Different features of spatial dynamics and spread are studied to show how the infection propagates through space.

\vskip.1cm Part 3 reports on the contribution of research in the fields of virology (in Section 4),  immune competition (in Section 5),  and economics, focused also on social behaviour (in Section 6). These sections propose a critical analysis which accounts for the modelling achievements and indicates key modelling targets. Section 7 looks ahead to research perspectives suitable for deeper consideration of the various initiatives proposed in the preceding sections. This section provides an answer to the general key objective described below by tracing the guidelines of a forward look to a future vision of the systems approach adopted here.

Summary statement:

\vskip.1cm
\begin{quote}
\textit{The key target of this paper consists in designing a multiscale modelling approach and framework suitable for designing and producing a simulation device with the capacity to explore crucial aspects of the spread of a pandemic. More specifically, this should be capable of exploring the different scenarios corresponding to possible strategies to control a virus pandemic, so that crisis managers can select the most appropriate or most effective choices towards effective control within the shortest timeframe. At a later stage, once reliable data are available, these data will be properly selected and stored in a database, and then the simulation device can become predictive, while the model can be further refined and developed to include additional details of the virus such as mutations, selection and evolution.}
\end{quote}

\section{Towards a mathematical framework} \label{sec:2}

This section provides a general introduction to the technical and formal contents of our paper developed in Sections 3--7. The presentation is in three parts: first, a brief historical introduction to epidemiology is delivered, subsequently a concise report on the phenomenology of the biological system under consideration is presented, and finally a rationale towards the modelling approach is brought to the reader's attention in view of the mathematical formalization proposed in Section 3.

Firstly we provide a concise introduction to the contribution of mathematics to epidemics, subsequently an outlook to the phenomenological behaviour of the system is presented  enlightening some  heterogeneity and multiscale features of the dynamics, lastly we focus on the key features of the rationale towards the modelling approach developed in the next sections.

The mathematical modelling of infectious diseases goes back over 250 years to Daniel Bernoulli who developed a model to predict the number of deaths due to smallpox and the effect on mortality of inoculation~\cite{[DB1766]}. This problem was also considered by D'Alembert who produced an alternative model to Bernoulli~\cite{[JDA1761]}, in fact a critique of Bernoulli's approach which led to a heated debate between the two~\cite{[DietzHeester2002]}. Our modern day subject of ``theoretical epidemiology'' goes back almost 100 years to the seminal paper of Kermack and McKendrick in 1927~\cite{[KMcK1927]}. Kermack and McKendrick proposed and developed the first ``compartmental'' model of disease dynamics and disease spread by considering a population of individuals sub-divided into three separate classes - susceptible, infectious and recovered - with the transmission of the disease/infection being dependent upon the number of interactions between individuals and the underlying rate of infection. This is the first \textit{SIR model} of disease transmission. In formulating their model in this way, the so-called {\it reproduction number} or $R_0$ was defined and introduced.

In the mid-1970s, interest in the subject was rekindled~\cite{[Bailey1975]} with both theoretical developments of the models and systems~\cite{[ODiek1977]} and prescient applications to real epidemic outbreaks such as the cholera epidemic in Bari, Italy, in the early 1970s by Capasso~\cite{[VCap1977],[VCap1979]}. Modelling the spatial spread of epidemics, although not much focussed on in general (using reaction-diffusion equations and the theory of travelling waves~\cite{[VCap1]}), has been applied to the spread of rabies~\cite{[JDM1985]} which led to an estimate of the speed of the invasive wave of infection. Applied to historical data from Europe,  this approach has also been used to estimate the spread of the black death/bubonic plague through Europe in the 14th century~\cite{[JDM2003]}. Subsequent work by, e.g. Diekmann and others, has formalised much of the theory behind epidemic models and introduced structured population models~\cite{[DH2000],[DHT2012]}.

In the 1980s and into the 1990s, with the onset of AIDS/HIV, work by Anderson \& May and others~\cite{[AMI1979],[AMII1979],[RMA1982],[MA1987],[Nowak1991],[NowakMay1991]} highlighted the role of mathematical modelling in predicting possible outcomes of disease spread in large populations. Building on this, ground-breaking work by Perelson and others~\cite{[Perelson1996],[Perelson1999],[Perelson2002]} on viral load, virion clearance rate and the role of the immune system/response helped lead to breakthroughs in treatment for AIDS/HIV.

Applications of predictive and insightful mathematical modelling to other diseases include the work of Grenfell and co-workers which has helped to quantify the dynamics of measles epidemics, in a range of settings and the impact of vaccination strategies against measles and other childhood infections~\cite{[Grenfell1995],[Grenfell1996],[Grenfell2001],[Grenfell2013]}. Prior to the current COVID pandemic, other applications of epidemiological modelling to disease outbreaks over the past 30 years or so have included: bovine spongiform encephalopathy (BSE), Foot-and-Mouth disease (FMD), SARS, MERS, and the flu pandemic among others~\cite{[Heesterbeek2015]}.


The virus, Corona-virus disease 2019 (COVID-19) is caused by the infection of the  \textit{SARS--CoV--2} virus, a corona-virus. The so called Corona-viruses are part of a large family of viruses that cause illness ranging from the common cold  to more severe diseases including Severe Acute Respiratory Syndrome (SARS) and COVID-19. The  \textit{SARS--CoV--2} virion is made up of four structural proteins known as Spike, Envelope, Membrane and Nucleocapsid.

The World Health Organization (WHO) upgraded the state of  \textit{SARS--CoV--2} infection from epidemic to pandemic in March 2020. Apparently, the epidemic first started in China, followed by South Korea and then Italy. Progressively many (almost all) countries around the world have have been invaded and have reacted by implementing different strategies including the imposition of mobility and physical locks-downs and border limitations. The behaviour of all countries to counteract the pandemic appears to be quite heterogeneous, as it is the national organization which cares about infected patients. This heterogeneity is also confirmed by the diffusion of the disease inside each country and the distribution of the level of the pathology.

As we shall see in the next sections, accounting for heterogeneity of the populations is not an essential requirement for the modelling approach, but it can contribute to a deeper understanding towards the contribution that economics and sociology can provide to all decision makers, from crisis managers all the way up to policy makers.

Although the knowledge of this topic is still evolving, some specific features can be extracted, in view of a deeper analysis to be developed in sections 4--6, with the aim of developing some reasonings on the modelling rationale towards the mathematical formalization developed firstly in Section 3 and subsequently also in Section 7. The amount of scientific literature which has appeared in the last few months after the appearance of Covid-19 is impressive and, in some cases, not easy to follow due to a contrast of view points which can be related to the non-exhaustive knowledge in the field.

Therefore, we need to issue a warning to the reader that we do not claim completeness, while we refer to the report~\cite{[CFM20]}, which is bimonthly updated and it is available at the website of the Italian Lincei Academy. The description is delivered in the following, where some of the Items have been extracted from~\cite{[CFM20]}.

\begin{itemize}

\item \textit{Virus transmission:}  \textit{SARS--CoV--2} is mainly transmitted through the respiratory route 10-12 via respiratory droplets, up to 1 millimetre in diameter, that an infected person expels when she/he coughs or sneezes. As the virus multiplies, an infected person may shed copious amounts of it.

\vskip.2cm \item \textit{Binding and cell attack:} The large Spike protein  forms a sort of crown on the surface of the viral particles and acts as an anchor allowing the virus to bind to the Angiotensin-Converting Enzyme 2 (ACE2)  receptors on the host cell. After binding, the host cell
transmembrane proteases (TMPRLRSS2 and Furin) cut the Spike proteins, allowing the virus surface to approach
the cell membrane, fuse with it and the viral RNA enter the cell.

\vskip.2cm \item \textit{Virus proliferation:} Then, the virus hijacks the cell machinery and the cell dies releasing millions of new viruses thus generating a virus infection.  COVID-19 starts with the arrival of  \textit{SARS--CoV--2} virions to the respiratory mucosal surfaces of the nose and throat that express high levels of ACE-2 receptors on the surface .

\vskip.2cm \item \textit{Immune system actions:} When the virus manages to overcome the barrier of the  mechanisms and the mucus secreted by goblet cells from a first effective reaction, a rapid release of danger signals activates the reaction of the host's innate immunity.
Corona viruses are successful at suppressing various mechanisms in an immune response, but a  protective immunity can be however developed.

\end{itemize}

This brief description induces a concept  of the \textit{amount of infection} roughly related to the quantity of initial virus corresponding to the initial infection. Therefore, it is wise to account for the heterogeneous level of infections and of the heterogeneous  ability of each individual to express his/her immune defence.

In addition, the modelling approach should include some ability to predict the trends  of  infected individuals towards full recovery and of those who do not succeed in surviving. The overall modelling should also be deemed to  support the strategic choice of hospitalization of infected individuals based on the estimate of the progression of the pathology.

Subsequently, the focus moves to the \textbf{contagion in space}, where it occurs by contact thus creating a population of infected individuals who move in space and propagate the epidemic in a globally connected world. The role of space  can be studied by treating the following sequential steps of the modelling approach:

\begin{enumerate}

\item \textit{Dynamics in the case of spatial homogeneity}, where the virus is transmitted from infected to healthy individuals by short-range interactions. In general contagion depends on the infectivity features of each individual and on the distances they maintain homogeneously in a crowd, namely on the local density.

\vskip.2cm \item \textit{Contagion in a crowd}, where individuals move in a spatial domain, namely with the loss of space homogeneity. Contagion depends on the local distance between individuals which depends on time and space. Contagion probability is somehow related to local density and infectivity.

\vskip.2cm \item \textit{Spatial dynamics} in exogenous  networks, where individuals  move by transportation across nodes of a physical network, while the dynamics within each node follows the dynamics outlined in the preceding Items.

\end{enumerate}

The various reasonings given above indicate that we are studying a complex system which presents multiscale features, where the microscopic scale, in short micro-scale, corresponds to virus particles and immune cells, which induces the dynamics at the higher scale of individuals who carry an infection (meso-scale), while collective behaviours are observed at the macro-scale of all individuals. As mentioned previously, an ability to develop an immune defence is heterogeneously distributed over the whole population.

Heterogeneity and multiscale features address the modelling approach to the so-called kinetic theory of active particles~\cite{[BBGO17]} which has already been applied to various branches of the so-called behavioural sciences~\cite{[Ball12],[KS19]}. For instance, the immune competition~\cite{[BD06]}, collective learning~\cite{[BD19],[BDG16]}, and the dynamics of crowds and swarms~\cite{[ALBI19],[ABGR20]}, where understanding the dynamics of the collective behaviour leads to a deeper understanding of the contagion dynamics.

The key features of the aforementioned mathematical approach, presented as a mathematical theory in~\cite{[BBGO17]}, can be summarized in the following sequence:

\begin{enumerate}

\item Subdivision of the overall system into functional subsystems, in short FSs.

\vskip.2cm \item Representation of the state of each FS by probability distributions over the micro-state of interacting entities.

\vskip.2cm \item Deriving a general mathematical framework deemed capable of capturing the complex features of living systems constituted by very many interacting entities.

\vskip.2cm \item Modelling interactions and deriving specific models by inserting these models into the said mathematical framework.

\vskip.2cm \item Simulation of the dynamics at the micro-scale by computing macroscopic quantities by weighted moments of the probability distribution functions mentioned in Item 2.
\end{enumerate}

An important step of the overall approach is the modelling of interactions which needs a multidisciplinary vision as it links knowledge in biology to advanced mathematical methods as interactions are, generally, nonlinearly additive, nonlocal, and stochastic. This rationale, once transferred into a mathematical framework, goes well beyond classical SIR models. In addition, the problem is multiscale as each individual is a carrier of an internal competition, at the micro-scale, between the virus and the host immune system.

Indeed, it is a broad vision although we acknowledge that recent developments, for instance~\cite{[LMSW20]} have effectively contributed to a deeper understanding of the COVID-19 pandemic.

\section{Multiscale modelling from particles to populations}\label{sec:3}

We consider a population of a large number of individuals, where a small fraction is infected and transfers the virus to healthy individuals by short range interactions. In general, contagion depends on the frequency of contacts, on the level of the infection within each individual, and on the level of physical  protection used by individuals aware of the risk of contagion. The frequency can be related to the distance, generally called \textit{social distance}, while the dynamics within each individual includes a competition between the virus and the immune system. Figure~1  shows how, in a crowd, the social distance can differ due to local aggregations. Individuals move according to certain walking strategies~\cite{[BGO19]}, while a special case (useful nonetheless to model the dynamics of contagion) refers to a crowd  homogeneously distributed in space.
\begin{figure}\label{het-crowds}
\begin{center}
\includegraphics[width=0.7\textwidth]{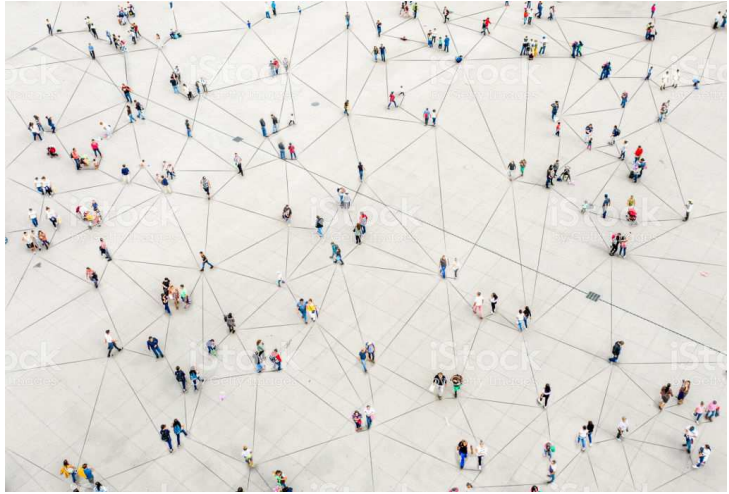}
\end{center}
\begin{center}
\caption{A crowd with aggregation of multiple groups.}
\end{center}
\end{figure}

This section is devoted to developing a modelling approach suitable to describe the dynamics over time and space of the whole population which includes healthy, infected, and recovered individuals. The model is also required to describe the growing number of those who could not recover  and therefore die. The study is developed along some sequential steps devoted to the following topics:  Modelling contagion under the assumption of spatial homogeneity; Simulations of the contagion dynamics linked to individual competition against the undesired host virus;  Modelling the contagion dynamics in complex spatial environments. These topics are treated in the next subsections, while the last subsection is devoted to a critical analysis and further developments of the modelling strategy.

\subsection{Modelling contagion, progression and recovery under the assumption of spatial homogeneity}

We consider a population consisting of $N_0$ individuals homogeneously distributed in space. A small number of them $\ve\, N_0$ is initially infected, while  $(1 - \ve)\, N_0$ is considered healthy. The general framework supporting the modelling approach is defined by a selection of key features somehow related to the general description given in Section 2 with the additional aid of the technical report~\cite{[CFM20]}. We suppose that modelling contagion dynamics, followed by the competition for survival within each individual, can be developed according to the following rationale:
\begin{enumerate}

\item Individuals are viewed  as  \textit{active particles}, in short a-particles, namely particles which are carriers of an internal state, called \textit{activity}. The level of infection  of each  a-particle can progress in time due to a prevalence of the virus aggressiveness over the immune defence or can regress due to the successful actions of the immune defence.  The ability to express an individual immune defence is heterogeneously distributed throughout the population.

\vskip.2cm \item Contagion depends on the level of the infection as well as on the physical distance between individuals which is a constant parameter in the case of spatial homogeneity.

\vskip.2cm \item Dynamics within each individual transmission depends on the competition between a proliferative virus and the immune system. Hence, the dynamics involve individuals which are carriers of an internal dynamics.

\end{enumerate}
In general, medical actions either to weaken the virus or to activate the immune system can contribute to drive the competition towards full recovery. On the other hand, the model might consider both the dynamics induced by isolation and death of individuals whenever the action of the virus has a gain over the immune defence. Additional features of the modelling approach are:

\vskip.2cm \noindent -- The overall population is subdivided into four sub-populations labeled by the subscripts $i =1,2,3,4$. The abbreviation $i$-FS is used to denote the $i$-th population viewed as a functional subsystem.

\vskip.2cm -- \noindent The micro-state of a-particles includes two variables $u \in [0,1]$ and $w \in [0,1]$ corresponding, respectively, to the progression of virus invasion and to the level of activation of the immune defence. In this sense, $u=0$ represents the absence of the viral infection, while $u>0$ characterizes the presence of the disease, where increasing values of $u$ towards $1$ correspond to more aggressive states. Similarly, $w=0$ and $w=1$ correspond, respectively, to the lowest and highest immune system activation. If discrete variables are used one has
$$
\bu = \{u_1=0, \ldots, u_j= \frac{j-1}{m-1}, \ldots, u_m=1\},
$$
and
$$\bw = \{w_1=0, \ldots, w_k=\frac{k-1}{n-1} , \ldots, w_n=1\},
$$
where $u_1 = 0$ corresponds to the healthy level and $w_1 = 0$  to the lack of immune defence.

\vskip.1cm  The representation of each population is given by a probability distribution over the microscopic state, where the use of discrete  variables is suggested by the difficulty to achieve a detailed measure of their continuous properties, while finite sets allow one to identify or cluster groups of a-particles. The following subdivision into FSs and representation is proposed:
\begin{itemize}

\item $i=1$: Corresponds to healthy individuals  with distribution  $f_1^{1,k}(t, u_1, w_k)$, where  $t$ is the time belonging to the interval $[0,T]$.

\vskip.1cm \item $i=2$: Corresponds to infected individuals with distribution $f_2^{j,k}(t,  u_{j}, w_k)$, with $1<j<m$.

\vskip.1cm \item $i=3$: Corresponds to individuals recovered from the infection with distribution  $f_3(t)$, namely infected individuals that succeed in reaching back to the state $j=1$. Here, the dependence on $w_k$ is not any longer relevant with respect to the specific dynamics under consideration.

\vskip.1cm \item $i=4$: $f_4(t)$ is the number of individuals of the infected population who do not succeed to recover, that are infected individuals who reach the state $j=m$. Here again, the dependence on $w_k$ is no longer relevant with respect to the specific dynamics under consideration.

\end{itemize}

Therefore, an individual with micro-state $(u_j,w_k)$ can be either healthy ($i=1$) if $j=1$, or infected ($i=2$) if $2\leq j < m$. If an individual from 2-FS reaches the states $(u_1,w_k)$ or  $(u_m,w_k)$, then he/she recovers ($i=3$) or dies ($i=4$), respectively.

\vskip.1cm

The time-dynamics of the four FSs can be obtained by using the general mathematical structures of the kinetic theory of active particles, see Chapter 5 in~\cite{[BBGO17]},  and subsequently by inserting into these structures appropriate models of interactions. However, some further developments of the theory are necessary to account for the multiscale features of the class of systems under consideration. Therefore, a general structure is reported as it provides the conceptual basis for these developments.
\begin{eqnarray}\label{S}
{} && \frac{d}{dt} f_{ij}^r = G_{ij}^r(\bof) -  L_{ij}^r(\bof) \nonumber\\[3mm]
{} && \hskip1cm  = \sum_{s=1}^m \sum_{h,k,p,q=1}^n \eta_{hk}^{pq}(r,s)(\bof)\cA_{hk}^{pq}(hk \to ij)(\bof) f_{hk}^r \, f_{pq}^s    \nonumber\\[3mm]
{} && \hskip2cm -  f_{ij}^r  \sum_{s=1}^m \sum_{p,q=1}^n   \ \eta_{ij}^{pq}(\bof)\,f_{pq}^s,
\end{eqnarray}
where (\ref{S}) refers to the dynamics of a large number of active particles whose state is defined by the probability $f_{ij}^r$ to find an active particle from the $r$ functional subsystem with micro-state $ij$. The subscripts $h,k$ and $p,q$ denote the micro-states corresponding to the $r,s$  FSs which by interactions lead to the dynamics of $f^r$. In addition, $\eta_{hk}^{pq}$,  $\eta_{ij}^{pq}$, denote  the interaction rates, and $\mathcal{A}_{hk}^{pq}$ the transition rate into the micro-state $i,j$ of the r-FS. The time dynamics are then ruled by a \textit{gain} term of particles which at time $t$ gain the state $(i,j)$ and a \textit{loss} term related to particles which lose such a state.

\vskip.1cm

The modelling of interactions can be developed according to the following assumptions:

\begin{enumerate}

\item $i=1$: Active particles belonging to 1-FS interact with a-particles from 2-FS and can become, with some probability, infected. The rate of infection depends on the physical interaction rate $\eta_0$, supposed to be constant, and to the level of progression $u_j$ of the infected individuals as the probability of infection grows with $u_j$.

\vskip.2cm \item $i=1,2$: The interaction rate depends on the social distance. Interactions do not modify the levels of the immune defence, while particles which move from 1-FS to 2-FS  take the value $u_2$ and start their competition to survive the attack from the immune system. The model allows the study of different scenarios  corresponding to various possible situations.

\vskip.2cm \item $i=2$: The dynamics within 2-FS are induced by the competition between progressing particles of the virus and the immune defence. Viral particles progress (proliferate) thanks to foraging of the surrounding tissues, while the immune defence counteracts the progression by inducing a regression. Interactions occur with a constant interaction rate $\mu_0 \not= \eta_0$. In addition, the dynamics should account for the inflow from 1-FS and outflows from 2-FS to 3-FS and 4-FS.

\vskip.2cm \item $i=2,3,4$: A-particles from 2-FS move to 3-FS if the immune defence succeeds to obtain a regression down to $u_1$. A-particles from 2-FS move to 4-FS if the immune defence does not succeed to obtain a regression and the virus progression reaches the highest value $u_m$. A-particles recovered from the infection are not subject to a new infection.

\end{enumerate}

The dynamics described in the Items above are shown in Figure~\ref{transfer_diagram}, while Figure~\ref{inside} depicts the progression and competition dynamics within each individual, corresponding to the block $i=2$ which is red as it specifically refers to the low scale of cells and virus particles, while the other blocks which are blue, correspond to individuals.


\begin{figure}
\tikzstyle{decision} = [diamond, draw, fill=blue!20,
    text width=4.5em, text badly centered, node distance=3cm, inner sep=0pt]
\tikzstyle{block1} = [rectangle, draw, fill=blue!20,
    text width=5em, text centered, rounded corners, minimum height=4em]
    \tikzstyle{block2} = [rectangle, draw, fill=red!60,
    text width=5em, text centered, rounded corners, minimum height=4em]
\tikzstyle{line} = [draw, -latex']
\tikzstyle{cloud} = [draw, ellipse,fill=red!20, node distance=3cm,
    minimum height=2em]
\begin{center}
\begin{tikzpicture}[node distance = 3cm, auto]
    \node [block2] (FS2) {Infected \\ $i=2$};
    \node [block1, left of=FS2] (FS1) {Healthy \\ $i=1$};
    \node [block1, above right of=FS2] (FS3) {Recovered \\ $i=3$};
    \node [block1, below right of=FS2] (FS4) {Dead \\ $i=4$};
    \path [line] (FS1) -- node  {} (FS2);
    \path [line] (FS2) |- node  {} (FS3);
    \path [line] (FS2) |- node  {} (FS4);
\end{tikzpicture}
\end{center}
\caption{Transfer diagram of the model.  Boxes represent functional subsystems and arrows indicate transition of individuals.} \label{transfer_diagram}
\end{figure}
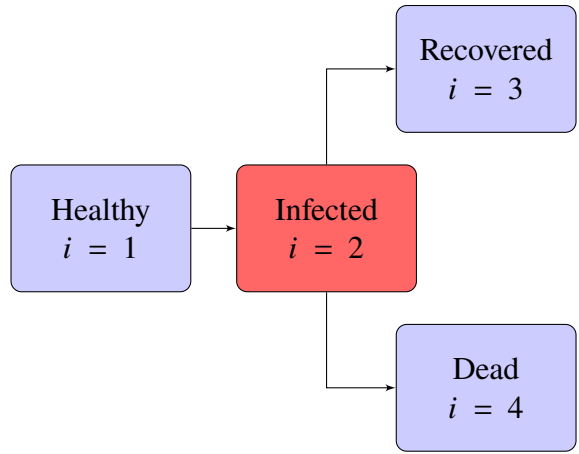


\begin{figure}
\begin{center}
\begin{tikzpicture}[>=stealth',auto,node distance=1.5cm,
  thick,main node/.style={circle,draw,font=\sffamily\Large\bfseries}]
  \tikzstyle{block1} = [rectangle, draw,
    text width=5em, text centered, rounded corners, minimum height=1em]
\draw (0,0) -- (10,0) ; 
\foreach \x in  {0,1,2,3,4,5,6,7,8,9,10} 
\draw[shift={(\x,0)},color=black] (0pt,3pt) -- (0pt,-3pt);
\foreach \x in {0} 
\draw[shift={(\x,0)},color=black] (0pt,0pt) -- (0pt,-3pt) node[below] (Rec)
{$u_{1}$};
\node [block1,below of=Rec]  {Recovered};
\foreach \x in {1} 
\draw[shift={(\x,0)},color=black] (0pt,0pt) -- (0pt,-3pt) node[below] (Hea)
{$u_{2}$};
\node [block1,above of=Hea]  {Healthy};
\foreach \x in {6} 
\draw[shift={(\x,0)},color=black] (0pt,0pt) -- (0pt,-3pt) node[below]
{$u_{j+1}$};
\foreach \x in {4} 
\draw[shift={(\x,0)},color=black] (0pt,0pt) -- (0pt,-3pt) node[below]
{$u_{j-1}$};
\foreach \x in {5} 
\draw[shift={(\x,0)},color=black] (0pt,0pt) -- (0pt,-3pt) node[below]
{$u_{j}$};
\foreach \x in {6} 
\draw[shift={(\x,0)},color=black] (0pt,0pt) -- (0pt,-3pt) node[below]
{$u_{j+1}$};
\foreach \x in {9} 
\draw[shift={(\x,0)},color=black] (0pt,0pt) -- (0pt,-3pt) node[below]
{$u_{m-1}$};
\foreach \x in {10} 
\draw[shift={(\x,0)},color=black] (0pt,0pt) -- (0pt,-3pt) node[below] (Dea)
{$u_{m}$};
\node [block1,below of=Dea]  {Dead};
\foreach \x in {5.5} 
\draw[shift={(\x,.3)},color=black]  node[above]
{$\beta\,u_j$};
\foreach \x in {4.3} 
\draw[shift={(\x,.35)},color=black]  node[above]
{$\gamma\,w_k$};
\draw[*-] (4.92,0) -- (6.08,0);
\draw[-*] (3.92,0) -- (5.08,0);
\draw [->,red] (5,0) to [out=90,in=90] (6,0);
\draw [->,red] (5,0) to [out=90,in=90] (4,0);
\draw [->,red] (1,0) to [out=90,in=90] (0,0);
\draw [->,red] (9,0) to [out=90,in=90] (10,0);
\draw [->] (10,-.6) to [out=90,in=90] (10,-1.5);
\draw [->] (0,-.6) to [out=90,in=90] (0,-1.5);
\draw [->] (1,.8) to [out=90,in=90] (1,0);
\end{tikzpicture}
\end{center}
\caption{Dynamics within infected population.}
\label{inside}
\end{figure}
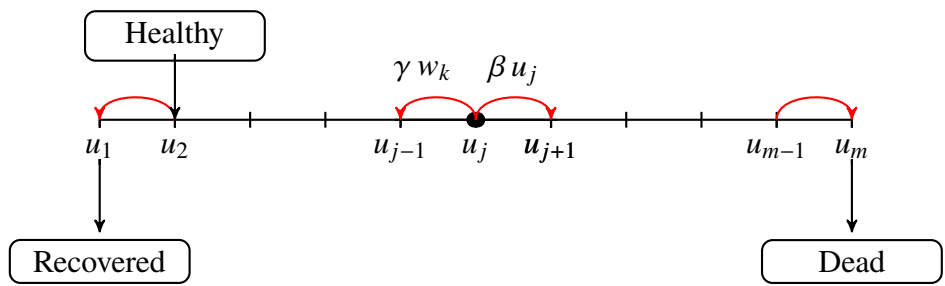


The technical difference, with respect to (\ref{S}), is that interactions involving subsystems 1, 3 and 4  only refer to the scale of individuals, namely a-particles are human beings, while interactions within 2-FS refer to micro-entities, namely virus particles and immune cells, and interactions between 2-FS and the other FS are mixed as they involve two scales. In addition, if birth and death rates due to other causes  are neglected, then the sum of the number size of the four populations is equal to $N_0$ for every time $t$.

The model, which is presented in the following, takes into account only  interactions which have an influence on the dynamics of the overall system, namely transitions from 1-FS to 2-FS, within 2-FS, and transitions from 2-FS into 3-FS or 4-FS.  Then, simulations should put in evidence different scenarios depending on different distributions corresponding to the levels of the defence ability of the immune system.  Possible developments of the model are discussed in the last subsection to account for additional features which are not included in this model.

Bearing in mind all the above, let us consider, in sequence, the dynamics related to each FS.

\vskip.2cm  \noindent \textbf{Infection dynamics:} A healthy individual from 1-FS with state $u_1$ interacts, with rate $\eta_0$, with an infected individual from 2-FS with state $u_j, \, j >1$, and becomes infected with a probability  which depends on a parameter $\alpha^*$ and on the level of infection of the  individual from 2-FS. The dynamics refers to $f_1^{1,k}(t)$ and it is governed only by the loss term:
\begin{equation}\label{1-FS-loss}
\partial_t f_1^{1,k}(t) = - L_{1}^k(t) = -  \, \sum_{s=1}^n  \sum_{j=2}^{m-1}  \alpha \, u_j \, f_1^{1,k}(t) \, f_2^{j,s}(t),
\end{equation}
for $k = 1, \ldots, n$ and $\alpha = \eta_0 \, \alpha^*$. The interaction rate $\eta_0$ depends on the local density of individuals. It grows with increasing value of the density thus reaching the maximal value in the case of maximal packing density (a technical value is 6-7 individuals per square metre). The use of the subscript $0$ refers to the spatially homogeneous case, where the density is the same throughout the spatial domain. Steady cases, but with different crowding levels, as shown in Figure 1, can be taken into account by a space-dependent interaction rate so that
 $\alpha = \alpha(\bx)$, where $\bx$ denotes the position. Further developments are discussed in the last subsection.

\vskip.2cm \noindent \textbf{Dynamics of infected individuals:} Each infected individual is the carrier of a struggle between virus particles and immune system. The virus takes advantage from the foraging of surrounding tissues and increases its micro-state from each $j$-level to the higher $j+1$-level depending on a parameter $\beta^*$ and on the $j$-th level. The immune system acts to decrease the  $j$-level to the lower $j-1$-level depending on a parameter $\gamma^*$ and on the $k$-th level. Individuals, whose virus progression levels reach the values $u_1$ and  $u_m$ move, respectively, to 3-FS and 4-FS. The dynamics refers to $f_2^{j,k}(t)$ and it is governed by both gain and  loss terms:
\begin{equation}\label{2-FS}
\partial_t f_2^{j,k}(t) =  G_{2}^{j,k}(t) - L_{2}^{j,k}(t),
\end{equation}
Detailed calculations yield:
\begin{equation}\label{2-FS-gain}
 G_{2}^{j,k}(t) = \delta_{2j}\,  L_{1}^k(t) +  \beta \, u_{j-1} \,  f_2^{j-1,k}(t) +   \gamma \, w_k \, f_2^{j+1,k}(t),
\end{equation}
for $j = 2,\ldots,m-1$ and $k = 1, \ldots, n$, $\beta = \mu_0 \, \beta^*$,  $\gamma = \mu_0 \, \gamma^*$, $\delta$ is the Kronecker delta function, and where the three terms correspond, respectively, to inflow from 1-FS, progression of the virus and regression of the virus due to the immune action. In addition
\begin{equation} \label{2-FS-loss}
 L_{2}^{j,k}(t) =   \beta \, u_{j} \,  f_2^{j,k}(t) +   \gamma \, w_k \, f_2^{j,k}(t)
 \end{equation}
 for $j = 2,\ldots,m-1$ and $k = 1, \ldots, n$, while the two terms correspond to the outflow of recovered persons who move into 3-FS and dead individuals who move into 4-FS.

\vskip.2cm  \noindent  \textbf{Dynamics of recovered individuals:} The dynamics are caused by the inflow from 2-FS into 3-FS:
\begin{equation}\label{3-FS-gain}
\partial_t f_3(t) =    \gamma \,\sum_{k=1}^n w_k \, f_2^{2,k}(t).
\end{equation}

\vskip.2cm  \noindent \textbf{Dynamics of dead individuals:} The dynamics are caused by the inflow from 2-FS into 4-FS:
\begin{equation}\label{4-FS-gain}
\partial_t f_4(t) =  \beta \, u_{m-1} \,  \sum_{k=1}^n f_2^{m-1,k}(t).
\end{equation}
\vskip.2cm
Collecting all equations of the differential system yields
\begin{equation}\label{swarm-s}
\begin{cases}
 \displaystyle \partial_t f_1^{1,k}(t) = - \alpha \, \sum_{s=1}^n  \sum_{j=2}^{m-1} \, u_j \, f_1^{1,k}(t) \, f_2^{j,s}(t),\\[5mm]
 \displaystyle \partial_t f_2^{j,k}(t) = \alpha \, \sum_{s=1}^n  \sum_{j=2}^{m-1} \, u_j \, f_1^{1,k}(t) \, f_2^{j,s}(t)\, \delta_{2j}\, + \beta  u_{j-1} \,  f_2^{j-1,k}(t) \\[3mm]
 {} \hskip1cm  + \gamma \, w_k \, f_2^{j+1,k}(t) - \beta\, u_{j} \,  f_2^{j,k}(t) - \gamma \, w_k \, f_2^{j,k}(t),\\[5mm]
 \displaystyle \partial_t f_3(t) =  \gamma \,\sum_{k=1}^n w_k \, f_2^{2,k}(t), \\[5mm]
  \displaystyle \partial_t f_4(t) = \beta\,  u_{m-1} \,  \sum_{k=1}^n f_2^{m-1,k}(t),
\end{cases}
\end{equation}
where for $(\ref{swarm-s})_1$  one has $k = 1, \ldots, n$, while for $(\ref{swarm-s})_2$ one has $j = 2,\ldots,m-1$ and $k = 1, \ldots, n$.
\vskip.2cm
It can be rapidly verified that model (\ref{swarm-s}) is a particular case of (\ref{S}) corresponding to specific models of interactions. It is the same at both scales and accounts for heterogeneity at each scale.

\subsection{Simulations of selected case studies}

A mathematical model has been proposed in the preceding subsection, with the rationale behind the model being proposed and described in Subsection 3.1, and the technical details given in Subsection 3.2. We do not naively claim that the complex biological dynamics have been exhaustively captured by our model, but we remark that it includes some important features such as the  ability of the immune system which is heterogeneously distributed throughout the population.

Important features still need to be considered, for instance, modelling the spatial dynamics, typical of crowds, which introduce a dynamically varying heterogeneity in the contagion dynamics. Therefore, a more detailed study of the modelling developments is postponed firstly to the next subsection mainly to account for the role of spatial dynamics and in Section 7 to account for the hints of Sections 4, 5, and 6 devoted to virology, immune competition and economics.  Some sample simulations are presented in this subsection simply to enlighten the role of the parameters of the model.

Let us study the behaviour of the system corresponding to different choices of the parameters.  We consider, at time $t=0$, only one infected individual in a population of $10$ million people ($\varepsilon=10^{-7}$). Simulations are developed for $m=n=5$, $\beta = 0.1$ and $ \gamma=0.2$, while three values of $\alpha=0.4, 0.3, 0.25$, are considered.

\begin{figure}[ht!]
\includegraphics[width=0.45\textwidth]{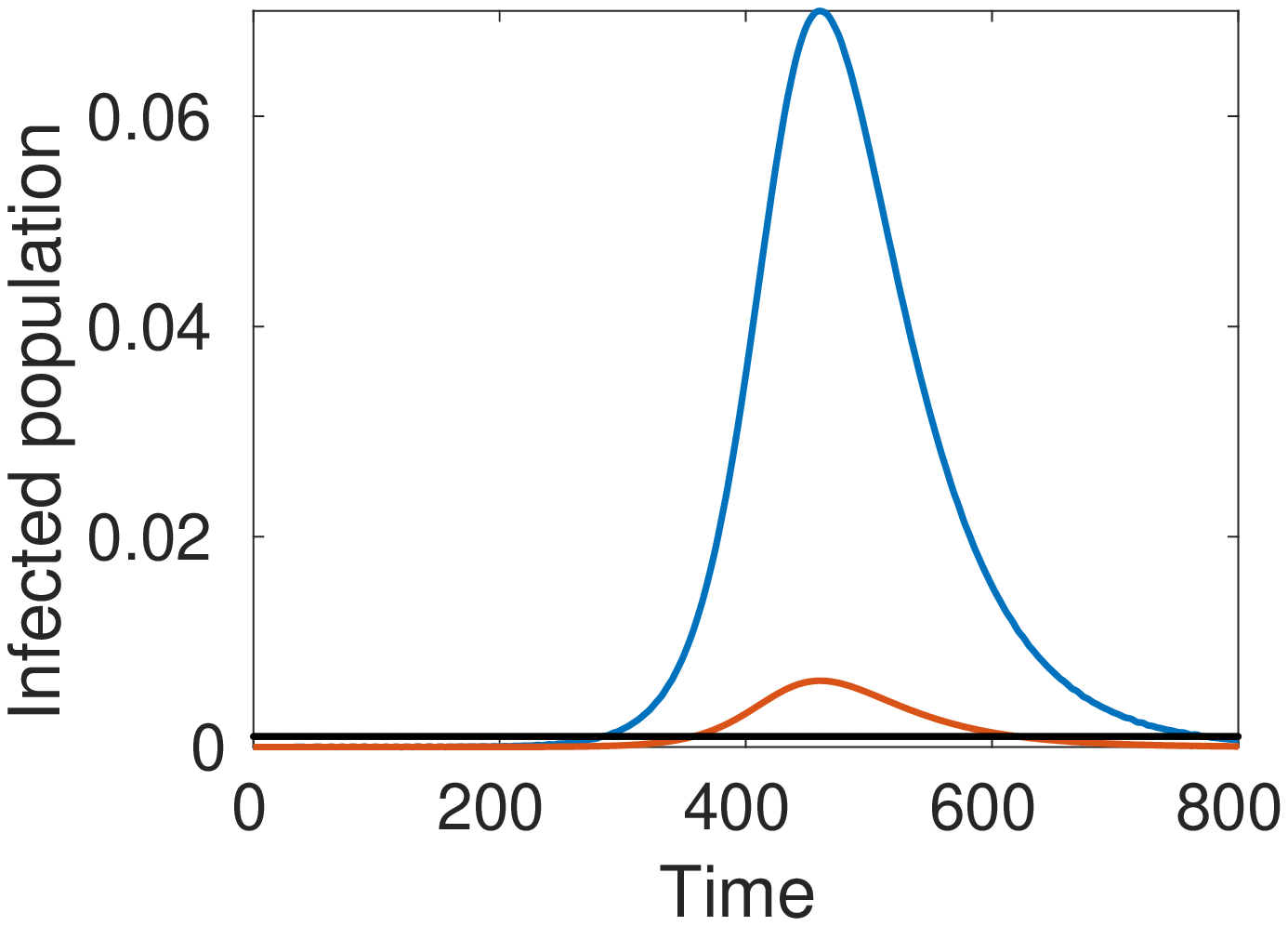} \qquad
\includegraphics[width=0.45\textwidth]{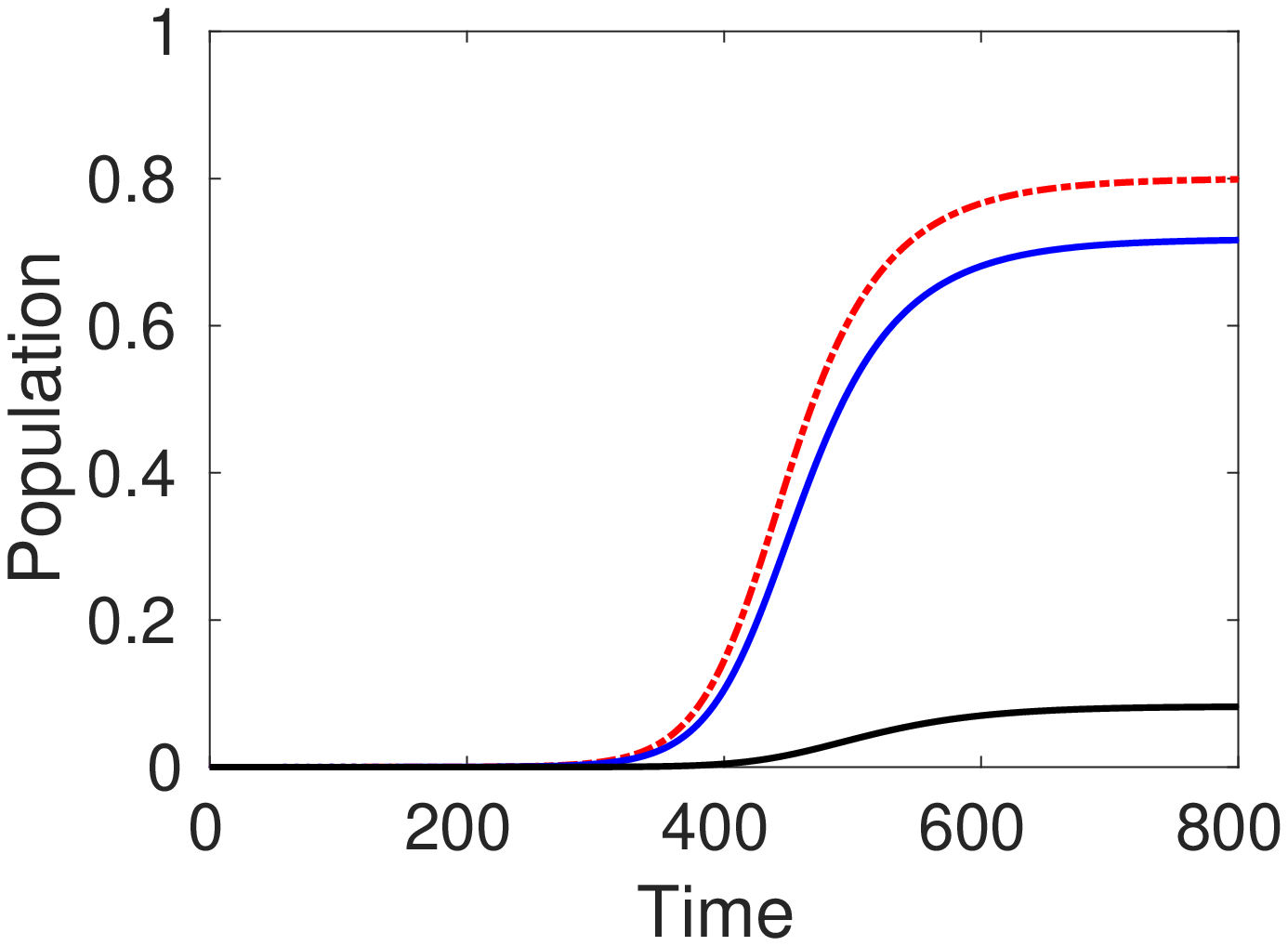}
\caption{$\alpha = 0.4$. Left: Total number of active cases (blue), active cases requiring hospitalization (red) and the number of available beds (black). Right: Cumulative infected population (red), recovered (blue) and dead (black) vs time.  } \label{varying-alfa1}
\end{figure}

\begin{figure}[ht!]
\includegraphics[width=0.45\textwidth]{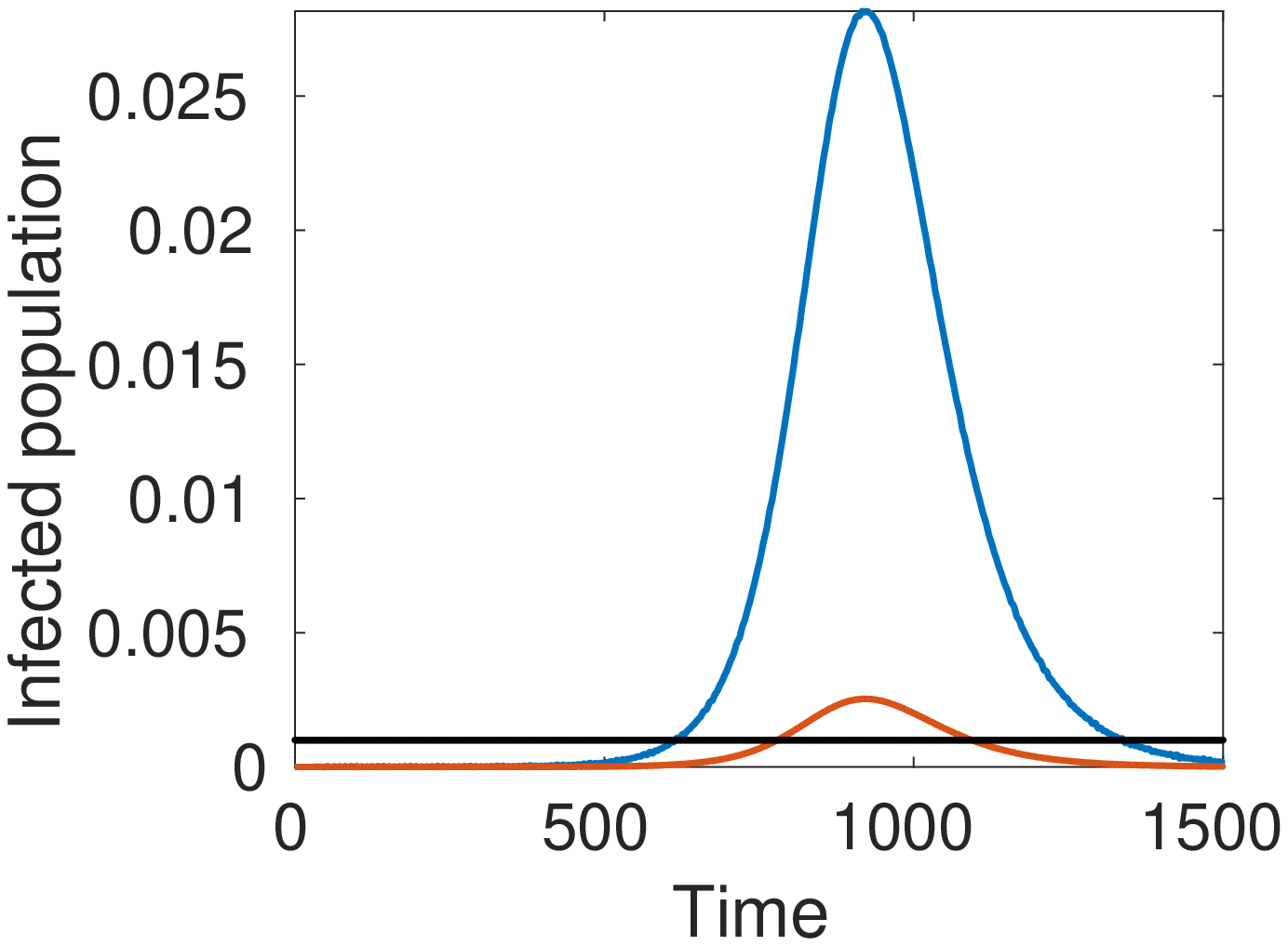} \qquad
\includegraphics[width=0.45\textwidth]{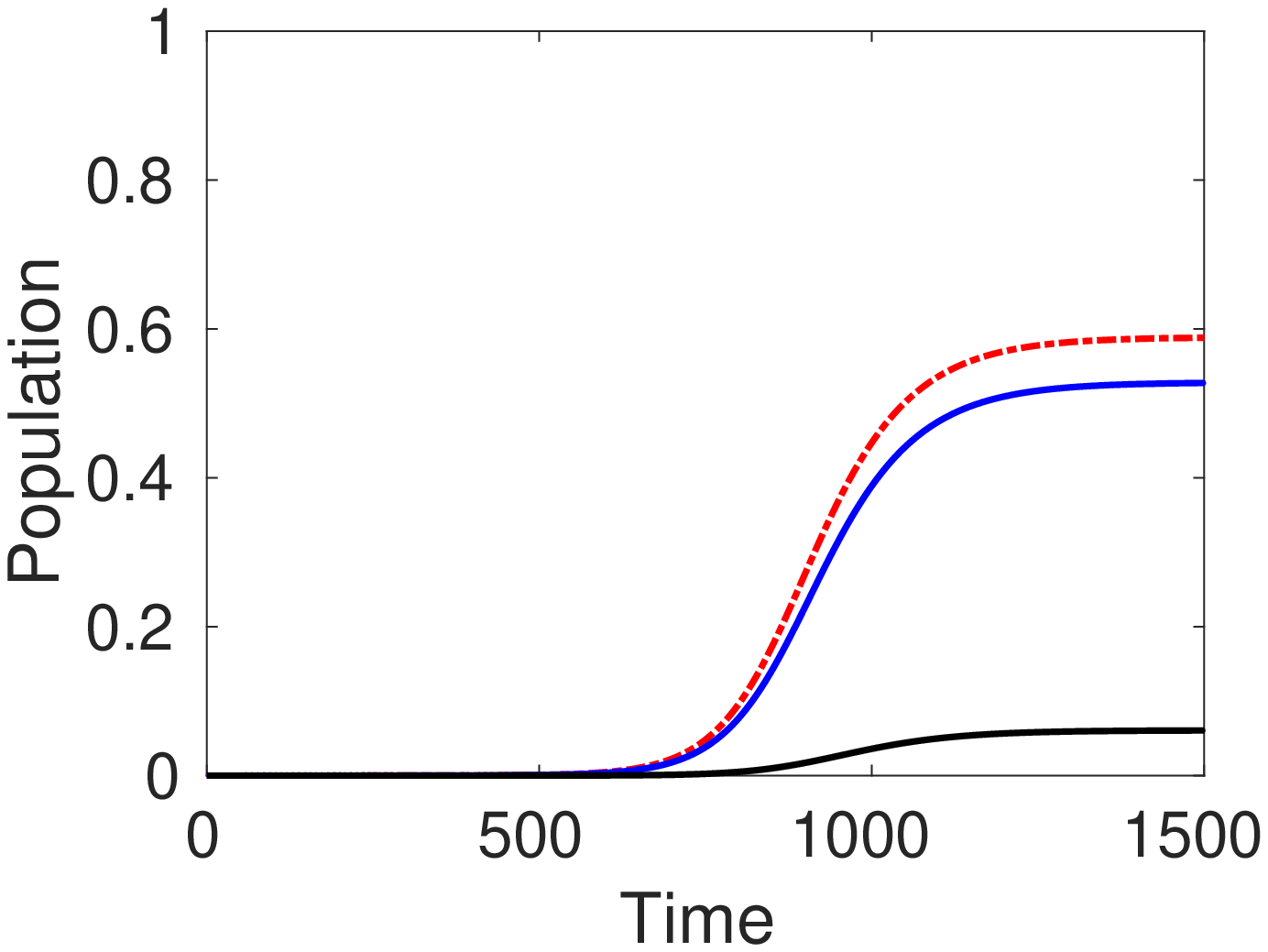}
\caption{$\alpha = 0.3$. Left: Total number of active cases (blue), active cases requiring hospitalization (red) and the number of available beds (black). Right: Cumulative infected population (red), recovered (blue) and dead (black) vs time. } \label{varying-alfa2}
\end{figure}

\begin{figure}[ht!]
\includegraphics[width=0.45\textwidth]{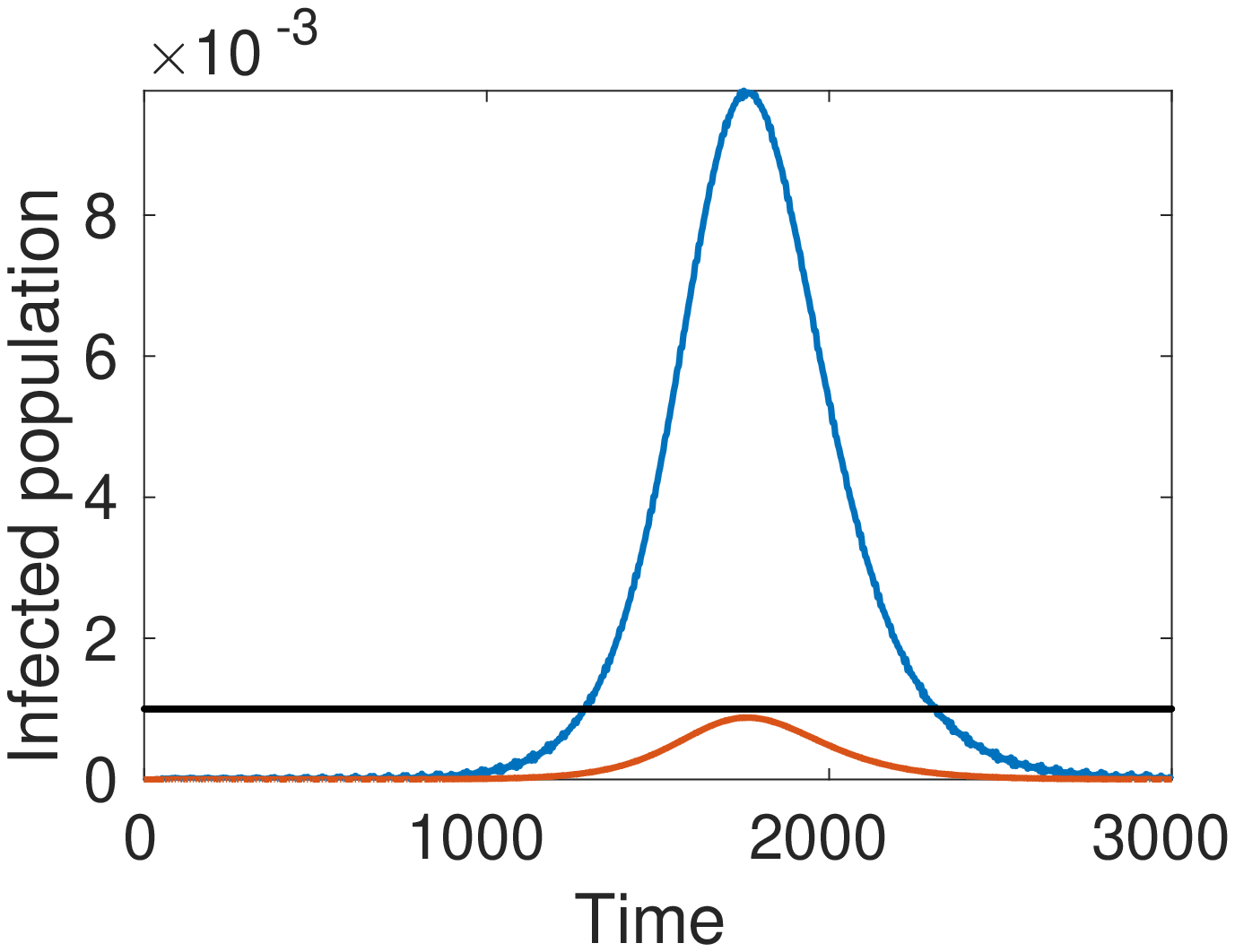} \qquad
\includegraphics[width=0.45\textwidth]{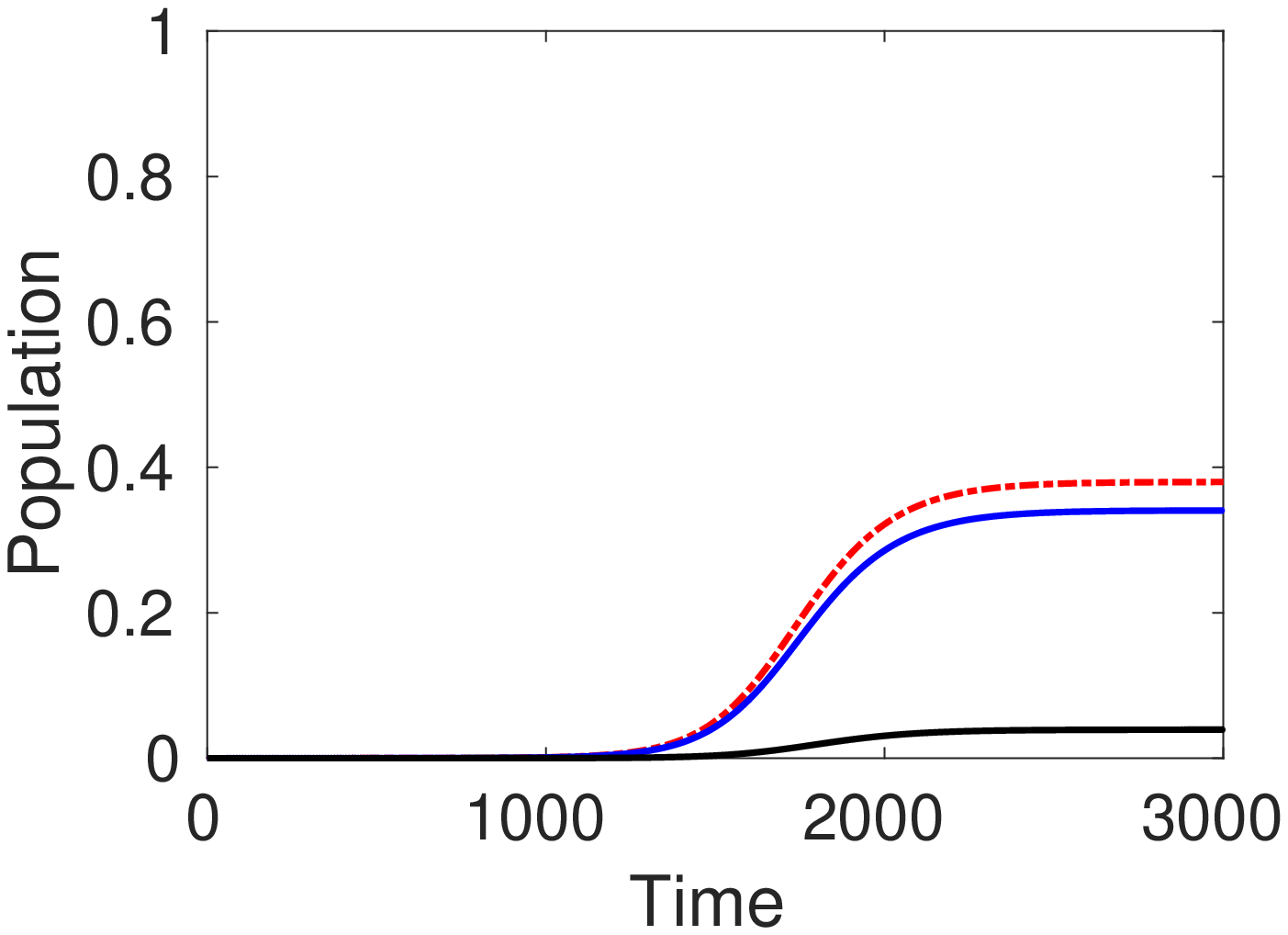}
\caption{$\alpha = 0.25$. Left: Total number of active cases (blue), active cases requiring hospitalization (red) and the number of available beds (black). Right: Cumulative infected population (red), recovered (blue) and dead (black) vs time. } \label{varying-alfa3}
\end{figure}

As expected, simulations show that the larger $\alpha$ is, the greater is the increase in the number of infected people, while  reducing interactions reduces contagion, see Figures~\ref{varying-alfa1}-\ref{varying-alfa3}.

An additional key factor considered in the simulations is the pressure on the local health-care system. The red line representing active cases requiring ICU admission can be compared with the black line representing the number of available ICU beds. For instance, recent data shows that, in Italy, about 12\% of  \textit{SARS--CoV--2} positive cases required ICU admission and that, if in practice about 1 per thousand of the patients are infected at the same time, then the total ICU capacity of the country would be saturated, as  reported in \cite{[CFM20]}.

\begin{figure}[ht!]
\includegraphics[width=0.45\textwidth]{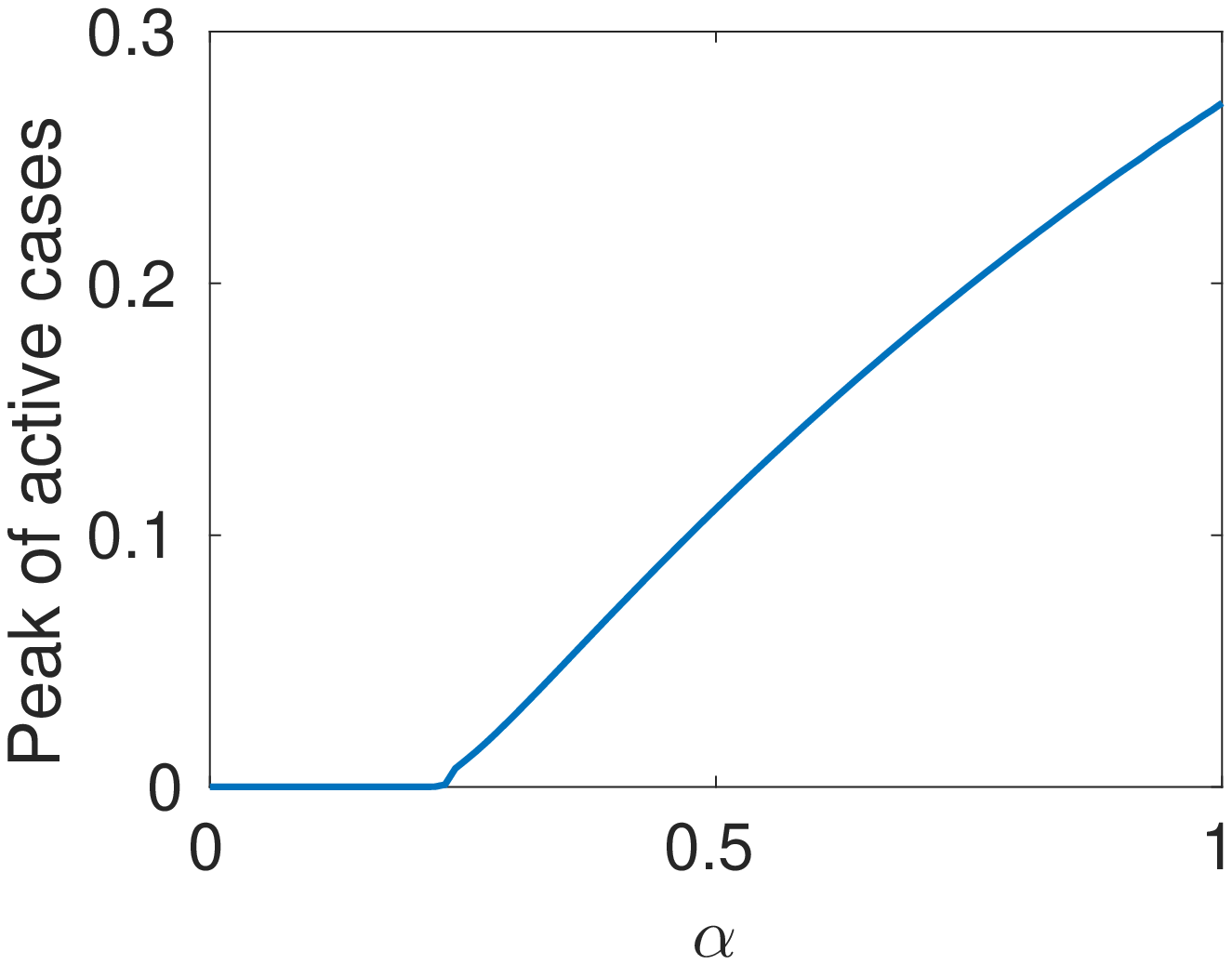} \qquad
\includegraphics[width=0.45\textwidth]{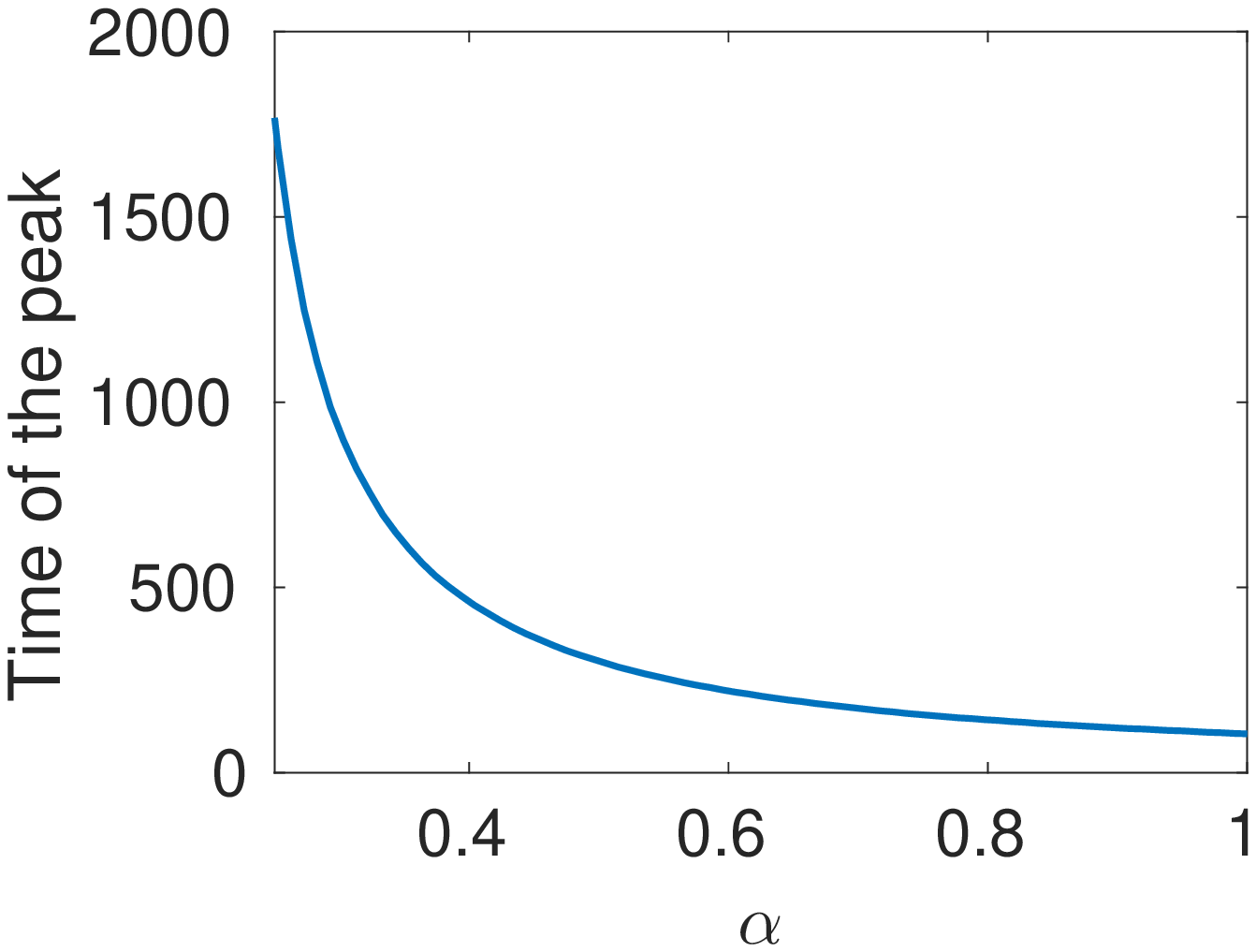}
\caption{Left: Proportion of people infected at the same time when the peak of infection is reached as a function of $\alpha$. Right: Time at which this peak occurs as function of $\alpha$} \label{varying-alfa-4}
\end{figure}

A natural question can be posed at this stage, namely how does the epidemic peak depend on model parameters?  Moreover, we argue that the epidemic can exist in a population in some equilibrium state in which the rate of newly infected people is compensated by those who recover. Figure~\ref{varying-alfa-4} shows the magnitude of the epidemic peak in the population (namely the  maximum number of people who are infected simultaneously) and the time at which this peak occurs, as a function of $\alpha$. On the left side of Figure~\ref{varying-alfa-4}, we see that the magnitude of the peak is an increasing function of $\alpha$. It is shown  that the peak could even reach a level of almost the $30\%$ of the whole population when $\alpha$ is close to $1$.

On the other hand, if $\alpha$ is small enough, then an outbreak of the epidemic is prevented. As mentioned, these dynamics correspond to equal rates of infected and recovered. Additionally, the right hand plot of Figure~\ref{varying-alfa-4} shows that lower values of $\alpha$ effectively delay the peak, and social distancing can thus be considered as a strategy to prepare and equip the health system, if necessary. A deeper analysis of this feature is further treated in Section 7.

\begin{figure}[ht!]
\includegraphics[width=0.45\textwidth]{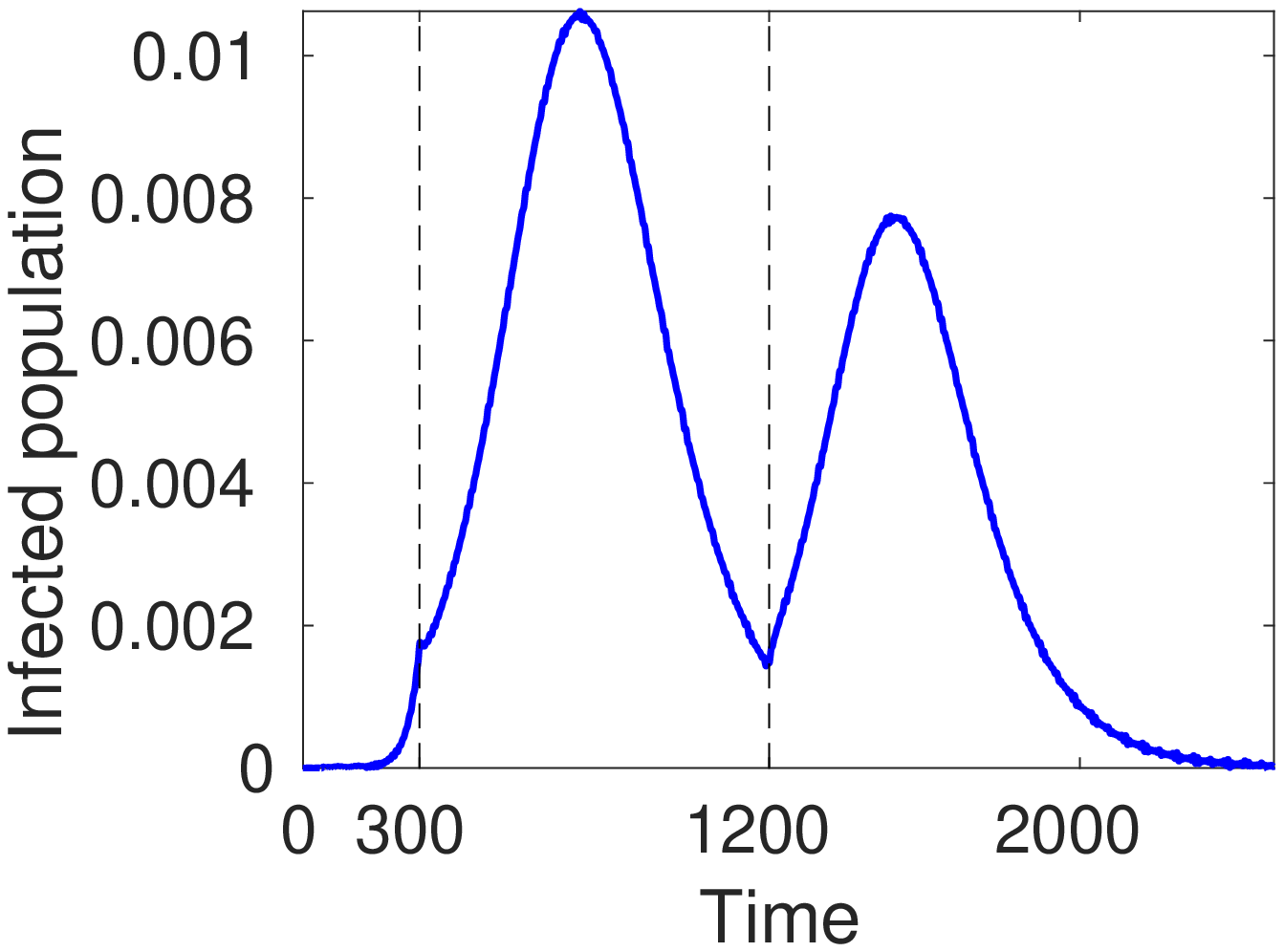} \qquad
\includegraphics[width=0.45\textwidth]{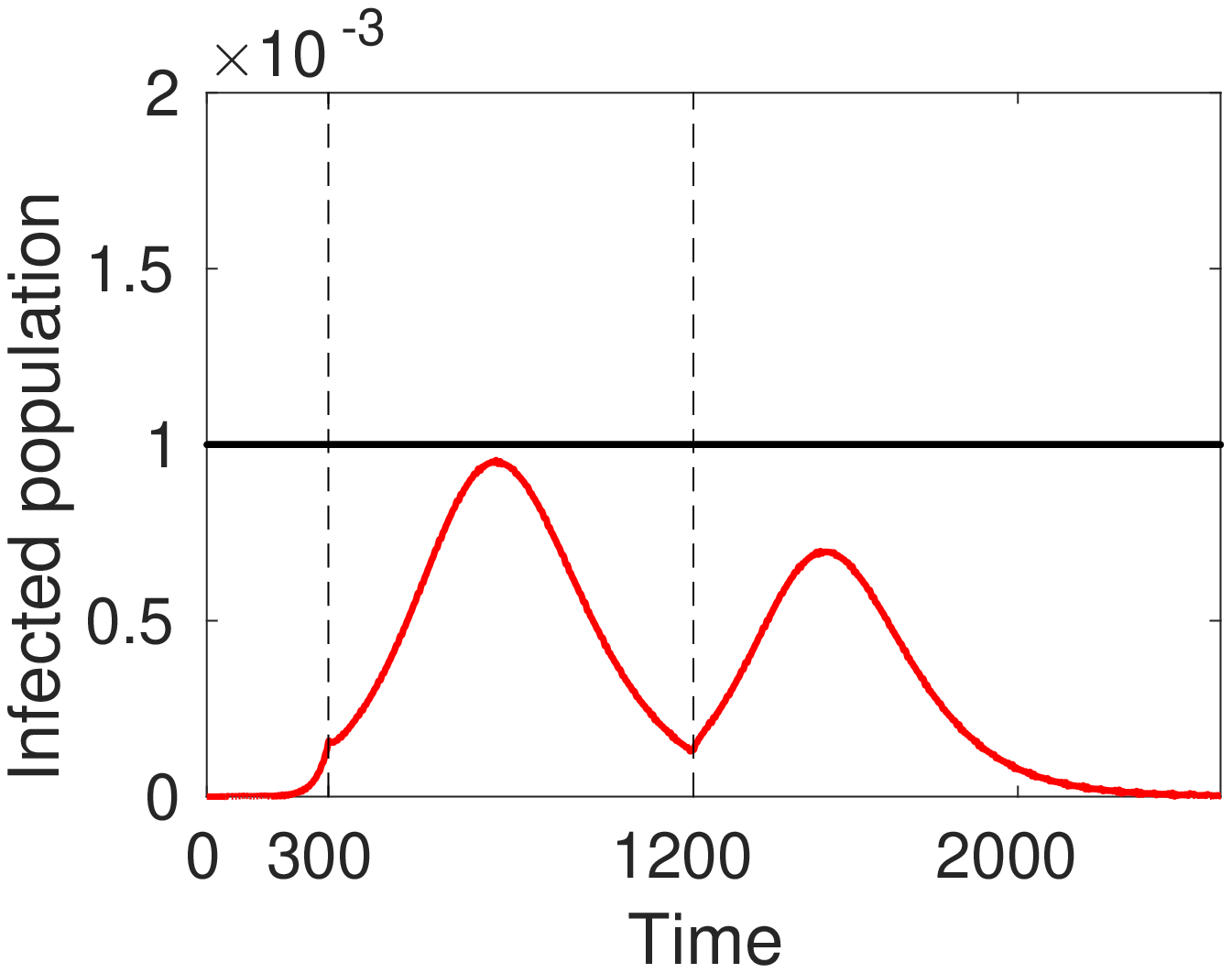}
\caption{$\alpha = 0.4$ for $t<300$, then reduced to $\alpha = 0.25$ at $t=300$, and set back to the ``normal/initial'' state at $t=1200$.    Left: Total number of active cases. Right: Estimated number of patients requiring ICU admission in relation to system capacity. } \label{social-distance-1}
\end{figure}

\begin{figure}[ht!]
\includegraphics[width=0.45\textwidth]{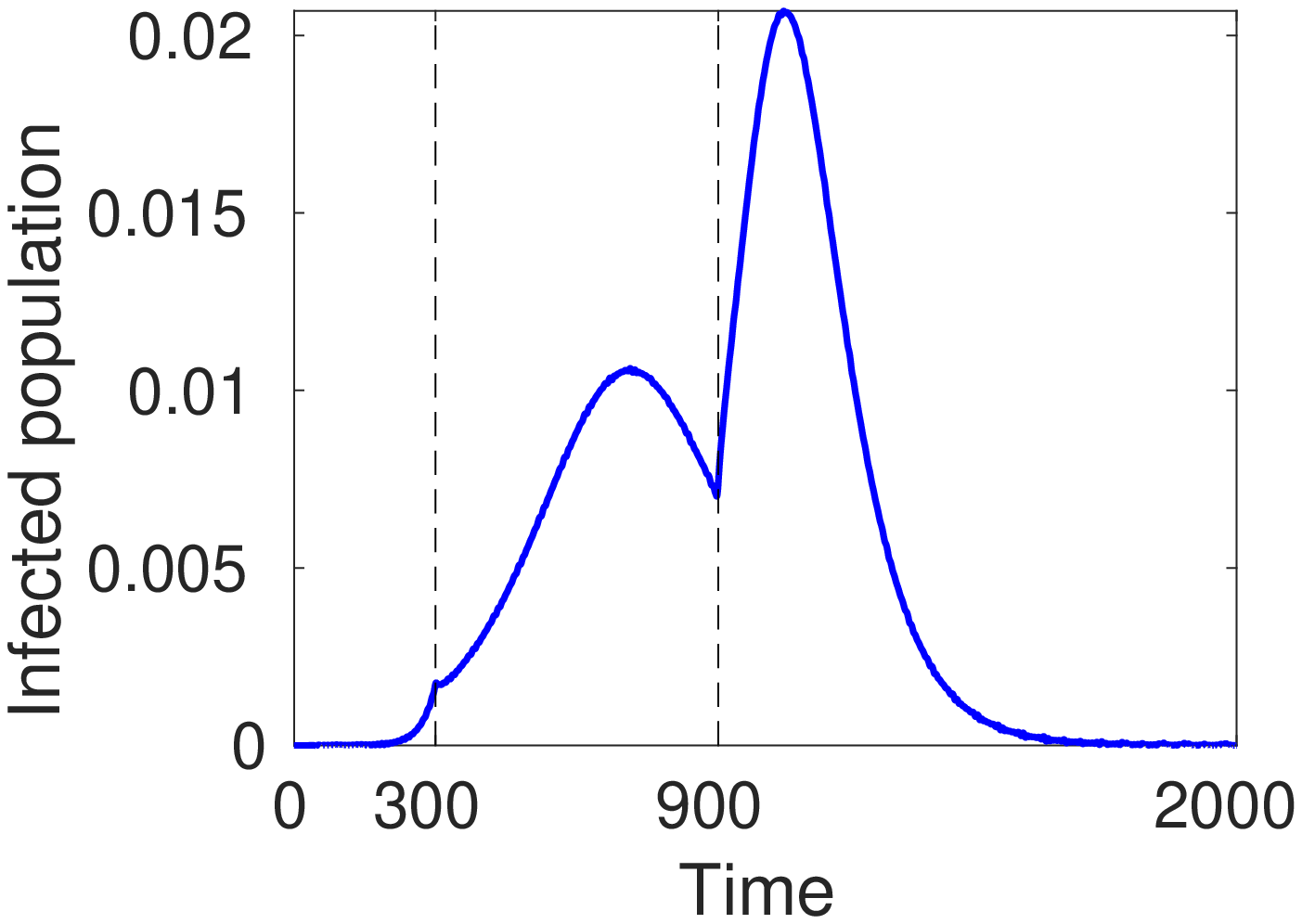} \qquad
\includegraphics[width=0.45\textwidth]{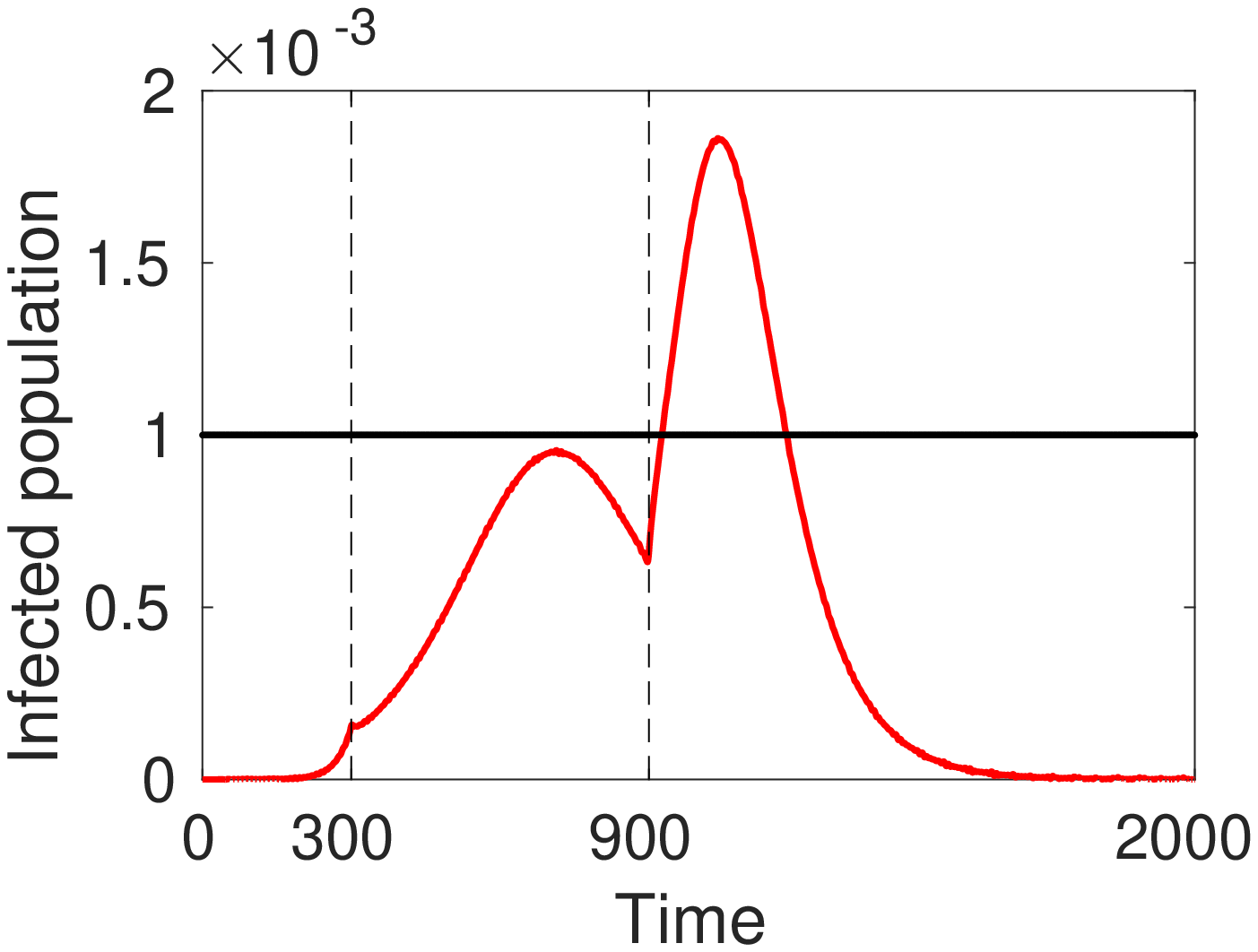}
\caption{$\alpha = 0.4$ for $t<300$, then reduced to $\alpha = 0.25$ at $t=300$, and set back to the ``normal/initial'' state at $t=900$.    Left: Total number of active cases. Right: Estimated number of patients requiring ICU admission in relation to system capacity.} \label{social-distance-2}
\end{figure}

Finally, let us study the effect of implementing social distancing at a given time. Consider for instance the dynamics shown in Figure~\ref{varying-alfa1}, where $\alpha=0.4$ for the whole time interval. Figures~\ref{social-distance-1} and \ref{social-distance-2}  show two different scenarios of implementation of social distancing. Notice that it contributes to the flattening of the infection peaks but, if not managed properly, it could be useless (for instance Figure~\ref{social-distance-2} shows that a premature lifting of social distancing can actually lead to the infection exceeding the capacity of the health system.)

Notice that the total population  can be calculated as follows:
$$
N(t) = \sum_{k=1}^n \left(f_1^{1,k}(t) + \sum_{j=2}^{m-1} f_2^{j,k}(t)\right) + f_3(t) + f_4(t),
$$
which keeps a constant value equal to $N_0$ due to the conservative nature of the transitions.

The large number of dead people (for instance almost $10\%$ of the population, see the black line in Figure~\ref{varying-alfa1} right-hand plot)  should not be surprising as this specific model does not include the effects of hospitalization and subsequent health care on individuals (i.e. the prospect of a recovery through treatment). Therefore the simulations refer to a ``disaster situation'', where infected individuals are not isolated.  Hospitalization contributes not only to the care of infected people, but also to reducing the spread of the infection.

Model (\ref{swarm-s}) can be  technically modified to include the effects of hospitalization on the dynamics, by which further developments of the infection are reduced. These specific dynamics can be modelled simply by adding a fifth population, viewed as a FS, of hospitalized individuals. The dynamics of this population can be modelled by extracting a fraction of infected individuals from 2-FS in different ways: for instance, hospitalizing individuals who present a high level of the infection, corresponding to high values of $j$, or simply a fraction of the infected independently of the said level.

Therefore, it is interesting to show how hospitalization influences the contagion. A simple model can be developed by supposing that individuals who show some clinical evidence/symptoms larger than a critical state, say $u \geq u_c$ are hospitalized. These patients undergo clinical treatments, which are not studied in this section, but our model shows how the contagion evolves accounting for dynamics where hospitalized individuals do not contribute to infecting healthy people (of course besides those within each hospital).

Let $\phi = \phi(t)$ be the population of hospitalized individuals. Then, the model can be given by  a technical  modification  of Eq.~(\ref{swarm-s}) as follows:
\begin{equation}\label{swarm-s-hosp}
\begin{cases}
 \displaystyle \partial_t f_1^{1,k}(t) = - \alpha \, \sum_{s=1}^n  \sum_{j=2}^{m-1} \, u_j \, f_1^{1,k}(t) \, f_2^{j,s}(t),\\[5mm]
 \displaystyle \partial_t f_2^{j,k}(t) = \alpha \, \sum_{s=1}^n  \sum_{j=2}^{m-1} \, u_j \, f_1^{1,k}(t) \, f_2^{j,s}(t)\, \delta_{2j}\, + \beta  u_{j-1} \,  f_2^{j-1,k}(t) \\[3mm]
 {} \hskip1cm  + \gamma \, w_k \, f_2^{j+1,k}(t) - \beta\, u_{j} \,  f_2^{j,k}(t) - \gamma \, w_k \, f_2^{j,k}(t),\\[5mm]
 \displaystyle \partial_t f_3(t) =  \gamma \,\sum_{k=1}^n w_k \, f_2^{2,k}(t), \\[5mm]
  \displaystyle \partial_t \phi(t) = \beta\, \sum_{j=c}^m u_j \,  \sum_{k=1}^n f_2^{j,k}(t),
\end{cases}
\end{equation}
where the second equation is now valid for $j=2,\ldots,c-1$. Individuals reaching the state $u_c$ are immediately removed from 2-FS and contribute as a gain in the last equation.

\begin{figure}[ht!]
\includegraphics[width=0.45\textwidth]{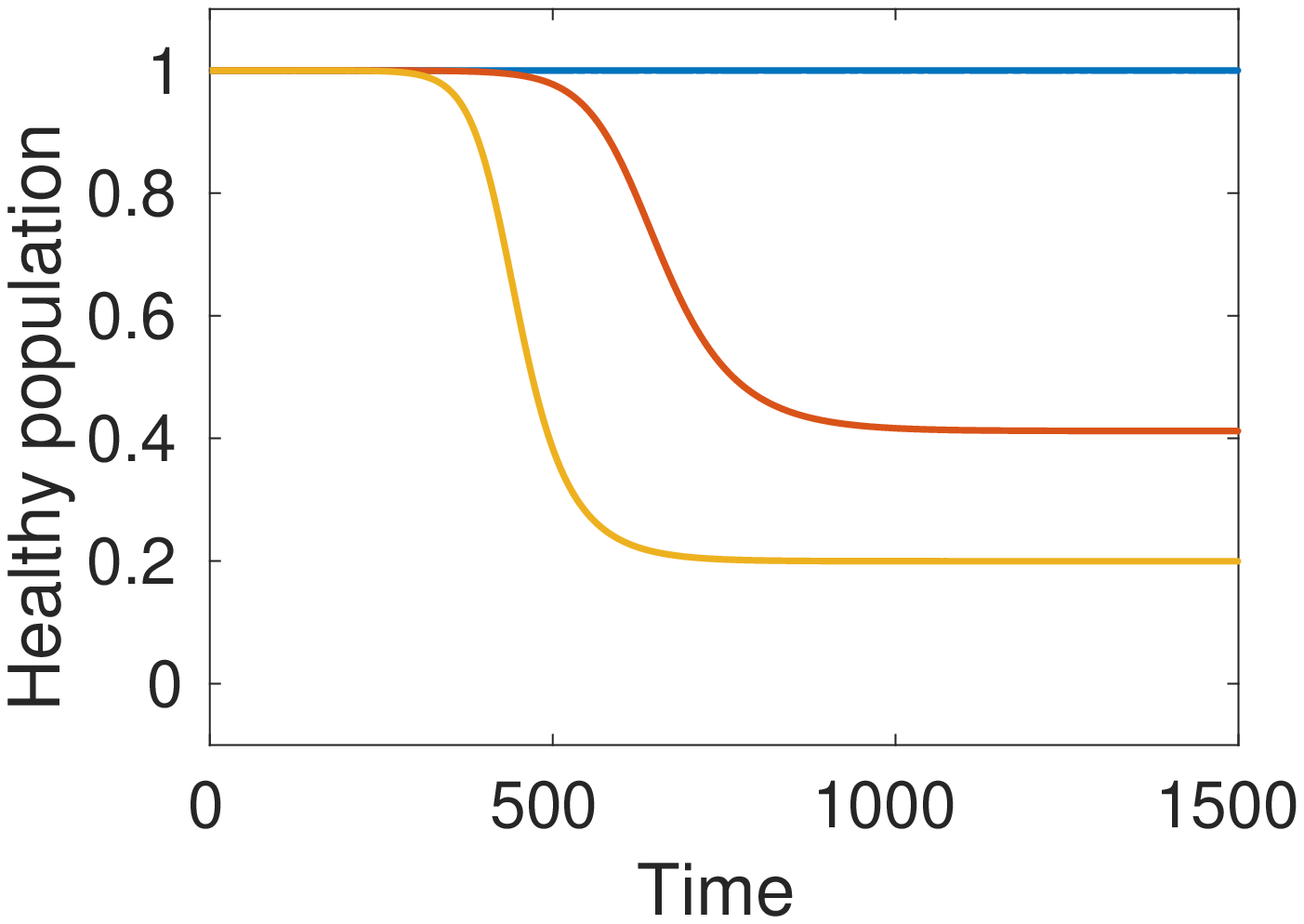} \qquad
\includegraphics[width=0.45\textwidth]{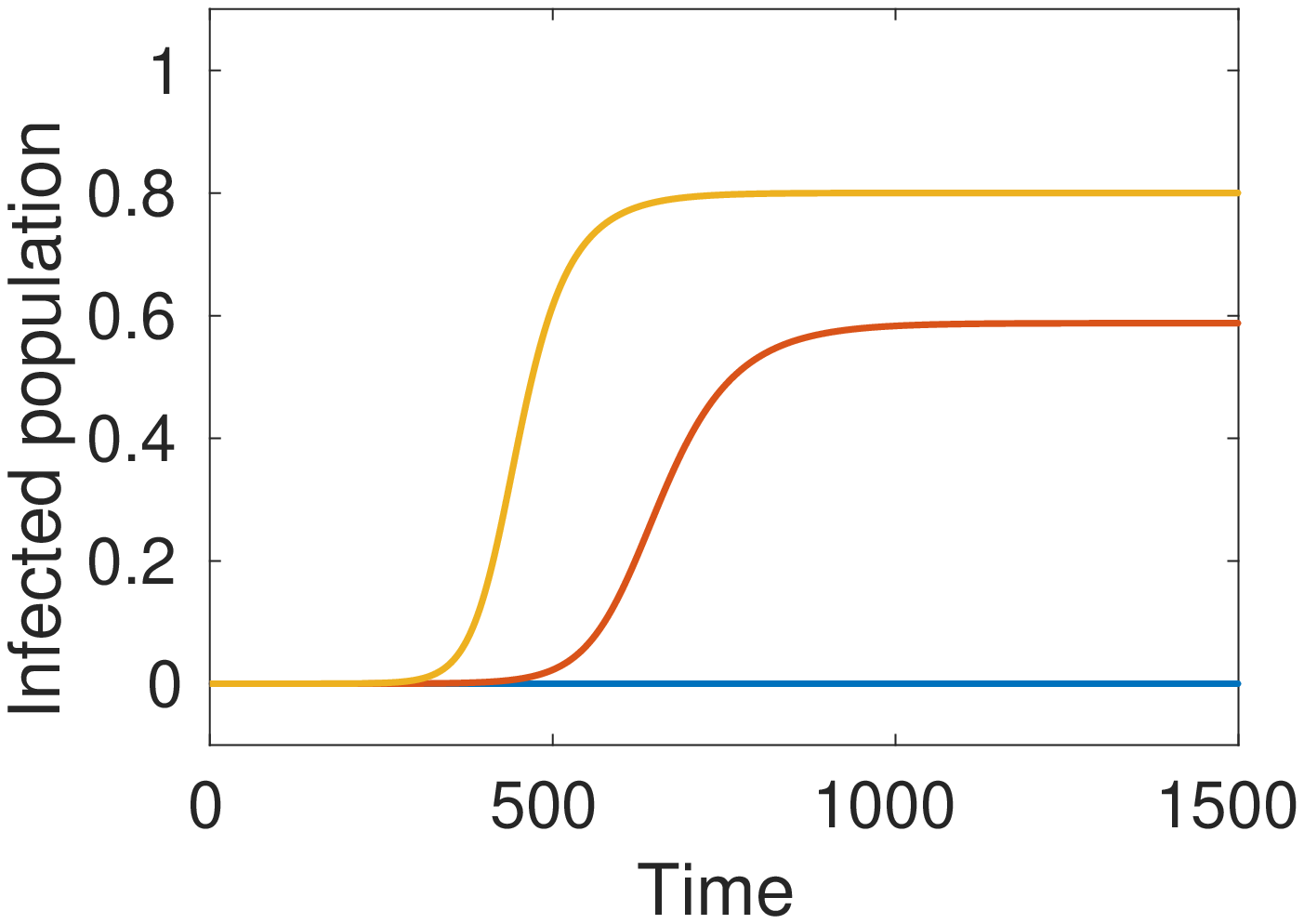}
\caption{Healthy and cumulative infected populations for different values of $u_c$, namely $c=5$ (yellow), $c=4$ (red), $c=3$ (blue), by simulations with $m=5$.} \label{hospitalization}
\end{figure}

Figure~\ref{hospitalization} shows how the healthy and infected population depend on the critical state $u_c$ at which people are isolated.

\subsection{Reasonings on spatial dynamics and the contagion}

Some reasonings on the modelling of the dynamics of contagion propagation through space are developed in this subsection. Spatial homogeneity in and of itself already provides useful information and insight. However, it is only an approximation of the physical reality of how the infection is transmitted spatially and additional work is needed to study more fully the dynamics in space and the related  pattern formation that is created over the spatial domain. This is not an easy task, and it should focus not only on  the dynamics in open areas, but also consider movement in complex venues across interconnected areas and networks. In both cases, the venue might include restricted aggregation areas, namely zones of high contagion diffusion.  This study poses challenging problems which require new ideas but at the same time opens interesting research perspectives. This subsection presents some of these perspective ideas towards possible research programs which should also include computational simulations.

The literature in the field of mathematical biology, specifically the dynamics of multicellular systems, already provides some perspective ideas which can contribute to a modelling strategy for the complex system under consideration. As an example, an analytic and computational study is proposed in~\cite{[BPTW19]} for the transport of an epidemic population model by a reaction-diffusion-chemotaxis system. A  study of the modelling of the movement of invasive cancer cells through the extracellular matrix has been developed in~\cite{[SMC20]} focusing on the multiscale features of the dynamics. The second part of~\cite{[SMC20]} presents a model of cell dynamics at the micro-scale for the cancer cell movement through the extracellular matrix accounting for the different way by which cells can move, namely mesenchymal and ameboid.  The authors specifically apply this technique to the phenomenon of cancer invasion, but some concepts can contribute  also to a deeper understanding of the problems studied in our paper. An additional problem, which cannot be neglected, is the derivation of models at the macro-scale from the underlying description at the micro-scale by asymptotic methods~\cite{[BBC16],[BC19]} somehow inspired by Hilbert's sixth problem~\cite{[DH1902]}.

A multiscale vision requires that the approach to modelling spatial dynamics should take into account both the literature in the field of biology to understand how cells and viruses move over and through human tissues, and of the literature on crowd dynamics to understand how individuals move in venues and across territory. This section  focuses on some selected topics and provides, for each of them, some perspective ideas to develop a modelling strategy to be further extended in an appropriate research program. Firstly, we focus on the contagion dynamics  in crowds corresponding to different levels of awareness of the risk of contagion, and subsequently,  the pattern formation of infected areas in a heterogeneous territory/domain are studied accounting also for the role of transportation networks.

\vskip.5cm \noindent $\bullet$ \textbf{Contagion dynamics in crowds:} Consider a crowd moving in an open domain where the distance between individuals varies in space and time.  A small number individuals in the crowd are initially infected, and our target is to study the dynamics of infection.

\begin{figure}\label{Fig-movement}
\begin{center}
\includegraphics[width=0.7\textwidth]{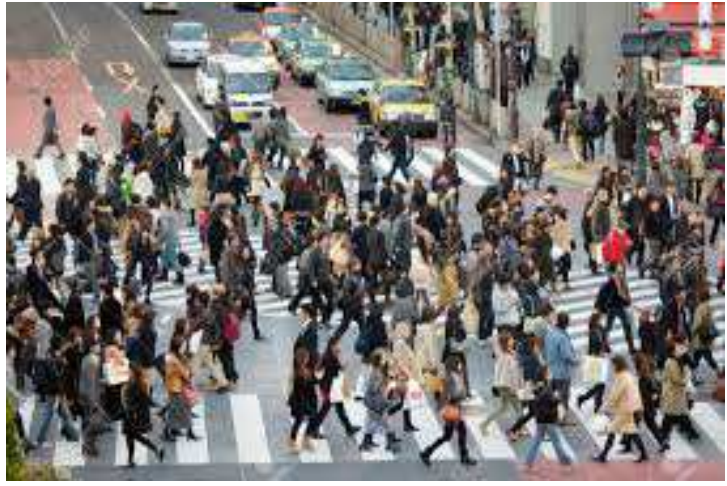}
\end{center}
\begin{center}
\caption{A crowd during city traffic.}
\end{center}
\end{figure}

Applications of  kinetic theory methods to crowd dynamics indicates that a key  modelling approach has been introduced in~\cite{[BBK13]} for a crowd supposed to move along a finite number of discrete velocity directions. This model has been subsequently developed in~\cite{[BG15]} to include continuous velocity dependence, the role of emotional states, and a self-organization ability homogeneously shared by the whole crowd. Derivation of a macroscopic equation from the underlying description at the micro-scale has already been developed in~\cite{[BB15]}. Further developments and related computational schemes have been proposed in~\cite{[KQ19]}. The spatial propagation of emotional states by a consensus dynamics towards a commonly shared status in one space dimension has been studied in~\cite{[BRSW15],[WSB17]}, where the term \textit{contagion} has been used to identify the dynamics by which individuals have a trend to share a common emotional state. Recent research activity is specifically focused on congestion problems in crowds~\cite{[LLH20]}. Congestion in urban areas might be induced by traffic as shown in Figure~\ref{Fig-movement}.

Focusing on contagion problems,  the dynamics in more than one spatial dimension has been studied in~\cite{[BGO19]}, where  different types of social interactions have been modelled. The coupling between  a kinetic model of crowd dynamics~\cite{[KQ19]} and infection transmission has been proposed in the pioneering paper~\cite{[KQ20]}. This is an important difference with respect to models of evacuation dynamics where the key social state is the level of stress which promotes aggregation of walkers rather than rarefaction which is different from the awareness of the contagion that induces walking strategies by which individuals try to maintain a safe distance from each other.

The study of crowd dynamics related to evacuation problems has brought about the derivation of quite sophisticated models which depict a detailed computation of the  local density and the mean speed of the flow. Indeed, both quantities are important in the modelling approach as high densities correspond to lack of safety, while high speed and optimal search of walking trajectories contributes to rapid evacuation. However, the study of contagion dynamics needs to be developed in a technically different framework. Hence we show how a  modelling approach  can be developed  to account for the technical  modifications induced by the awareness of the risk of contagion.

In more detail, let us consider the dynamics where individuals move in two dimensional space with direction $\theta$ and speed $v$:
\begin{equation}
\bv = v (\cos \theta\, \bbi + \sin \theta \,\bbj) = v\, \bnu,
\end{equation}
where $\bbi$ and $\bbj$ are orthogonal unit vectors of a plane frame, $\theta$ denotes the velocity direction, and $\bnu$ is a unit vector which defines the velocity direction. Individuals in the crowd are viewed as  a-particles whose \textit{microscopic state}  is defined by position $\bx$,  velocity $\bv$,  and activity  $u$.

Dimensionless quantities can be used  by referring the components of $\bx$ to some spatial scale $\ell$, while the velocity modulus is divided by the limit velocity $V_\ell$ which depends on the quality of the environment, $V_\ell$ is the speed which can be reached by a fast pedestrian in free flow conditions, while $\ell$ is the diameter of the circular area containing the domain where the crowd is initially localized.

The overall system is subdivided into $n$ functional subsystems, while their mesoscopic (kinetic) representation  is obtained from the statistical distribution at time $t$, over the microscopic state:
\begin{equation}
	f_i = f_i(t,\,\bx,\,\bv, u) = f_i(t,\,\bx,\,v, \bnu, u),
\end{equation}
where $\bx \in \mathbb{R}^2$, $v\in [0,1]$, $\theta \in [0,2\pi)$,  $u \in [0,1]$.
If $f_{i}$ is locally integrable then $f_{i}(t,\,\bx,\,\bv, u)\,d\bx\,d\bv\,du$ is the (expected) infinitesimal number of pedestrians  of the i-th FS whose micro-state, at time $t$, is comprised in the elementary volume $[\bx , \bx + d\bx] \times [\bv , \bv + d\bv] \times [u , u + du]$ of the space of the micro-states, corresponding to the variables space, velocity and activity.

We refer to~\cite{[BBGO17]} to propose the following structure  which consists in an integro-differential system suitable for describing the time dynamics of the distribution functions $f_i$:
\begin{eqnarray}
\label{structure}
&& \left(\p_t  +  \bv \cdot \nabla_\bx \right)\, f_i(t, \bx, \bv, u)  = J_i[\f](t, \bx, \bv, u) \nonumber \\[3mm]
&& \hskip.5cm = \sum_{k=1}^n \,  \int_{D^2} \eta_{ik}[\f](\bx,\bv_*,\bv^*, u_*, u^*) \, \cA_{ik}[\f](\bv_* \to \bv, u_* \to u|\bv_*,  \bv^*, u_*, u^*) \nonumber \\
 && \hskip3truecm \times \, f_i(t, \bx, \bv_*, u_*) f_k(t, \bx,  \bv^*, u^*)\,d\bv_*\,d\bv^* \, du_*\, du^* \nonumber \\[3mm]
 && \hskip.5cm -  f_{i}(t, \bx, \bv,u)  \sum_{k=1}^n \int_{D} \eta_{ik}[\f](\bx,\bv,\bv^*,u, u^*;\alpha) \, f_{k}(t, \bx,  \bv^*, u^*)\, d\bv^* \, du^*,
\end{eqnarray}
where  the integration domain is $D= D_\bv \times D_u$, where $D_\bv$ and $D_u$ correspond, respectively, to the domains of $\bv$ and $u$. In addition, the following quantities model interactions:

\vskip.1cm \noindent  $\eta_{hk}[\f](\bx, v_*,  v^*, \bnu_*, \bnu^*, u_*, u^*)$, is the \textit{interaction rate} modelling  the frequency by which a candidate (respectively test)  $h$-particle in $\bx$ develops contacts, in $\Omega_s$, with a field $k$-particle.

\vskip.2cm \noindent  $\cA_{hk}[\f](\bnu_* \to \bnu,  v_* \to v, u_* \to u|v_*,  v^*, \bnu_*, \bnu^*, u_*, u^*)$ is the  \textit{transition density} modelling the probability density that a candidate $h$-particle in $\bx$ with state $\{v_*, \bnu_*, u_*\}$ shifts into the state $\{v, \bnu, u\}$ of the $i$-test particle due to the interaction with  a field $k$-particle in $\Omega_s$ with state  $\{v^*, \bnu^*, u^*\}$. Both $\eta$ and $\cA$ can depend on the local density.

\vskip.1cm   Interactions correspond to  a decision process by which each a-particle modifies its activity and decides on its mechanical dynamics depending on the micro-state and distribution function of the neighbouring  particles in its  interaction domain. This process modifies the velocity direction and speed. Three types of a-particles are involved in the interactions. The \textit{test particle}, the \textit{field particle}, and the \textit{candidate particle}. Their distribution functions are, respectively $f_i(t,\bx, \bv, u)$,  $f_k(t, \bx,  \bv^*, u^*)$, and  $f_h(t, \bx,  \bv_*, u_*)$. The test particle is representative, for each FS, of the whole system, while the candidate particle can acquire, in probability, the micro-state of the test particle after interaction with the field particles. The test particle loses its state by  interaction with the field particles.

The derivation of the mathematical models is obtained by inserting into (\ref{structure})  models of interactions specialized to account for each specific physical situation under consideration.  The strategy expressed by individuals is that they first modify the dynamics of the emotional state, then they select the walking direction and  finally, the walking speed.

The modelling approaches known in the literature, see~\cite{[ABGR20]} as a reference, suggest that each individual interacting with others in his/her sensitivity domain first selects the walking direction by a weighted choice from the following directions: A trend towards the target $\theta^T$,  attraction towards the main stream $\bxi$, and the search of paths with less congested local density $\theta^V$. In more detail, the selection depends on the parameter $u$ and is weighted by the local density $\rho$, where increasing values of  $u$ correspond to a trend towards the stream $\bxi$ with respect to the trend towards the target, while the local density increases the trend towards vacuum or empty zones. Therefore, the model depend, for each FS, on the micro-state quantity $u$ which differs in each FS and on two local macro-scale quantities, namely $\rho$ and $\bxi$.

The selection is modelled by theoretical tools of stochastic game theory, where the output of the interaction is govened by the local density and a parameter modelling the level of individual stress in the crowd. In detail, increasing the value of the said parameter increases the attraction toward the main stream against trajectories across less crowded areas, while decreasing the density increases the attraction towards the target.

Somehow different is the case of a crowd of individuals trying to avoid contagion, where the contrast is between the trend towards the target and the search for less crowded areas. These dynamics can be modelled by a parameter $\sigma \in [0,1]$ modelling the level of awareness of the risk of contagion, where $\sigma = 0$ and $\sigma =1$ denote, respectively dominant attraction to the target and dominant attraction from the search for empty space/vacuum. Therefore, modelling should include this parameter  of individual awareness of the risk of infection which can modify in a significant way the dynamics of the crowd with respect to the dynamics in absence of such an awareness. The role of $\sigma$ is that the directional strategy is dominated by the search for low concentration trajectories rather than attraction towards the stream which generate high concentration. Then, modelling  infective contagion in crowds, as shown in~\cite{[KQ20]},  can be developed by coupling a selected model of crowds to the contagion model presented in the preceding subsection.

The modelling approach can therefore be technically developed according to the following rationale:

\begin{enumerate}

\item A model of crowd dynamics is selected corresponding to a binary mixture of infected and healthy individuals.

\vskip.2cm \item Interactions can even disregard the aforementioned attraction towards the main stream, while some technical assumptions can simplify the  model, for instance by constant speed and variable velocity directions.

\vskip.2cm  \item The contagion dynamics modelled by Eq.~(3.2) should be based on an appropriate modelling of the contagion term $\alpha$ depending on the local density.

\vskip.2cm  \item If the domain where the crowd moves includes walls and/or obstacles, nonlocal boundary conditions should be implemented as shown in Section 4 of~\cite{[ABGR20]}.

\vskip.2cm  \item  A technical generalization consists in the modelling of the crowd across interconnected domains, where flow conditions depend on the geometry and quality of each domain of the venue.

\end{enumerate}

A simplification of the mathematical structure, mentioned in Item 2, can be obtained by assuming that the speed is equally shared by all walkers. This hypothesis, with obvious meaning of notations, yields:
\begin{eqnarray}\label{structureB}
{} && \hskip.5cm \big(\partial_t  + \bv \cdot \nabla_\bx \big) \vf_i  (t, \bx, \theta)\nonumber \\[2mm]
{} && \hskip1cm =\eta (\rho) \,\sum_{h=1}^n \int_0^{2\pi}  \int_0^{2\pi} A(\bvf)(\theta_* \to \theta; \theta_*, \theta^*, \kappa)\, \vf_i (t, \bx, \theta_*)\,  \vf_h (t, \bx, \theta_*)\,d\theta_*\,d\theta^* \nonumber \\[2mm]
{} && \hskip1.5cm - \eta (\rho) \, \vf_i (t, \bx, \theta)\,\sum_{h=1}^n \int_0^{2\pi}  \vf_h (t, \bx, \theta^*)\,d\theta^*,
\end{eqnarray}
where the crowd has been divided into $n$ different groups to include infected individuals. In this case the representation of the crowd is achieved as follows: $\vf_i, \hskip.2cm  i = 1, \ldots, n$, corresponding to the i-th  walking group with level of contagion $i = 1, \ldots, n$, where $i=1$ correspond to healthy people, while $\bvf$ denotes the whole set of $\vf_i$.

The design of computational codes towards the simulation of the spatial dynamics of $\vf_{i}$ can be developed by different techniques depending on the mathematical structure of the specific model used for the simulations. Different examples can be found in the pertinent literature, for instance, finite differences have been used in~\cite{[BBK13]}, while Monte Carlo particle methods~\cite{[PT13]} and splitting methods have been developed, respectively, in~\cite{[BG15]} and in~\cite{[KQ19]}.

If the domain where the crowd moves includes walls and/or obstacles, nonlocal boundary conditions should be implemented as shown in Section 4 of~\cite{[ABGR20]}, where this topic is treated at each modelling scale, namely microscopic, mesoscopic (kinetic), and macroscopic (hydrodynamical). A technical generalization, which consists in the modelling of the crowd across interconnected domains is discussed below.

\vskip.5cm \noindent $\bullet$ \textbf{On the role of  networks:} The propagation of infective states can occur in complex venues and in transportation networks. We define a \textit{complex venue} as a small-sized network of interconnected areas, where individuals walk to reach a certain target without using  transportation systems. We define a \textit{globally connected world} as a network whose nodes are connected by transportation means, say train, bus, airplanes, etc. Each node of a globally connected world is constituted by a subnetwork of complex venues, typically it is a town.

The modelling of the contagion dynamics in complex venues can be developed using in each area the approach defined above in this subsection, but accounting for the specific physical and geometrical features of each area. The modelling approach to the dynamics in large networks is an open problem and here we simply present some perspective ideas towards a rationale to tackle the said approach.
\begin{enumerate}
\item The globally connected world is subdivided into an exogenous network constituted by interconnected nodes.

\vskip.2cm \item Each node is subdivided in a network of interconnected complex venues.

\vskip.2cm \item The overall dynamics are modelled by coupling the various networks, where the input and output flows in each node are described by models of  migration dynamics across nodes in the line of the modelling approach to migration phenomena and networks interaction proposed in~\cite{[KNO13],[KNO14],[KT20]}.

\vskip.2cm \item  A possible simplification to reduce the computational complexity consists in local averaging of the dynamics in complex venues and even in nodes.

\end{enumerate}

\subsection{Critical analysis towards modelling perspectives}

A mathematical computational model  has been proposed in this section. Focusing on the descriptive skills of the model, the main features, selected among various ones, can be summarized as follows:

\begin{itemize}
\item  It describes the dynamics of the virus transmission, from infected to healthy individuals, depending on an encounter rate related to the confinement distance for a population characterized by a heterogeneous distribution of the ability of the immune system.

\vskip.2cm \item The model predicts the time evolution of the number of healthy, infected, recovered and dead individuals. These dynamics are related to a model of competition internal to each individual between virus particles and immune cells.

\vskip.2cm \item Inside each individual, progression and recovery are modelled, resulting in the evolution of the time dynamics of the number of recovered and dead individuals.

\vskip.2cm \item Explorative ability of the hospitalization policy which can be related to the level of progression of the pathology. Simulations can show how planning hospitalization on the basis of the level of the pathology can influence the output of the dynamics.

\vskip.2cm \item Explorative investigation on the confinement strategy referring both to the different levels of confinement and to the time interval of the application of this action.

\end{itemize}

In addition, we have proposed a rationale for the modelling of spatial pattern formation, due to the spread of the infection throughout a given spatial domain, in a small world as well as in the network of a globally connected large world.

The achievements reported in the items above cannot, however, be considered the end of the story, as it is necessary to look ahead to additional work to be developed within dedicated research programs which can take advantage of the flexibility of the mathematical framework and of the computational tools proposed in this paper. For instance, new specific features of the dynamics can be included in the general framework also accounting for the perspective ideas presented in the next sections related to virology, immunology and economics.

Focusing on the modelling approach, we mention that it goes far beyond SIR models and recent developments, as the contents of this section shows how the predictions of  our model can contribute  to the planning of health care. Indeed, the model is derived within a multiscale vision, where the dynamics at the scale of particles is linked to that of populations whose dynamics depends on that at the low scales. The description of the dynamics at the higher scale of populations provides information useful to the planning of hospitalization as it provides not only numbers but also levels of the infection and the type of hospitalization to be addressed to specific levels of the infection. These features are enlightened by the simulations proposed in Subsection 3.3.

Definitely, challenging research perspectives should look at the small-scale, namely  at the dynamics inside each individual by  a more detailed  description of the dynamics of virus progression which might include also darwinian mutations, as well as by specializing the different actions that the immune system can develop to counteract the virus progression. In addition, the transfer of the dynamics from the small to the large world requires additional work to be technically related to the heterogeneous features of the territory.

A detailed presentation of research perspectives is proposed in Section 7 accounting, as mentioned, not only of the contribution of this present section, but also of that of the next Sections 4, 5, and 6. In particular, Section 7 reports a representation of the overall flow which is extended, with respect to Figure 2, to account for spatial dynamics and the role of hospitalization.

The final goal consists in developing a systems approach towards pandemic diseases suitable to lead  an explorative model with descriptive ability to contribute to depict a broad panorama of simulations useful to support the selection  medical and biological strategies as well as to the strategic indications that might be delivered within the framework of economical sciences.

\section{Reasonings on virology problems}\label{sec:4}

Viruses constitute one of the most abundant species on the planet, and play important roles in all kingdoms of life. Phages, viruses infecting bacteria, are essential for areas as diverse as the ecosystem of the oceans and gut health, and epidemics caused by plant and animal viruses make severe impacts on agriculture and human health.  \textit{SARS--CoV--2}, the causative agent of COVID-19, is a prominent example \cite{RB1}. Like SARS-CoV and MERS, which caused outbreaks in 2003 and 2012, respectively, it is a betacoronavirus. However, in contrast to these viruses it has evolved properties that make it far more dangerous, such as its ability to spread between hosts with ease, and in many cases stay asymptomatic for a significant time after infection. Mathematical modelling of individual viral particles can play a key role in understanding how changes in viral genomes due to mutation result in dramatically different properties of the virus.

\vskip.2cm \noindent $\bullet$ \textit{Modelling of viral geometry.}
Viral genomes encode instructions for the production of the proteins required to build progeny virus (the structural proteins), as well as proteins with a range of other functions in the viral life cycles. This includes virally encoded polymerases required that catalyze genome translation and transcription, and in larger and more complex viruses (such as HIV) also proteins to counteract host defense mechanisms. The genetic information is stored in the form of ribonucleic acid (RNA) or deoxyribonucleic acid (DNA). Smaller viruses, with genomes ranging between about 1k to 30k nucleotides (nts) in length, mostly have RNA genomes, whilst larger viruses predominantly use DNA and can have genomes up to a size of 2.5M nts. However, all viruses face the same challenge of protecting their genomes between rounds of infection. For this, they use protein containers, called viral (nucleo)capsids, and/or a lipid membrane, within which the genetic material is packaged. In RNA viruses, genome packaging usually occurs concomitant with capsid assembly, whilst in larger and more complex viruses, additional molecular machinery (packaging motors) is required in order to package the genome into a preformed capsid.

For RNA viruses, such as SARS-CoV-2, that package their genomes during particle assembly, an understanding of viral geometry is important. This is because the geometric shapes of the structural proteins and their assembly properties are intimately linked \cite{RT9a,RT9b}, and also affect other aspects of the viral life cycle, such as the structural transitions that in some viruses are important for infection \cite{RT9c}. In particular, in order to attribute as little coding sequence as possible to the viral capsid - a phenomenon called the principle of genetic economy - viruses encode blueprints of only a minimal number of distinct proteins (as small as one for the simplest form of virus), that are then repeatedly synthesized from the same genome segment, thus delivering multiple identical copies for virus assembly. As multiple identical proteins form the same types of interactions with each other, this results in protein assemblies with symmetry. As Crick and Watson noted in a seminal paper \cite{CW}, they must either form spherical shells with the symmetries of the Platonic solids - as is the case in icosahedral viruses, which constitute the vast majority of viruses - or be rod shaped. RNA viruses with large RNA genomes, such as coronaviruses, typically exhibit helical geometries. This is, perhaps, as their long genomes would be difficult to compact inside a spherical container. Coronaviruses, by contrast, wrap their 30k nt long genomes around a helical core formed from the nucleocapsid (N) protein \cite{tubular}. The nucleocapsid complex, formed from genomic RNA and N protein, is enveloped by a lipid membrane, the viral envelope, that is studded with three types of glycoproteins: the Spike protein (S), which is important for receptor binding and thus virus entry into a host cell, as well as the membrane (M) and envelope (E) proteins.  While these components are common to all coronaviruses, genetic changes in new emerging variants can lead to substantially different infection dynamics. For example, the S protein of  \textit{SARS--CoV--2} contains a novel, short sequence of amino acids which enables the virus to enter the cell more rapidly than previously circulating coronaviruses \cite{RB2}.  Understanding how changes to the protein structure affect the dynamics of virus assembly and virus-host cell interactions can be key to the identification of effective drug targets.

Any repeat organisation of protein units can be modelled via lattice theory \cite{RT3,RT10}. The first such theory has been proposed by Caspar and Klug in 1962, where they classified the surface architectures of icosahedral viruses formed from clusters of 5 (pentamers) and 6 (hexamers) identical protein subunits. Extensions of this theory have provided models also for viruses that fall out of this scheme, such as the noncrystallographic architectures of the cancer-causing papillomaviruses \cite{RT1}, and tubular variants formed from the same proteins \cite{RT2}. A mathematical approach based on Archimedean lattices, that embody the concept of identical local interactions for different types of proteins, has provided an overarching framework for the modelling of icosahedral viruses that accommodates Caspar-Klug theory as a special case \cite{RT11}.  A dual view, from a mathematical point of view, is to use affine extended symmetry groups to model material boundaries in icosahedral viruses \cite{RT4,RT9c,RT6,RT7}. This approach is also directly applicable to the core architecture of coronavirus, which is given by a translation and rotation operation that in combination define the 70 degree angle and the helical pitch of 140 Angstrom for a unit of 16 N proteins \cite{tubular}.

\vskip.2cm \noindent $\bullet$ \textit{Implications for viral life cycles.}

Models of viral geometry provide the opportunity to the analysis of the intracellular dynamics of a viral infection. Models of intracellular dynamics include reactions, or equations, for all processes inside the cell pertaining to viral replication. They typically include the copying of the viral genome (transcription) and protein production (translation), as well as reactions describing virus assembly \cite{RB7}. In general, virus assembly is described by a single reaction. However, this misses the intricate interdependence of its different functions, both as a template for replication and as a packaging substrate. Insights into viral geometry enables refinement of these reactions \cite{Eric_PNAS}, and therefore provides a much more realistic view of intracellular dynamics, revealing aspects of viral life cycles that had previously been overlooked.
Model outcomes can then be included into infection models at different scales, including intercellular models of within host dynamics and between host dynamics, and thus provide a foundation for a deeper understanding of viral infection dynamics.  Models of the within-host progress of viral infections can be used to inform pharmaceutical interventions \cite{RB3,RB4} or provide insights into viral phenomenology \cite{RB5}.  The long genomes of coronaviruses encode at least 19 distinct proteins, which in turn results in a complex interaction network.  Viral proteins frequently perform multiple roles in the viral life cycle, creating degeneracy in the network.  The assembly of infectious virus particles requires non-uniform amounts of the constituent proteins, with many more structural proteins required than non-structural ones \cite{numbers}.  Many viruses, including coronaviruses, therefore produce shorter fragments of their genomes, known as subgenomic fragments, to control this process.  Identifying these subgenomic fragments \cite{RB6} is a key step for building viral life cycle models, as each fragment should be a node in the corresponding interaction network.  Accurate modelling of viral life cycles is built upon comprehensive knowledge of the interactions of viral and cellular proteins and genetic material, which is still incomplete at this point and an active area of research for  \textit{SARS--CoV--2}.

\vskip.2cm \noindent $\bullet$ \textit{Implications for viral evolution.}

Viruses occur as populations of genetically related viral strain variants called quasispecies \cite{RB8}. Their distribution is important for their ability to adapt to environmental conditions, as well as to different hosts. In the case of  \textit{SARS--CoV--2}, it is likely that the virus has spread from animal hosts, perhaps at a wet market in China, as its genetic sequence exhibits high similarity with coronaviruses infecting bats and pangolins \cite{D_prox}.

The accumulation of mutations over time allows for the reconstruction of the spread of the infection through the population, an essential step in developing public health interventions \cite{RB9}.  Any virus accumulates mutations as it passes through a population, as replication inside each host brings the chance for novel mutations to occur. However, the majority of mutations do not significantly affect the phenotype (and therefore fitness) of the virus, but some mutations confer selective advantages to the virus and therefore become fixed in the population. Such mutations are often interpreted as adaptation by the virus to environmental changes, but there are publications cautioning that this is likely an over-interpretation of the data, particularly for widely circulating viruses \cite{RB10}.

\vskip.2cm \noindent $\bullet$ \textit{Implications for therapy.}

Mutations also enable viruses to circumvent challenges from anti-viral therapy. Mutation rates are much higher in RNA viruses (typically $10^{-5}$ substitutions per nucleotide per cell infected), as opposed to DNA viruses (typically $10^{-7}$ substitutions per nucleotide per cell infected) \cite{RB11}, making the development of long-lasting vaccines and antiviral therapies difficult.  The rapid evolution of the RNA genome of the influenza virus requires the development of a new vaccine every winter \cite{RB12}, whereas the smallpox vaccine has effectively eradicated the disease \cite{RB13}.  Better understanding of the mutational landscape of  \textit{SARS--CoV--2} will enable predictions of the efficacy of vaccination efforts\cite{RB15,RB18}.   Many existing antiviral therapies are developed against widely shared viral mechanisms, such as polyprotein cleavage and genome replication, however these are often susceptible to rapid adaptation by the virus \cite{RB14}.  It is therefore essential to better understand the geometric constraints on viruses, as these can point to evolutionarily conserved features that could serve as more stable and effective drug targets.

Due to its large genome size compared to other RNA viruses, and the vast frequency of infection, and thus opportunities for mutation, during a pandemic, understanding the consequences of mutations, and finding evolutionarily stable targets, is particularly important for viruses such as  \textit{SARS--CoV--2}. Mathematical modelling can support the discovery of such novel therapeutic targets in many ways. Traditionally, this occurs through the analysis of individual viral components and their dynamic properties. For example, since the start of the COVID-19 pandemic, normal mode analysis and quantum mechanical computations have been used to better understand potential drug targets on viral proteins. This includes structural proteins, most importantly S-protein, which is involved in ACE2 receptor binding \cite{A_spike,B_ace2}, as well as non-structural proteins, such as proteases \cite{C_allo}.  The rapid emergence of COVID-19 has also spurred the development of  novel methodologies to support the discovery of antiviral therapies, in particular \textit{in silico} screening  techniques of multiple compounds against  \textit{SARS--CoV--2} proteins \cite{RB16,RB17,Guowei}.

Once effective antiviral therapies have been developed, mathematical modelling can provide valuable insights into effective treatment \cite{RB19} and mechanisms of drug resistance \cite{RB20,RB21}. The modelling of viral life cycles based on viral geometry adds a new dimension to the modelling. It provides a framework in which the merits of different forms of antiviral strategies can be compared and synergies explored \cite{RT8}, thus revealing novel opportunities for antiviral therapy.

\section{Reasonings on immunology problems}\label{sec:5}

A systems approach to model, by a differential system, the dynamics of the  \textit{SARS--CoV--2} in a population has been developed in Section 3 within a general framework of  a multiscale vision which includes the competition inside each individual between a progressing viral infection and the immune system. This competition initiates when the virus is transferred from infected to healthy individuals by a dynamics  modeled at the higher scale. This section provides a description of the immune competition at the low scale and  outlines the relevant features of the aforementioned dynamics according to the present state of the art.

\vskip.2cm \noindent $\bullet$ \textit{The contagion.}

The contagion by the  \textit{SARS--CoV--2} occurs mainly by air, breathing respiratory droplets released by an infected person when coughing, sneezing or speaking. The virus load in the droplets increases as the infection progresses~\cite{[Service20]}. Respiratory droplets finally land on various surfaces where the virus maintains its infective capacity for various times~\cite{[VBEA20]}. Therefore,  \textit{SARS--CoV--2} can also be transmitted by contact between a susceptible person and the infected one or touching contaminated surfaces. In this case, the infection takes place if the susceptible person touches the mucosa of the mouth, nose or eye after capturing the virus~\cite{[WXEA20]}. The probability of infection rests on the viral load carried by the droplets expelled through respiratory emissions and the persistence over time of virus infectivity depending on the environmental conditions.

With the arrival of the \textit{SARS--CoV--2} virions  on mucosal surfaces a quantitative and time-sequence competition between the virus ability to infect and host defense mechanisms is putted in motion. The final outcome (inhibition of the viruses, minor and asymptomatic infection, a severe and fatal disease) rests on a series of confrontations between the virus infective ability and distinct immune defense mechanisms~\cite{[MDN20]}.

\vskip.2cm \noindent  $\bullet$  \textit{First line of defense: the mucosal barrier.}

The number of infecting  \textit{SARS--CoV--2} particles is a first and crucial variable in the competition between the virus and host defense mechanisms~\cite{[MDN20]}. The virus load arriving on mucosal surfaces has to deal with the very effective barrier made by the mucus. Mucus is a complex mixture of glycoproteins continuously produced by goblet cells in the mucosal membranes and by particular glands. It contains salts, \textit{lactoferrin}, enzymes and antibodies (secretory IgA and IgM)~\cite{[BJEA15]}. This viscoelastic gel covers the mucosa lining the nose, throat and lungs. By harnessing and neutralizing viral particles, it prevents a direct contact of the viruses with the surface of the epithelial cell of the mucosa.
Mucus production is regulated mainly by two \textit{lymphokines} (IL-13 and IL-22) secreted by sentinel lymphocytes associated with the mucous membranes. IL-13 is mainly produced by Innate Lymphoid Cells (ILC), IL-22 by T-helper lymphocytes17 (Th17). The overproduction of mucus gives rise to phlegm~\cite{[TGEA2020]}.

Mucus is continually transported by the coordinate beats of the cilia of the hairy cells in the mucous membranes, then it is swallowed and destroyed in the stomach. Mucus transport is fundamental for its protective action. Under normal conditions, the ciliary beat frequency is around 700 beats per minute. The intensity of the beats is negatively regulated by IL-13 and is lowered by environmental pollutants present in the breathed air, by the humidity and by the low temperature~\cite{[LP01]}.

The importance of the mucus barrier in hindering \textit{SARS--CoV--2} infection it is unexplored,  although it is well know the influence that environmental factors (humidity, temperature, air pollution, etc.) have on the defense against coronavirus infections that cause winter colds. It is thus probable that in the great majority of cases mucus and the beats of cilia clear the invaders. Nevertheless, when the delivered viral load is very high and mucus production and its transport are disturbed, a few virus particles might be able to sneak through and establish a direct contact with the surface of mucosal epithelial cells.

The peculiar current distribution of the \textit{SARS--CoV--2} pandemic, which somewhat has spared the warmer countries while has hit harder areas with high air pollution could depend, in addition to a different survival of the virus in a warmer and drier environment, on a more efficient barrier effect created by mucus and ciliated cells. On the other hand, circumstances requiring speaking aloud in a low temperature and humid environment favor  \textit{SARS--Cov--2} super--spread~\cite{[KUP20]}.

\vskip.2cm \noindent  $\bullet$  \textit{Overcoming the barrier.}

An efficient barrier defense confronted with a low viral load probably efficiently stops the infection. In all cases, however, the barrier drastically reduces the number of viral particles that manage to reach the surface of the cells of the mucosa. If any  \textit{SARS--CoV--2} particles get through the mucus layer, they can reach the cell membrane of mucosal cells. In the event that a physical contact takes place, the  \textit{SARS--CoV--2} Spike glycoprotein, that forms a sort of crown on the surface of  \textit{SARS--CoV--2} particles, anchors the virus to the cell surface.

The Spike glycoprotein forms homotrimers protruding from the virus surface. Each Spike monomer is made by an external S1 domain and a S2 domain connected with the virus membrane. The distal S1 domain comprises the Receptor Binding Domain (RBD), a molecular subunit that engages with high affinity a region of the N-terminal domain of human Angiotensin-Converting Enzyme (ACE2). ACE2 is an ectoenzyme that is normally highly expressed on the surface of the cells of human respiratory and intestinal epithelia. It is also expressed on the surface of many other kind of cells, even if at lower density~\cite{[HKEA20]}.

Following Spike-ACE2 binding, other cell ectoenzymes (Transmembrane Protease Serine 2 TMPRLRSS2 and Furin) present on the surface of human cells cut the Spike protein separating S1 from S2 domain. As a consequence of the cut, the S2 domain, connected to the virus membrane,  exposes particular sequences of amino acids (fusion peptides) that facilitate the fusion between the viral capsid and the membrane of the human cell~\cite{[CYR20]}.
Thanks to this fusion, the RNA of the virus enters the cell where it is directly translated into proteins by the human ribosomes. New virions are assembled as reported in the previous Chapter and the cell die releasing millions of new viruses.

\vskip.2cm \noindent  $\bullet$  \textit{Second line of defense: intracellular reaction against viral RNA.}

The invaded cell is endowed with several mechanisms to sense and block viral invasions.
In the cell cytoplasm, the  \textit{SARS--CoV--2} RNA are recognized and destroyed by RNA helicases. The peculiarities of viral RNA are also is recognized by RIG-1-like receptors and Toll like receptors (TLR) present in the cell cytosol. Upon activation, these receptors induce signaling cascades leading to the phosphorylation of transcription factors ultimately conducting to transcription of Type I interferon (IFN), a cytokine that activates antiviral programs in the invaded cell and induces the expression of families of transmembrane proteins that inhibit virus entry in nearby cells~\cite{[VBEA20]}.

In addition, the signals transduced by RIG-1-like and Toll like receptors trigger the assembly of the inflammasomes, cytoplasmatic multimeric complexes. Once assembled, the inflammasome activate caspases which permit the production of high amounts of important pro-inflammatory cytokines (IL1, IL18, $\dots$). The secretion of the mature form of these cytokines promotes the release of additional cytokines and the induction of an innate immunity inflammatory response. Moreover, inflammatory caspase 1 cleaves gasdermin D to cause cytokine release and pyroptotic cell death, a kind of programmed cell death that occurs most frequently upon viral infection. The death of virus infected cell is an effective way to block the progression of the viral infection~\cite{[MJ20]}.

Faced with this sophisticated series of receptors and intracellular reaction mechanisms,  \textit{SARS--CoV--2} has elaborated several mechanisms to avoid viral RNA recognition or to antagonize with receptor signal transduction. The first protein coded by the  \textit{SARS--CoV--2} RNA penetrated inside the mucosal cell, ORF1ab, is a chain of 16 proteins joined together. Two portions of this long protein make the cuts that free the different proteins. The activity of several of them is to interfere with intracellular reaction mechanisms: NSP1 protein prevents the cell from assembling antiviral proteins; NSP3 protein alters the regulation of cell protein, thus reducing  cell ability to put in motion the antiviral mechanisms; NSP10 and NSP16 proteins protect viral RNA from destruction; NSP13. Other virus coded protein, such as ORF6 interfere with the signals activating cell reaction to viral RNA; ORF9b protein suppress intracellular signaling thus limiting antiviral defenses nonspecifically~\cite{[CZ20]}. Other viral proteins interfere with the signaling downstream of IFN release: ORF3a, ORF6 viral proteins alter various steps of the signal transduction pathway that bridge the IFN receptor to the STAT proteins that activate transcription of IFN-stimulated genes~\cite{[VBEA20]}.

\vskip.2cm \noindent  $\bullet$  \textit{Third line of defense: the inflammatory reaction.}

If the infecting viral load is enough to overcome the mucosal barrier and sneak through intracellular defense mechanisms, the spread of  \textit{SARS--CoV--2} RNA becomes evident in nostril, pharynx and eye mucosal surfaces. Here, viral infection is confronted by multiple reaction mechanisms of innate immunity. The intensity, efficacy and features of this third line of defense decides the natural history of infection: whether the virus spread will be efficiently blocked in upper airways and how many viruses will reach the lungs~\cite{[MDN20]}.

Humoral elements of innate immunity, including the complement and coagulation systems, soluble proteins of innate immunity such as the Mannose Binding Lectin, natural antibodies and cross-reactive antibodies induced by previous infections by different viruses are immediately confronted with the virus spreading and the local cell damages. The damages associated with the viral invasion, alarm signals and cytokines induce the dilatation of local post capillary venule with the subsequent exit of plasma and of reactive leukocytes. The reactivity of leukocytes will be driven by multiple alarm signals, cytokines and interferons released by infected and dying epithelial cells.

Metatranscriptomic sequencing performed to profile the inflammatory reaction shows a markedly elevated expression of several chemokine genes. CXCL8 chemokine gene overexpression is directly connected with the recruitment of neutrophils into the inflamed tissues, while the upregulation of CCL2 and CCL7 chemokine genes plays a central role in the recruitment of monocytes/macrophages. Consonant with this pattern of gene expression, neutrophils, activated dendritic cells, monocytes are the most abundant cells present in the reactive infiltrate [14].

  \textit{SARS--CoV--2} infection triggers the expression of several IFN-inducible genes. While this suggest that a robust IFN response is ongoing, the IFN gene is not upregulated. This discrepancy, probably due to the interference exerted by  \textit{SARS--CoV--2} proteins on the IFN gene expression and  on the signaling cascade of IFN receptors, has a crucial importance in shaping the efficacy of the inflammatory reaction and the evolution of the viral infection~\cite{[VBEA20],[ZREA20]}. This confrontation between host's innate immunity and  \textit{SARS--CoV--2} decides whether infection will be blocked in upper airways or the virus will reach the lungs, making the patient sick or very sick~\cite{[MDN20]}. Epidemiological data suggest that this reaction is extremely effective as the vast majority of  \textit{SARS--CoV--2} infected people are asymptomatic or minimally symptomatic.

\vskip.2cm \noindent  $\bullet$  \textit{Pathogenic fallout of innate immunity reactions.}

The feature of initial local viral spread, the genetic characteristics and health status of the infected person decides whether the innate immunity reactions will lead to the containment of the  \textit{SARS--CoV--2} infection or will worsen its evolution. For example, IFN is protective in the very early stages of the inflammatory response, while later it becomes pathogenic~\cite{[VBEA20]}. The expression of ACE2 on the cell membrane is modulated by IFN, and its upregulation in airway mucosal cells favors the expansion of  \textit{SARS--CoV--2} infection~\cite{[ZAEA20]}.

The intracellular recognition of viral RNA and the presence of  \textit{SARS--CoV--2} coded ORF3, ORF8b, and Envelope proteins triggers the leukocyte assembly of the inflammasome and the consequent massive production of IL-1 and IL-18. Similarly, the intracellular presence of  \textit{SARS--CoV--2} coded NSP9 and NSP10 proteins trigger IL-6 production~\cite{[VBEA20]}. These three cytokines together induce anti-viral programs and markedly enhance the inflammatory response.  However, elevated serum concentrations of IL-6, IL-1 and IL-18 may result in a systemic ``cytokine storm'' involving the secretion of vascular endothelial growth factors (VEGF), monocyte chemoattractant protein-1 (MCP-1), IL-8 and additional IL-6. The vascular permeability and vessel leakage are increased leading toward hypotension and acute respiratory distress syndrome (ARDS)~\cite{[ZAEA20]}.  In this way, if the  \textit{SARS--CoV--2} is able to reach the lung alveolar cells, innate immunity mechanisms driven by the cytokine storm may dramatically contribute to disease severity~\cite{[VBEA20]}.

\vskip.2cm \noindent  $\bullet$  \textit{Fourth line of defense: The T and B cell adaptive response.}

T cells play a central role in the healing of numerous infectious diseases. CD4 T cells are required for the activation of the production of high affinity antibodies by B cells.  Moreover, they modulate and guide a more efficient inflammatory response. CD8 T cells are able to find and kill virus infected cells before they become factories of millions of new viral particles.
Given the central role T cells have in the healing of viral infections, it is not surprising that an overall reduction of T cells in peripheral blood is associated the aggravation of  \textit{SARS--CoV--2} infection. T cell reduction during  \textit{SARS--CoV--2} infection is likely due to a number of factors, including the cytokine storm~\cite{[VBEA20]}.

A persistent and robust T reactivity to the peptides of Nucleocapsid, Membrane and Spike  \textit{SARS--CoV--2} proteins is evident in patients recovering from  \textit{SARS--CoV--2} infection.  Often a preferential specific reaction against Spike peptides was found. While the induction of a robust T cell reactivity is essential for virus control, a too high CD4 T cell reactivity may facilitate the development of lung immunopathologies and exacerbate the cytokine storm~\cite{[VBEA20],[MJ20]}.

While the reaction of CD8 T cells is directed to the killing of virus infected cells in which millions of new viral particles are assembling, the action of antibodies is against each viral particle. An antibody binding to the proteins exposed on the external surface of a virus severely limits it capacity to infect the target cells. In most cases the antibodies are mainly directed towards the Spike protein. In some cases, it has been possible to identify antibodies that specifically bind the RBD, the region of the Spike glycoprotein that binds to the ACE2 enzyme on human target cells. Often, the appearance of virus specific IgA and IgM and Ig (serum conversion) is associated with a complete virus clearance:  the  \textit{SARS--CoV--2} RNA is no more detected in the swabs.

 The effectiveness of antibodies in neutralizing viral infectivity is also shown by the success of therapies based on the transfer of plasma obtained from convalescent patients. Moreover, in several cases it has been possible to obtain monoclonal antibodies that interact with RBD and that block  \textit{SARS--CoV--2} infectivity starting from B cells obtained from patients in convalescence. Antibodies play a role not only in the healing but also in preventing  \textit{SARS--CoV--2} re-infections~\cite{[KUP20B]}.

\vskip.2cm \noindent  $\bullet$  \textit{The dark side of the adaptive responses.}

In many cases of  \textit{SARS--CoV--2} infection, the subsequent activation of adaptive responses corresponds to healing. However, the situation is not always so clearly defined. Some people recovering from  \textit{SARS--CoV--2} infection seem unable to produce a significant antibody response. Moreover, there are various reports on high antibody titer associated with a more severe clinical outcome. These observations highlight that in certain cases antibodies can facilitate viral infectivity, a phenomenon known as Antibody-Dependent Enhancement (ADE)~\cite{[VBEA20],[OMEA20]}.

When the various mechanisms of immune reactivity fail to control virus spread and the infected person starts to produce and shed copious amount of  \textit{SARS--CoV--2}, the virus may diffuses to all the body districts infecting the cells that express the ACE2 receptor enzyme. The expression of these receptor enzymes is particularly abundant on the cell membrane of both lung alveolar epithelial cells and enterocytes of the small intestine~\cite{[WCKM20]}. If the virus infects and replicates in lung alveolar or cells in the cell of intestinal mucosa, the immune attack may be responsible of major pathophysiological complications. The deposition of immunocomplexes made by antibodies and  \textit{SARS--CoV--2} virions triggers Complement activation, vasodilatation, inflammatory reactions, coagulation and tissue damage. Local cytokine and chemokines contribute to the local vasodilatation, recruitment of innate immunity cells and serum antibodies extravasation~\cite{[MDN20],[VBEA20]}.

The massive killing of virus infected cells leaves the lung alveoli stew of fluid and dead cells. Some patients slowly recover while others develop a poor oxygenation index and a condition called acute respiratory distress syndrome (ARDS) requiring intensive care therapy and mechanical ventilation. Similar lesions occur in the intestine.  The spread of  \textit{SARS--CoV--2} throughout the body also promote blood clots, hearth attack, and cardiac inflammation. At this stage, almost half of the patients slowly recover whereas the other die~\cite{[WCKM20]}.

\vskip.2cm \noindent  $\bullet$  \textit{Additional reasonings.}

  \textit{SARS--CoV--2} is a sophisticated single stranded RNA virus endowed with the capacity to infect in a short time over 6 million people around the world and to kill over 370,000 of them~\cite{[COR]}, causing  dramatic health, social and economic problems.
Epidemiological studies suggest that the number of people who have been infected with the virus is probably much higher~\cite{[DAY20]}. Therefore, even the infection by this extraordinary virus appears to be managed by the immune system in well over 90$\%$ of cases. Paradoxically, much of the virus lethality come from a ``friendly fire'' due to an excessive immune reactivity~\cite{[VBEA20],[WCKM20]}.

Immune defense mechanisms that are activated to block  \textit{SARS--CoV--2} infection are here schematically distinct in four lines. Each one is based on peculiar and distinct reaction mechanisms and follows a different strategy. Nevertheless, these four lines of reaction are highly interconnected with a very integrated transition from one line of defense to another.

In the vast majority of cases, each of these lines of defense wins its confrontation with the virus. Even if, as to now, it is not yet possible to know what the percentage of complete victory is, each line of defense proves to be extremely effective. Probably, the percentage with which the defensive system wins against the infectious ability of  \textit{SARS--CoV--2} is particularly high as regards the first lines of defense. The elevated lethality of the virus in elderly people due to the natural immunodeficiency connected with aging (the immunosenescence~\cite{[PAW18]}) highlights how important the defense role played by the immune system is in the control of  \textit{SARS--CoV--2} infection.

\section{Reasonings on economy and social problems}\label{sec:6}

In this section we intend to uncover the socio-economic implications of the spread of COVID-19. After distinguishing between the direct impact of the pandemic and the indirect one of the lockdown, we discuss the potential limitation of generalised lockdown policies facing localised contagion dynamics. We emphasize the role of heterogeneity in order to properly account for the contagion dynamics itself. Moreover we suggest the explicit account of clustered networks as a potential modelling extension. Additionally, we stress the importance of the timing of the lockdown and the need to introduce early warning indicators. We conclude by outlining a series of policy actions to be put in place beyond lockdowns to deal with the spread of collective infectious diseases.

\subsection{The economic impact of the pandemic}
The explosion of the COVID-19 pandemic and the ensuing policy of social distancing undertaken by many countries have put the organization of the production and the economy as a whole under an unprecedented stress. Analysis of the impact of the pandemic on the labour market are now spurring with scaring projections in terms of number of jobs and related income losses. The ILO projects two hundreds millions losses worldwide. The Internationally Monetary Fund estimates a decline of world GDP by 3\%, a drop much more severe than the 2008-2009 crisis, while the Euro-Area estimated loss currently is $7.5$\% of GDP~\cite{[IMF2020]}. Both are conservative forecasts. However these numbers are expected to increase. Already after one month since the lock-down, around $11$ millions European workers have been hit by the consequence of the pandemic, with an increase of four millions unemployed people and with seven millions of short-term contract workers at risk, according to ETUC estimates~\cite{[ILO]}.

The direct and indirect impacts of the pandemic invest many realms of the economic analysis, from  the organization of production and global value chains (GVCs), to patent systems and appropriability conditions in the pharmaceutical sector, to the provision of health as a public good, up to the study of unconventional fiscal and monetary policies (see~\cite{[B2020]}). On top of that, implications in terms of the organization of the workplace and related of the work-process are going to be huge. Indeed, social distancing is expected to jeopardize business and employment opportunities in a labour market at the outset marked by strong inequalities.

To detect the economic impact of the  COVID-19 pandemic it is important to distinguish the \textit{direct} economic effects and the \textit{indirect} ones of the policy of social distancing. The \textit{direct} impact is and is likely to be \textit{per se} quite limited. Among developed countries, in Europe, mortality has been concentrated on the elderly population, with significant losses only in the range  of plus 70 years. Differently, in the U.S., whose healthcare system is largely private, impeding provision of health assistance for poor and more vulnerable communities, the impact of the virus is spreading remarkably across Blacks, Coloureds, and Latinos of relatively young age (less than fifty). Indeed, the existence of previous health diseases as diabetes and obesity, more diffused among the poorest population, has been recognised a factor aggravating the infectiousness of the virus.

Notwithstanding heterogeneity across countries, in general it is not likely to expect the pandemic impacting on labour supply to a magnitude recalling the Black Death or even the Spanish Flu (see~\cite{[BUW2020]}). Together, this pandemic, unlike other historical episodes such as the Plague of the 14th century, will not serve to alleviate income and wealth inequalities, by increasing the wages of a scarce labour force and reducing the value of real estates on sale for the death of their primary owner. On the contrary it is, and will be much more so, amplifying existing inequalities, ranging from access to hospitalization, possibility to work-remotely, benefiting of a stable income, risk of unemployment (see~\cite{[CGV2020]}). On top of that, even risks of contagion are strongly heterogeneous, much more concentrated among worker categories directly exposed, such as those in the health sectors, and relatively less concentrated for workers able to work remotely.

Granted the unequal effects of the direct and indirect impacts of the pandemic, the modelling effort has to focus on \textit{heterogeneity} to meaningfully capture also its economic consequences. In particular, given the concentrated \textit{direct} impact on the elderly population, one might not consider any direct economic consequence arising from the death of the elderly. This does not mean that deaths are acceptable because they do not impinge directly on the economic system. It means that we refrain from attributing any direct value-estimation to the life of human beings. Clearly, deaths are an enormous social and humanitarian cost which have to be considered as such well beyond any economic consideration grounded on dismissible cost-benefit analyses.

Therefore, the economic impact of the pandemic is largely \textit{indirect}:
\begin{itemize}
 \item The economic damage of the pandemic increases with the amplitude and severity of the lock-down.

 \vskip.2cm \item The economic damage, arising from the lock-down, unevenly hits the population, with low-income individuals more harshly affected than high-income ones.
\end{itemize}

Defining the all set of potential variables and transmission mechanisms affecting the economy via the lock-down is out of the scope of the paper. As mentioned, the economic impacts are quite diverse, including production, global value chains, business closures, job and income losses, fiscal burden of policy interventions, debt accumulation, financial instability and markets volatility, just to mention a few. All these mechanisms, diverse as they are, might interact via cascade and cumulative effects along propagation channels, all contributing to fuel the longest last and most severe crisis the economy has faced since the 1929. The end result will be various possible ``damage functions'', which we shall present below, combining the diverse economic effects and their potential interaction.

Let as define a damage function $\mathcal{D}(L(\alpha),\sigma(\omega_{i}/\omega_{max}))$ depending on the intensity of the lock-down and on a proxy of inequality of a given system. $L(\alpha)$, the lock-down, is governed by the parameter $\alpha$, discussed in Section 3, increasing with the amplitude and duration of the policy, represented by the reduction of $\alpha$, while $\sigma(\omega_{i}/\omega_{max})$ represents the spread between the actual and the maximum income level in the system, with low-income individuals hit harder than high-income ones, where $\omega_i$ defines the income level of individual $i$.

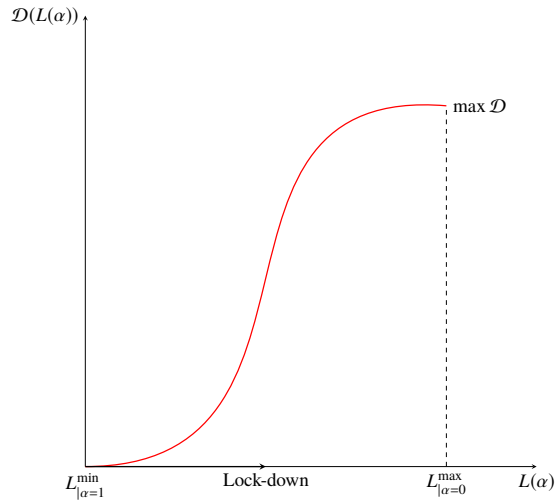
\begin{figure}
\begin{center}
\scalebox{0.60}{
   \begin{tikzpicture}[>=stealth]
   \draw[->] (0,0) -- (10,0) node [below] {$L(\alpha)$};
   \draw[->] (0,0) -- (0,10) node [left] {$\mathcal{D}(L(\alpha))$};
       \draw [thick, red] (0,0) .. controls (6,0) and (2,8.5) .. (8,8);
   \node at (8,0) [below] {$L^{\max}_{|\alpha=0}$};
      \node at (0,0) [below] {$L^{\min}_{|\alpha=1}$};
   \draw [->,thick] (0,0) -- (4,0) node [below] {Lock-down};
      \draw[dashed] (8,0) -- (8,8) node [right] {$\max{\mathcal{D}}$};
   \end{tikzpicture}}
   \end{center}
   \caption{The economic damage as a function of the lockdown}
   \label{fig:1}
\end{figure}

We can graphically sketch its functional form. Figure~\ref{fig:1} depicts a positive non-linear behaviour of the damage function vis-a-vis the intensity of the lock-down, meaning that the higher the intensity of the lock-down, given by a reduction of $\alpha$, ranging from $[0,1]$, the higher the economic damage, and non-linearly so. The lock-down policy is implemented by reducing the parameter $\alpha$, controlling for social distance. For low intensity of the lock-down, the economic damage is rather negligible (e.g. local lock-downs). Extending the lock-down after the first inflection point harshly hits the economy with more than proportional increments of the damage, while after very persistent lock-downs (second inflexion point), the cumulative damage is so high that it can only increase less than proportionally. The intersection between the curve and dashed vertical axis at $\alpha=0$ represents the maximum damage.

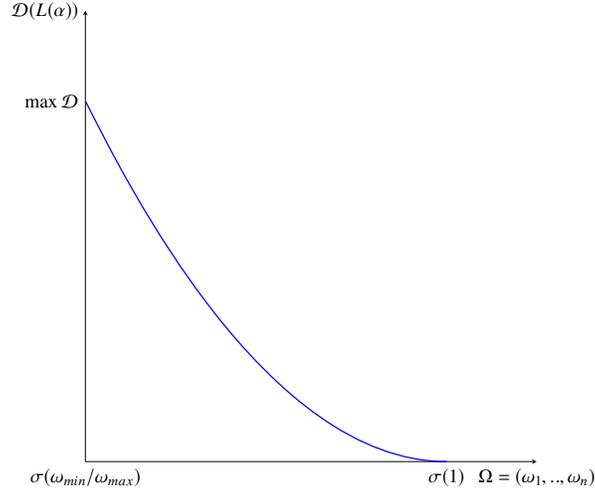
\begin{figure}
\begin{center}
\scalebox{0.60}{
   \begin{tikzpicture}[>=stealth]
   \draw[->] (0,0) -- (10,0) node [below] {$\Omega = (\omega_1,..,\omega_n)$};
   \draw[->] (0,0) -- (0,10) node [left] {$\mathcal{D}(L(\alpha))$};
       \draw [thick, blue] (8,0) parabola (0,8);
   \node at (8,0) [below] {$\sigma(1)$};
      \node at (0,0) [below] {$\sigma(\omega_{min}/\omega_{max})$};
      \node at (0,8) [left] {$\max{\mathcal{D}}$};
   \end{tikzpicture}}
   \end{center}
      \caption{The economic damage function across income levels}
   \label{fig:2}
\end{figure}

Figure~\ref{fig:2} presents a negative convex relationship with income level, meaning that the lower the level, the higher the impact of the economic damage. The damage reduces non linearly with income level so that losses scale down more than proportionally and relatively richer people turn out almost unaffected. The maximum of the damage is at the vertical intercept. Given the actual income distribution $\Omega= (\omega_{i+1}, \ldots,\omega_{i+n})$, the damage will be higher, the higher the ratio $\sigma$ between the minimum and the maximum income level, that is when $\omega_i \rightarrow \omega_{min}$, conversely it will be lower when $\omega_i \rightarrow \omega_{max}$.

Note a major implication of the foregoing argument: the relationships between human losses and economic ones not only is not linear but under some lock-down conditions might not be even monotonic. Tighter degrees of lock-down for sure imply higher economic losses but in some circumstances, as we shall see, might also yield \textit{higher rates of infection}.

\subsection{Generalised lock-down policies versus localised contagion dynamics}
The model developed in Section 3 allows to overcome the limitations of the most incumbent compartmental modelling approaches. Indeed, it accounts for infection rates endogenously affected by the state of the virus invasion in infected individuals $(u_j)$. This is a first major advancement in that one can account for the importance of the \textit{viral load} in determining the evolution of the disease after the individual enters the infectious status. The second major advancement relevant to the discussion here concerns the definition of an intra-body interaction process between the virus and the \textit{individual immunity} $(w_k)$. Now, considering the infection process resulting from the coupling of these two types of heterogeneity, this will not yield a single parameter $R_0$ but a distribution of  $\mathcal{R}_{0,t,i} = (R_{0,i+1}, ..., R_{0,i+n})$  across individuals, evolving in time, according to the state of the infection.

Such an analysis encompassing heterogeneity fundamentally differs from SIR compartmental structures modelled by differential equations (as pointed out long ago by~\cite{[RS2008]}), so that the level of $R_0$ parameter decreases with:

\begin{itemize}
\item The degree of individual heterogeneity in susceptibility to infection by the virus.

\vskip.2cm \item the clustering of the contact structure.
\end{itemize}

The foregoing model introduces individual heterogeneity and local interactions while considering the rate of interaction $\eta_0$ as constant and equal across all individuals (e.g., space homogeneity). Under that set-up, the simulation analysis points at a general positive impact of the policy of social distancing, implemented by the reduction of the $\alpha$ parameter, decreasing the peak of contagion and the actual number of infected individuals.

The current model specification can be extended in two directions:

\begin{enumerate}
\item The interaction dynamics across individuals is made to evolve in structured networks, say lattices which can represent homes, workplaces, schools, hospitals. This would imply a modification from $ \eta_0 \Rightarrow \eta_{i,j}$.

\item The level of infection, namely $u_j$, may be affected by the interaction process itself, $ u_j \Rightarrow u_j(\eta_{i,j})$, implying that the virulence increases with repeated contacts.
\end{enumerate}

The relevance of the two extensions is well supported by the evidence both on the infection dynamics and on its spatial heterogeneity. Indeed, the $R_0$ estimated in hospitals is by far higher than the one which might have been recorded on the streets. Contagion has spread even during the lock-down, mainly occurring at home and in hospitals. In the following we shall outline two potential venues to account for the above dynamics.

\vskip.2cm \noindent \textbf{Extension 1: occasional versus structured contacts}\\
Let us suppose the existence of different types of $\alpha$ parameters, one for occasional contacts, say occurring along a street, and one for structured contacts, occurring at home and in hospitals: $\alpha_o$ and $\alpha_s$. Under two different social distance parameters, the policy of the lock-down has opposing effects: from the one hand it reduces $\alpha_o$ by reducing occasional contacts, while, on the other, it increases $\alpha_s$ by increasing structured contacts. If this is the case, the overall contagion rate might not be reduced by lock-down policies. On the other hand, the lock-down will have a strong economic damage as discussed above.

Thus, one could envisage heterogeneous probabilities of interaction and different interaction dynamics, e.g. (i) homogenous/random contacts vs structured contacts in lattice, and (ii) presence or absence of the possibility of long-range infection (small world). This will result in a distribution of individual social interactions, say $(\alpha_{i+1}, ..., \alpha_{i+n})$. Understanding the extent to which the contagion dynamics is clustered also entail alternative policy measures. So, for example, \textbf{generalised lock-down policies might be ineffective under clustered dynamics of contagion}.

\vskip.2cm \noindent \textbf{Extension 2: coupling interaction with virulence}\\
Actual recorded contagion in schools and workplaces has been very low. Notwithstanding the limited robustness of such evidence due to the absence of a generalised testing strategy, a possible higher immunity response in the youth and adult population demands a second potential extension of the model, namely the possibility of making the virulence dependent also on the ``types'' of interacting agents with higher immunity rates in youth and young adults.

\subsection{Early warning indicators: when and how long}

Countries have reacted very differently in terms of the management of the  COVID-19 crisis: some countries, including South-Korea, Taiwan, New Zealand, Japan, Germany report case-fatality rates ranging from [1.4 - 5] $\%$, while some other countries like Italy, France, Spain, U.K., Belgium do report far higher rates, in the range [10 - 15] $\%$ (see~\cite{[JHU]}, data retrieved on the 26$^{th}$ of May). Sweden, with no lock-down policy at all records a fatality rate proximate to the Italian one [11.9 vs 14.3] $\%$. These numbers are clearly biased by data collection and testing strategies. However, country heterogeneity is a robust fact.

More than the amplitude of the lock-down, a variable strongly affecting the overall dynamics of the pandemic seems to be the \textit{timing} of the policy intervention. In general, \textbf{timely and selective closures have been more effective than delayed generalised closures}. This has been the case of South Korea, with massive contact tracing technologies, but also of Germany, which has undertaken a massive testing strategy and very early selective isolation. In both cases, the lock-down was not generalised.

Together, hospital capacity and the number of ICU beds has been the other crucial variable. However, while ICU beds in the very short-term might be considered as relatively fixed, this is not the case for early detection, which might even prevent hospitalization.

When the policy measure should be put in place? Is there any threshold indicator useful to understand the timing of the policy action? Together, when the lock-down should end? Which type of indicators should be used? These questions are extremely important but addressing them entails the appropriate understanding of the contagion dynamics. The saturation of ICU beds indicator strongly reduces rooms for policy actions. Being a ``downstream'' indicator, the system will detect the saturation only when the contagion is peaking, or at least accelerating.

Early warning indicators, upstream, are of crucial importance. And they involve monitoring systems communicating suspect phenomena, like, in the Corona viruses case, anomalous picks of pneumonia and persistent flue symptoms across patients. Early warning indicators are also crucial to avoid unnecessary long lock-down phases, exerting strong impact on the socio-economic systems but less so on the contagion dynamics.

\vskip.2cm \noindent \textbf{Extension 3: introduction of early warning indicators and the interaction with policies}\\
At the current stage the model includes in the dynamics of infection the social distance parameter as a crucial variable of policy action. The model might be extended by introducing an interaction from an early warning indication of the state of the system to the containment, hospitalization and treatment policies also able to reduce the pool of susceptible individuals, isolating and treating early detected cases, thus avoiding the spreading of the contagion.

\vskip.2cm \noindent \textbf{Extension 4: endogenizing the timing of the policy action}\\
The time of the policy action is currently exogenous to the model. However, the timing in which the social distance parameter is changed might become endogenous and depend on some state of the system as well, e.g. the fraction of infected/hospitalised/dead individuals. This ought to apply both to the beginning and the end of policy measures.

\subsection{Policy actions beyond the lock-down}

Overall, the spreading of the pandemic is exacerbating a series of old inequalities and vulnerabilities. If the common perception is that ``everybody is equal'' in front of the pandemic at closer inspection this is not true. What people do at work, their contractual framework, and their position in the internal organizational hierarchy strongly affect the possibility to remote working. Gender and geographical imbalances matter. The digital divide is exacerbating: access to high-speed internet connection and ICT-devices are the necessary condition to learn at the e-schools. However learning dramatically depends on the education level of the parents themselves. Schools are never been as unequal as nowadays.

The coupling of the pandemic and social distancing are making diverse risks conflating: health risk (exposition to social contacts are higher for low income occupations), income risk (probability of job losses is higher for temporary low-income occupations), employment risk (feasibility to remotely work is lower for low-income occupations).

In the following, we shall list a series of policy actions to be undertaken, beyond the lock-down, in a medium term perspective, to cope with the increasing risk of widespread, collective diseases. Overall, the health-management system deeply affects the contagion dynamics and therefore needs to be completely reorganised. Let us highlight some crucial reforms to be undertaken in our view:

\begin{itemize}
\item Increasing the overall public expenditure for the health system by strengthening local hospitals and laboratories: a capillary hospital system is able to cope with widespread diseases.

\vskip.2cm \item Reducing the public subsidies to private clinics, being the latter more interested in profit-seeking activities rather than general medical care assistance and provision of ICU beds for the general public.

\vskip.2cm \item Strengthening the role of GPs, implementing forms of communication and monitoring activities, fostering at home visits.

\vskip.2cm \item Increasing the public financing research.

\vskip.2cm \item Compelling the pharmaceutical sector to perform genuinely innovative $R\&D$ activities.

\vskip.2cm \item Revitalizing national-based laboratories to discover drugs away from the market system starting with vaccines.

\vskip.2cm \item Maintaining inventories of safety devices and instruments necessary to equip hospitals, protect workers and perform testing
\end{itemize}

\section{Looking forward to research perspectives}\label{sec:7}

A systems approach to  modelling  the contagion and spread over a given territory has been developed in our paper, where  a multiscale vision has been a
key feature of the approach which includes hospitalization dynamics, and recovery or/death of patients.

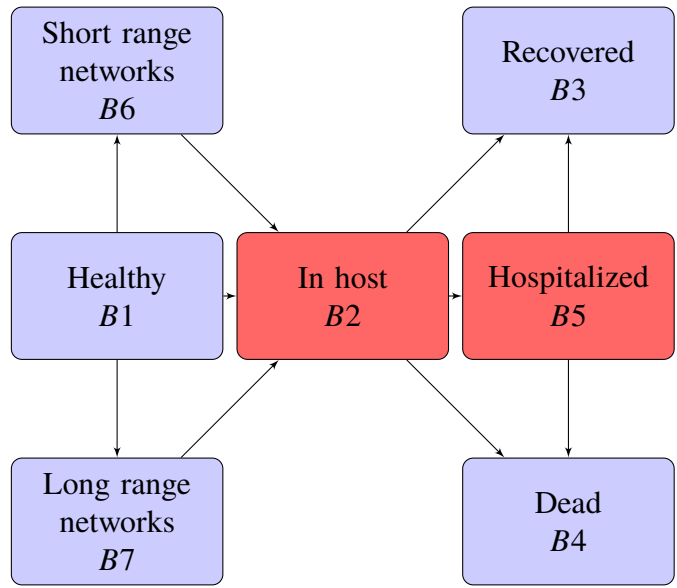
\begin{figure}
\tikzstyle{decision} = [diamond, draw, fill=blue!20,
    text width=4.5em, text badly centered, node distance=3cm, inner sep=0pt]
\tikzstyle{block1} = [rectangle, draw, fill=blue!20,
    text width=6em, text centered, rounded corners, minimum height=4em]
    \tikzstyle{block2} = [rectangle, draw, fill=red!60,
    text width=6em, text centered, rounded corners, minimum height=4em]
\tikzstyle{line} = [draw, -latex']
\tikzstyle{cloud} = [draw, ellipse,fill=red!20, node distance=3cm,
    minimum height=2em]
\begin{center}
\begin{tikzpicture}[node distance = 3cm, auto]
    \node [block2] (Inf) {In host \\ $B2$};
    \node [block1, left of=Inf] (Hea) {Healthy \\ $B1$};
    \node [block2, right of=Inf] (Hos) {Hospitalized \\ $B5$};
    \node [block1, above of=Hea] (Short) {Short range networks \\ $B6$};
    \node [block1, below of=Hea] (Long) {Long range networks \\ $B7$};
    \node [block1, above of=Hos] (Rec) {Recovered \\ $B3$};
    \node [block1, below of=Hos] (Dea) {Dead \\ $B4$};

    \path [line] (Hea) -- node  {} (Inf);
    \path [line] (Inf) -- node  {} (Hos);
    \path [line] (Inf) -- node  {} (Rec);
    \path [line] (Inf) -- node  {} (Dea);
    \path [line] (Hos) -- node  {} (Rec);
    \path [line] (Hos) -- node  {} (Dea);
    \path [line] (Hea) -- node  {} (Short);
    \path [line] (Hea) -- node  {} (Long);
    \path [line] (Long) -- node  {} (Inf);
    \path [line] (Short) -- node  {} (Inf);
\end{tikzpicture}
\end{center}
\caption{Transfer diagram of the model.  Boxes represent functional subsystems and arrows indicate transition of individuals.}
\label{Overall-system}
\end{figure}

An updated version of Fig.~2 is shown in Fig.~\ref{Overall-system} to account for all specific blocks and related flow chart of the systems approach. Specifically, for spatial dynamics and the role of hospitalization which, in agreement with the reasonings presented in Sections 4--6, can be viewed as an important feature of the systems approach. The figure shows the following blocks:

\vskip.2cm \noindent \textbf{Block 1:} The dynamics of contagion is at the level of individuals depending on the level of confinement only in the  case of spatial homogeneity, while a deeper understanding of these dynamics might account for crowd movement in complex venues.

\vskip.2cm \noindent \textbf{Block 2:} The dynamics are generated by the contagion and, subsequently, develop inside each individual depending on the interaction at the small scale between virus infection and immune particles, \textit{in host dynamics}. The modelling takes account the heterogeneous behaviour of individuals, as well as the heterogeneity, progression and competition inside each individual entity.

\vskip.2cm \noindent \textbf{Blocks 3,4:} Show the output of the interactions consisting in recovery or death of patients, where this final exit can go through the passage across the hospitalization which is related to the level of the pathology.

\vskip.2cm \noindent \textbf{Block 5:} Refers to the passage from Block 2 to an organized hospitalization dynamics. If the dynamics within Block 5 are properly modelled accounting for medical care, the number  patients which are recovered should increase, while that of dead persons should decrease.

\vskip.2cm \noindent \textbf{Block 6:} Refers to the dynamics by which the contagion spreads over a territory made of a sequence of  interconnected areas. The dynamics might include aggregation through endogenous networks.

\vskip.2cm \noindent \textbf{Block 7:} Studies the dynamics by which the contagion spreads over a territory  through  long range exogenous networks, where connections between nodes depend on the transportation system.

\vskip.2cm The systems approach has been somehow revisited in Sections 4,5 and 6 by the contribution from the sciences of virology, immunology, and socio-economics. Sections 4 and 5 mainly, but not exclusively, refer to Block 2, while Section 6 mainly refers to the whole systems  approach.

We do not naively claim that our paper has given an exhaustive description by a differential system of the highly complex dynamics under consideration. Indeed,  the multidisciplinary interaction has motivated a number of challenging research perspectives which, in turn are not straightforwardly transferred to mathematics as some of them need additional research within the specific fields the said hints come from. Therefore, keeping the framework represented in Fig.~\ref{Overall-system}, the modelling of the dynamics within each block can be further developed and improved.

The strategy to take advantage of the contribution from virology, immunology and economy  is consistent with the idea that these disciplines refers to milestones of the research on infections and possible pandemic. Indeed, virology can work to explain the onset of the virus by mutations and selection related to the environment where the virus can progress, immunology studies the complex interactions between virus infection and immune system in view of the development of vaccines and medical care, economics can develop studies on the impact on society of the various possible actions developed towards the control of the pandemic. Therefore, the contribution of mathematics should be developed with the aim of providing  useful inputs to the scientists and crisis managers involved in the multidisciplinary dialogue.

Bearing this specific target in mind let us develop a further study to investigate the lock-down and lock-up strategy by adding simulations those in Section 3 corresponding to Figures \ref{social-distance-1} and \ref{social-distance-2} related to the ICU hospitalization capacity in a territory. In detail, they show that locking-down too early can generate a second outbreak larger than the first one, with a subsequent number of individuals requiring a hospitalization larger than the effectively available beds. These previous simulations suggest the introduction of the key number:
$$
\kappa = \frac{\alpha \cdot \beta}{\gamma}
$$
which refers the intensity of the infection $\alpha \cdot \beta$ to the immune defence $\gamma$. Namely, increasing values of $\kappa$ denote an increasing level of the infection attack.
\vskip.2cm


\begin{figure}[ht!]
\begin{tabular}{cc}

\includegraphics[width=0.45\textwidth]{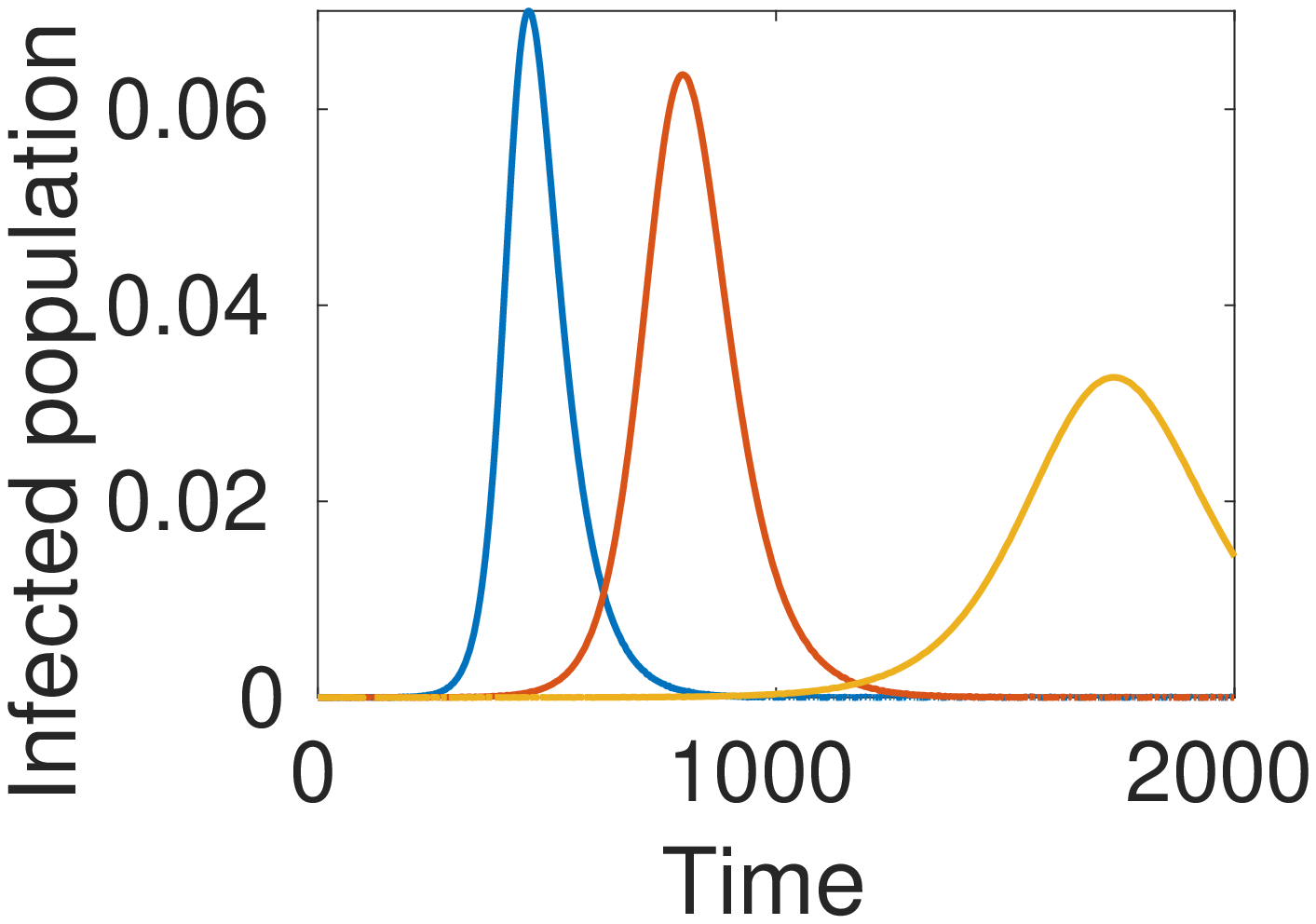} &
\includegraphics[width=0.45\textwidth]{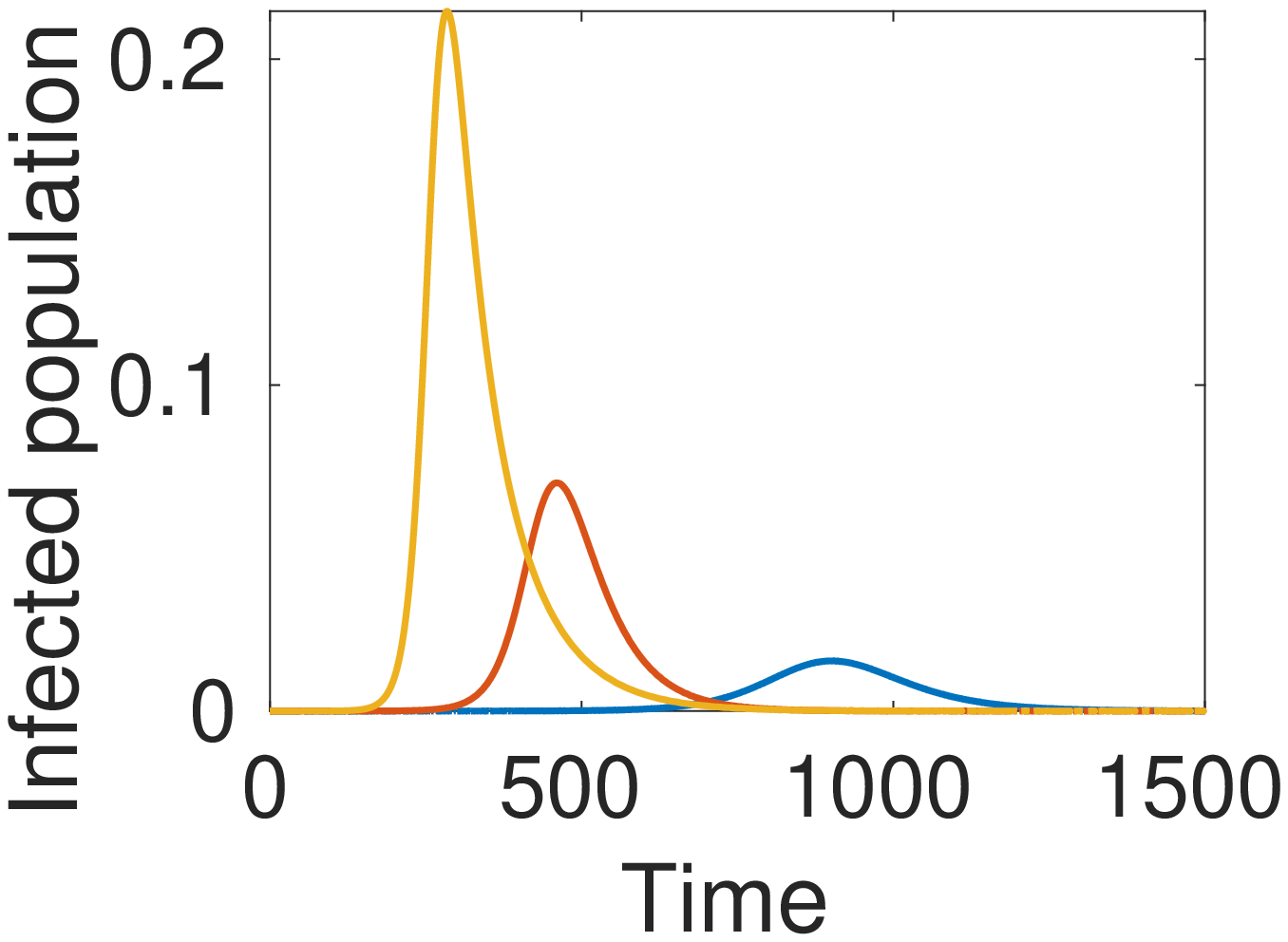} \\
(a) $\kappa=0.2$, fixed $\beta=0.1$ & (b) $\kappa=0.2$, fixed  $\alpha=0.4$\\
{} & {} \\
\includegraphics[width=0.45\textwidth]{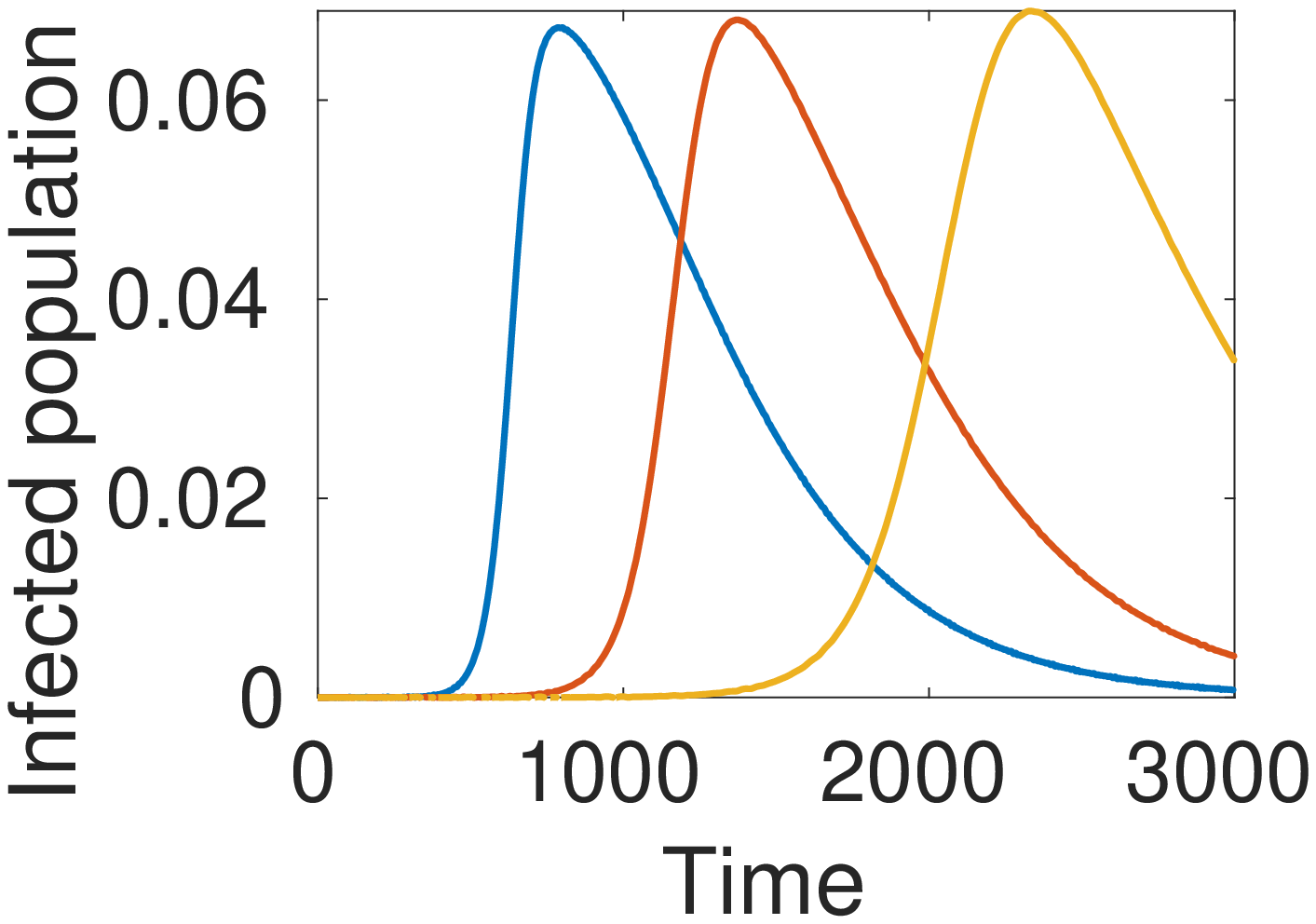} &
\includegraphics[width=0.45\textwidth]{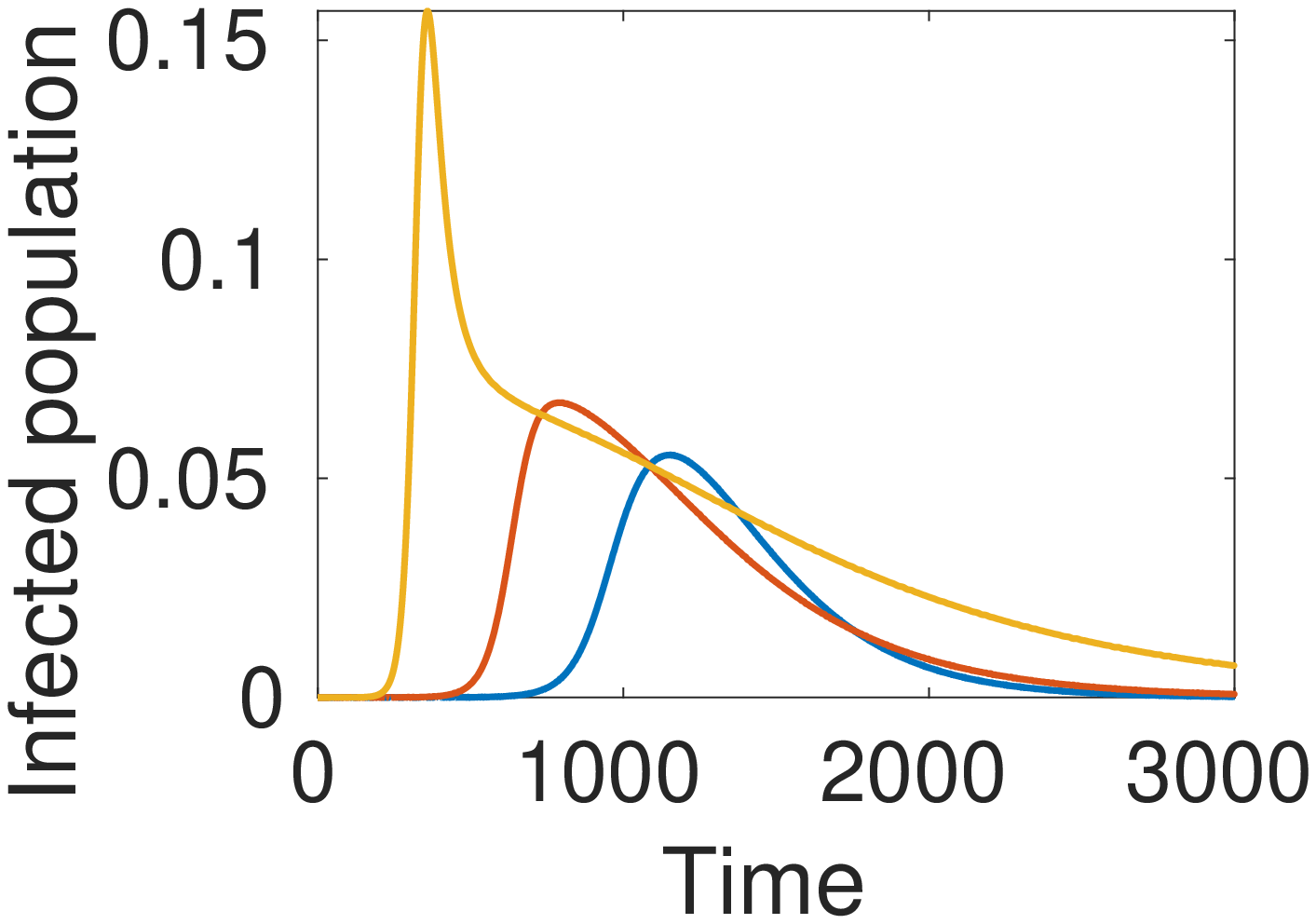} \\
(c) $\kappa=0.02$, fixed $\beta=0.01$ & (d) $\kappa=0.02$, fixed  $\alpha=0.4$ \\

\end{tabular}

\caption{Sensitivity to $\kappa$.\newline
 (a) Blue: $\alpha =0.4$, $\gamma=0.2$, Red: $\alpha =0.2$, $\gamma=0.1$, Yellow: $\alpha =0.1$, $\gamma=0.05$.\newline
  (b) Blue: $\beta=0.15$, $\gamma=0.3$, Red: $\beta=0.1$, $\gamma=0.2$, Yellow: $\beta=0.05$, $\gamma=0.1$.  \newline
  (c) Blue: $\alpha =0.4$, $\gamma=0.2$, Red: $\alpha =0.2$, $\gamma=0.1$, Yellow: $\alpha =0.1$, $\gamma=0.05$.\newline
  (d) Blue: $\beta=0.15$, $\gamma=0.3$, Red: $\beta=0.1$, $\gamma=0.2$, Yellow: $\beta=0.05$, $\gamma=0.1$. } \label{varying-kappa}
\end{figure}


The following new simulations study the sensitivity to the parameter $\kappa$ depending on different values of $\alpha$, $\beta$ and $\gamma$ however with the constraint of keeping $\kappa$ fixed, see Fig.~15. In more detail:

\begin{itemize}

\item $\kappa = 0.2$: (a) If $\beta = 0.1$ is kept constant, then decaying values of the infection rate $\alpha$ are followed by the same decay of the immune defence due to the constraint $\kappa = 0.2$; simulations show that, even if the immune defence decays, the peak decays and the infection spread is delayed.

\vskip.2cm \item $\kappa = 0.2$: (b)  If $\alpha = 0.4$ is kept constant, then decaying values of the virus progression rate $\beta$ are followed by the same decay of the immune defence due to the constraint $\kappa = 0.2$; simulations show that, even if the immune defence decays, the peak decays and the infection spread is delayed, by a faster decay-delay, but the initial infection is much higher than that of case (c).

\vskip.2cm \item  $\kappa = 0.02$: (c) The difference with respect to case study (a) is that the decay of the infection rate $\alpha$ only influences the delay.

\vskip.2cm \item $\kappa = 0.02$: (d) The dynamics is analogous to that of the case study (b) but the height of the peak is lower.

\end{itemize}

A very first, and rapid, biological interpretation is as follows:

\begin{quote}
\textit{The defence of the immune system applies an effective contrast to the virus progression. However the efficacy of the action is more relevant if the defence keeps a fixed value independently on the level of infection or progression. If, in addition, the defence increases with increasing values of $\alpha$ and $\beta$, the efficacy is even higher.}
\end{quote}

Let us now study how the explorative ability of models can be used by crisis managers who have the responsibility to take decisions about the strategy to reduce the damages generated by the spread of the infection. This additional investigation is motivated not only by biologists and clinicians, but also by economists who are charged with studying the complex interaction between the pandemic and socio-economic systems. It is a delicate task  which has to balance human safety on one side and the consequences on the wealth distribution over the society on the other. Accordingly, let us focus on the strategy to plan firstly the lock-down deemed to control the pandemic evolution and subsequently lock-open to activate production and marketing systems.

In more detail, two key variables play an important role, namely the selection of the locking and lock-open times $T_\ell$ and $T_d$, respectively, as well as the related intensity of the actions represented by the contagion coefficients $\alpha_\ell$ and  $\alpha_d$, respectively. The following specific case studies can be simulated by the model:

\begin{itemize}

\item  $T_\ell$ is the lapse of time, after the discovery of the infection, at which locking behavioural rules are imposed, corresponding to  $\alpha_\ell$ with the aim to keep the number of infected people under a threshold level.

\vskip.2cm \item  $T_d$ is the lapse of time from  $T_\ell$ to impose less restrictive locking rules (locking-down) corresponding to  $\alpha_d > \alpha_\ell$, also in this case with the aim of keeping the number of infected people below a threshold level.

\end{itemize}

\begin{figure}[h!]
\includegraphics[scale=.8]{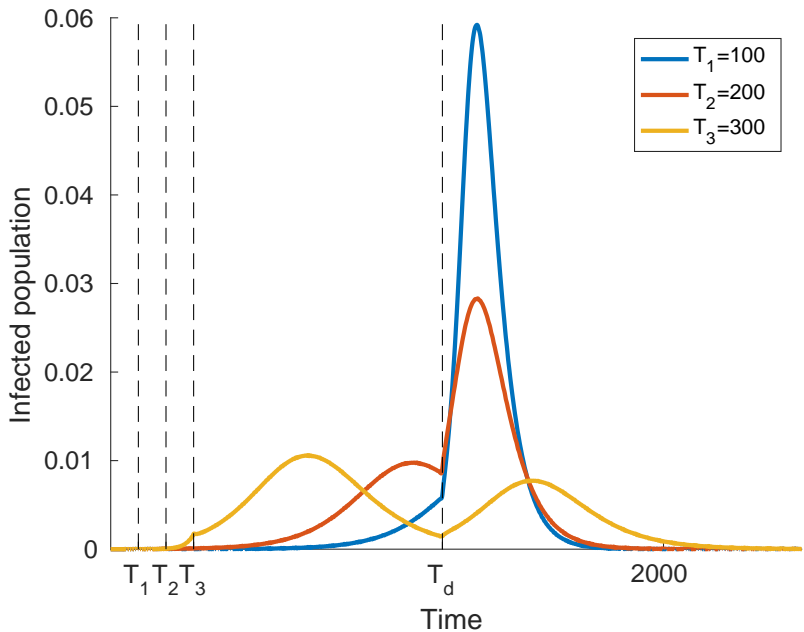}
\caption{Varying locking times. We take $T_\ell=100,200,300$, and fixed  lock-open time $T_d=1200$. $\alpha=0.4$ for $t\in [0,T_\ell)\cup [T_d,T_{max}]$ while $\alpha=0.25$ during the locking interval.}\label{locking-times}
\end{figure}

 Figure~\ref{locking-times} presents a simulation to investigate  the influence of the up-locking time, namely the delay to apply the up-locking, on the subsequent dynamics.
 
\begin{figure}[h!]
\includegraphics[scale=.8]{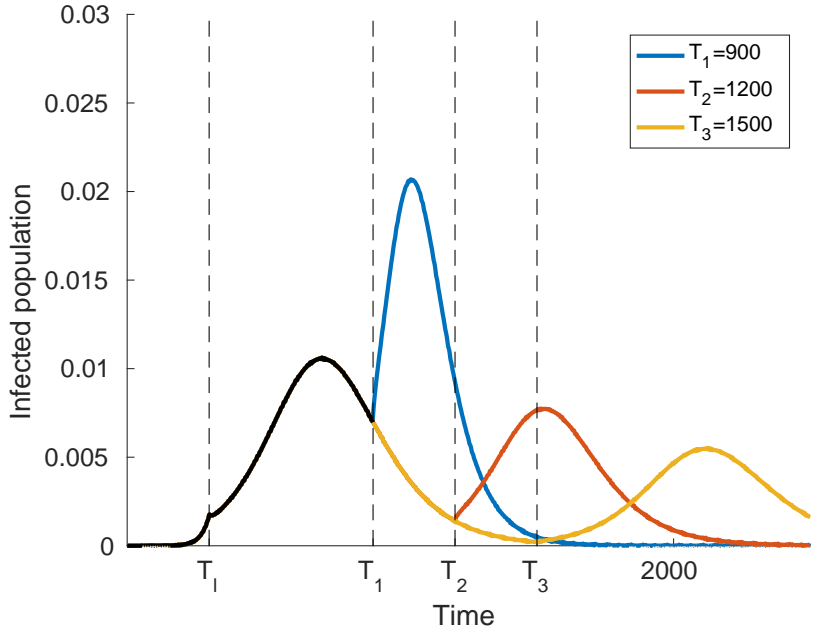}
\caption{Varying de-locking times. We take a fixed locking time $T_\ell=300$, and three different  lock-open times $T_d=900,1200,1500$. $\alpha=0.4$ for $t\in [0,T_l)\cup [T_d,T_{max}]$ while $\alpha=0.25$ during the locking interval. Notice that the three curves coincide until the first  lock-open time, and are consequently represented in black in that interval. }\label{delocking-times}
\end{figure}

Figure~\ref{delocking-times} studies the influence of the lock-down time. Simulations show how delaying $T_d$ reduces the peak, but increases the time interval of the persistence of the infection.

\begin{figure}[h!]
\includegraphics[scale=.8]{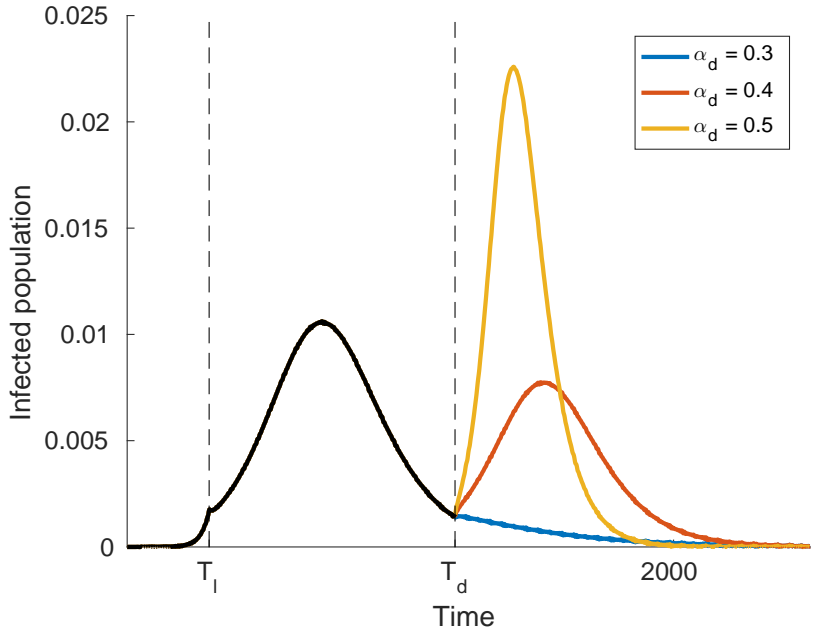}
\caption{Varying the de-locking value $\alpha_d$. We take fixed locking  and  lock-open times $T_l=300$ and $T_d=1200$, respectively. $\alpha=0.4$ initially for $t\in [0,T_l)$ then reduced to $\alpha = 0.25$ during the locking interval and finally we consider three different  lock-open values $\alpha_d = 0.3,0.4,0.5$.
Notice that the three curves coincide until the  lock-open time, and are consequently represented in black in that interval. } \label{alfad}
\end{figure}

Figure~\ref{alfad} studies the influence of the lock-down level at fixed values of $T_d$ and $T_\ell$. As expected, simulations show how a high level of lock-down can generate high level peaks.
persistence of the infection.


\vskip.1cm The task of crisis managers consists firstly in providing the  strategic plans to reduce and achieve effective values of   $\alpha_\ell$ and $T_\ell$. Subsequently, the problem consists in referring   $\alpha_d$ to  $T_d$. The strategy depends on a variety of social and political variables which go beyond the aims of our paper. Simulations delivered in Figures~\ref{locking-times}--\ref{alfad}  do not cover the whole variety of possible case studies. An exhaustive study can be, however, developed under the direct requirements of crisis managers. Therefore, we simply claim that modelling and simulations provide support to decision making, while the overall systems approach shown in Figure \ref{Overall-system} leads to study of all different interactive dynamics which appear in the system under consideration.

\vskip.2cm
Bearing all of the above in mind let us now propose a forward look to research perspectives by selecting, according to the authors' bias, five key problems which are presented in the following, mainly focusing on mathematical topics with the aim of building a bridge between mathematics and biology. However accounting for the hints,  warnings and forward looks on the perspectives proposed in Section 6 focused also on the interaction between  economic  strategies and social dynamics.

Indeed, it is not surprising that economists have been induced to develop a critical analysis on the strategies that governments might develop whenever they account for future scenarios and not simply for  present situations.  As mentioned in Section 1, the event studied in our paper cannot be classified straightforwardly as a ``black swan'', since it could have been predicted, and indeed was predicted by some, at least up to certain levels. Mathematics offers general explorative models that can depict, via computational simulations, the overall collective dynamics accounting for the possible actions that the science of medicine can offer. This contribution does not solve the problems of biology and cannot straightforwardly lead to devices of artificial intelligence, but it can help clinicians to develop a care strategy by offering them additional information to be used alongside their own clinical experience.

Reasonings on the key problems take advantage of the fact that our approach, which does not claim to be exhaustive, appears to be sufficiently flexible to include new specific features motivated not only by biology, but also by economics. 

The presentation of our critical analysis  refers to the blocks of Figure~\ref{Overall-system} and looks ahead to perspectives to be developed within the multidisciplinary and multiscale framework proposed in our paper (KP -- Key Problem).

\vskip.5cm \noindent \textbf{KP 1 - Virus progression and immune competition:} This key problem refers to the dynamics inside each individual (Block 2). The modelling approach of our paper has already introduced some important features of the system under consideration. For instance, progression of the virus by levels which correspond to hospitalization - followed by medical care - up to the extreme level corresponding to death. The competition between the virus and immune system is also treated accounting for the heterogeneous reaction ability of patients.

The output of simulations is definitely consistent with the systems vision, but further refinements of the model of the dynamics  in Block 2 should be developed in order to contribute to account for the actions of medical care which refer to Block 5. Focusing on the objective of designing models with enriched descriptive ability,  we remark that the main source towards this development is the literature cited in Sections 4 and 5, by which we can learn about firstly about the virus dynamics, subsequently the mechanics of contagion, and finally  the different sub-populations which play the game and the actions exerted by them. This literature also  indicates that the biological phenomena related to this specific competition are not fully understood, see~\cite{[CS20]}.

Despite this correct statement, we observe that the model of the in-host dynamics, corresponding to Eqs.~(3.3)--(3.5) by the dynamics depicted  in Figure~\ref{inside}, is  sufficiently flexible to include a more advanced interpretation of biological reality. In addition, further dynamics might be taken into account, for instance that by which the virus chases for foraging and attaches to cells of the lung tissue to expand by proliferation which ends with further destructive actions over the lung tissue. Analogous care should be focused to model the foraging cells offered by the host lung which might damage the overall mechanics of the lung in the breathing dynamics~\cite{[CPFBC13]}.

An additional, important refinement consists in subdividing the overall populations according to age, social and physical state. This subdivision should be useful not only to refine the modelling of interactions, but also to contribute to the decision strategy of crisis managers concerning locking, hospitalization and, social support in general.

As mentioned, the development of the in-host modelling can take advantage of Sections 4 and 5, starting from their contribution to understanding virus mutations by which   \textit{SARS--CoV--2} is responsible for COVID-19.  A pioneer paper devoted to the modelling of virus mutations followed by a learning dynamics can be found in~\cite{[DDS09]}. This paper  provides  some ideas to be developed towards a modelling approach to depict the complex mutation-learning dynamics specifically referred to  COVID-19.

\vskip.5cm \noindent \textbf{KP 2 - Medical care:} The dynamics of progression of the pathology for hospitalized patients has not been explicitly treated in our paper as the model simply describes how hospitalization can be related to a level of the pathology which cannot be cared by home confinement. The model computes the flow of infected people towards hospitals and compares it with the effective capacity of them in a given territory. The model also depicts different scenarios related to the different levels by which hospitalization is demanded.

This topic  deserves further attention towards a detailed study of the role of medical cares localized in  Block 5 of Figure~\ref{Overall-system}. The modelling approach is not a straightforward generalization of that developed in Block 2 as it requires a detailed modelling of therapeutical actions. Hence it requires additional interdisciplinary work from the medical sciences. The representation of Figure~\ref{inside} is still valid, but progression of the immune system and regression of the virus  should be related to well-defined medical actions. The contents of Section 4 and 5 contribute, as mentioned,  to model some of the possible therapeutic actions.

\vskip.5cm \noindent \textbf{KP 3 - Clustered contagion, heterogeneity and monitoring:}  As argued in Section 6 above, the contagion (or not) develops within and across network structures which affect who is in contact with whom, for how long, etc. A straightforward development ought to address the clustering of interactions in structured spaces - e.g. factories, schools, hospitals, elderly residences, families \textit{in primis}. It could well be for example that above some levels of diffusion, confinement does increase the spread of the disease in that it confines the interactions but increases their interaction rates. Additionally, a necessary extension entails the account for intrinsic heterogeneity among agents, e.g. young versus old, healthy versus affected by previous pathology, based in areas with clean versus polluted atmosphere. Finally, in terms of the medical treatment and crisis management, early detection, as home diagnostics by GPs, and continuous monitoring might allow to overturn congestion in hospitals.

The model already accounts for heterogeneous behaviours of individuals, but it does not yet refer to the aforementioned features. Accounting for them is possible within the framework of the model technically firstly by specializing the overall population into a variety of subsystems and subsequently by referring them to a network of aggregation sites. The simulation by agents proposed in~\cite{[TERNA]} moves precisely towards this target with the aim of providing a broad variety of scenarios to be used by crisis managers. However, it cannot be naively hidden that the price to   pay is an  higher level of complexity somehow related to the calibration of the model treated in the next key problem.

Recent empirical investigations have shown that a key role in the in host dynamics is exerted by the initial virus load in each individual which might be induced by repeated contacts, e.g. affecting medical staff in hospitals, or simply by lack of awareness to the risk of contagion. Therefore, further studies should be addressed to derive  models where the modelling of the contagion rate includes an additional information on the aforementioned virus load. The main variable in charge of the modelling appears to be the \textit{awareness to contagion risks} which can contribute to deeper vision of a lock down strategy. This strategy is not simply related to opening and closing of areas of possible aggregation, but also to the care put on physical distancing and protection.

\vskip.5cm \noindent \textbf{KP 4 - Calibration of models:} This key problem refers to the identification of the parameters, since only if the said parameters can be properly assessed can the model be used for predictive investigations. This specific aspect of modelling induces a search for a balance between enriching the complexity of the model and the difficulty of tuning models with an excess of parameters. A mathematical model has been proposed in Subsection 3.1 corresponding to the flow chart in Fig.~\ref{transfer_diagram}, block $i=2$, and the visualization of the progression dynamics and immune reaction shown in Fig.~2, block $i=2$, as well as the visualization of the progression dynamics and immune reaction shown in Fig.~3. The model uses the following parameters: $\alpha$ modelling the risk of infection related to the level of confinement; $\beta$  referred to the ability of the virus to progress; and $\gamma$ modelling the the ability of the immune system to induce regression  of the virus. These parameters have been compacted in the key number $\kappa$. In addition, the model accounts for the heterogeneous distribution in the population of the aforementioned ability which can be roughly related to the age of the population.

Some recent research achievements can contribute to the tuning of specific parameters. A valuable example is the probabilistic study of contagion risk developed in~\cite{[CT20]} corresponding to different environmental conditions and typology of interactions. Simulations have been developed with explorative aims, namely by a sensitivity analysis of the dynamical response depending on the parameters of the model. This output has shown to be useful to crisis managers towards the decision process about the possible strategies. An example has been  the study of the up-lock and lock-down strategy by simulations which have shown how far the system is sensitive to closing-opening plans and have shown how a wrong decision might lead to disasters. An exhaustive sensitivity analysis might definitely contribute to the design of an artificial intelligence  system.

\vskip.5cm \noindent \textbf{KP 5 - Dynamics over a globally connected world:} This topic may be the most challenging key problem and research perspective. It refers to Blocks 6 and 7, where the dynamics occur involving both healthy and infected people. Some indications have been given towards modelling and simulations accounting both for the dynamics in complex venues and for that involving transportation system. The full development of this hint leads to the completion of all the details of the overall systems approach, where the dynamics in each block contributes to define the global dynamics in the world.

\vskip.5cm

We are aware that the interdisciplinary horizons should be enriched involving additional experts in various research fields, e.~g.~medicine, informatics, physics and various others. Indeed, we simply wish that our paper is a very first step towards this forward-looking scientific dialogue. Therefore all reasonings proposed in this section allow us to go back to the key target posed at the end of Section 1 by revisiting the quotation, given in the last part of the section, by the following conclusive remarks.

\vskip.2cm
\begin{quote}
\textit{The key target of this paper has been the design of a multiscale modelling approach suitable to produce simulations  mainly with explorative ability. This objective has been pursued all along the paper by mathematical tools of the kinetic theory of active particles along  a systems approach suitable to account of the dynamics and interactions  in blocks sketched in Figure \ref{Overall-system}.}
\end{quote}

\vskip.2cm
\begin{quote}
\textit{The systems approach and the multiscale vision appear firstly in the specific models in each block, where the modelling needs the use more than one scale, and subsequently in the modelling of interactions between blocks.}
\end{quote}

\vskip.2cm
\begin{quote}
\textit{The critical analysis focused on the  key problems has shown how the various ingredients of the approach can be further developed and improved. This remark motivates further research activity to update the models of each block shown in Fig.~\ref{Overall-system}. This means that we consider our paper a receptor of future scientific contributions from the scientific community somehow motivated also by individual and social care.}
\end{quote}

\vskip.2cm Finally, going back again to mathematical topics, we recall that, concerning the modelling of smallpox, Daniel Bernoulli wrote (to Euler)~\cite{[DietzHeester2002]}: 

\begin{quote}
\textit{Dolus an virtus quis in hoste requirat!}~\cite{[Virgil]} (`What matters whether by valour or by stratagem we overcome the enemy?')
\end{quote} 

Concerning COVID-19 the correct stratagem, we believe, is to develop a novel multiscale mathematical framework informed by multidisciplinary data as outlined in the current paper.  The current ``crisis'' is presenting us, and all mathematicians, with an opportunity to develop new mathematical theories, ideas and techniques, not only to shed light on the spread of pandemics, but also to further develop the mathematics itself. 

It is perhaps fitting to close this paper with the final words of the visionary talk by David Hilbert in 1900 before the opening of the Mathematical Congress of Mathematicians in Paris~\cite{[DH1900],[DH1901],[DH1902]}:

\vskip.2cm
\begin{quote}
\textit{We also notice that, the farther a mathematical theory is developed, the more harmoniously and uniformly does its construction proceed, and unsuspected relations are disclosed between hitherto separate branches of the science. So it happens that, with the extension of mathematics, its organic character is not lost but only manifests itself the more clearly. \;$\dots$ The organic unity of mathematics is inherent in the nature of this science, for mathematics is the foundation of all exact knowledge of natural phenomena. That it may completely fulfil this high mission, may the new century bring it gifted masters and many zealous and enthusiastic disciples.}
\end{quote}

\section*{Acknowledgements}

\vskip.1cm \noindent Nicola Bellomo: Support of Granada University, modelling Nature Group, Spain,  and Hosting support of the Italian Research Council, IMATI, CNR, Pavia, Italy.

\vskip.1cm \noindent Reidun Twarock: Financial support via an EPSRC Established Career Fellowship (EP/R023204/1), a Royal Society Wolfson Fellowship (RSWF\textbackslash R1\textbackslash 180009), and a Joint Investigator Award to RT and Prof. Peter Stockley (110145 $\&$ 110146) is gratefully acknowledged.

\vskip.1cm  \noindent  Mark Chaplain acknowledges the assistance of the Rapid Assistance in Modelling the Pandemic project coordinated by the Royal Society. 

\vskip.1cm  \noindent  Giovanni Dosi and Maria Enrica Virgillito acknowledge support from European Union's Horizon 2020 research and innovation
programme under grant agreement No. 822781 GROWINPRO -- Growth Welfare Innovation Productivity.

\vskip.1cm  \noindent  Damian Knopoff: Support of CONICET (Grant Number PIP 11220150100500CO)

\noindent and Secyt UNC (Grant Number 33620180100326CB).

\vskip.1cm  \noindent John Lowengrub acknowledges partial support from US National Science Foundation grants DMS-1763272 and the Simons Foundation (594598QN) for a NSF-Simons Center for Multiscale Cell Fate Research, as well as grants DMS-1714973 and DMS-1936833. JL also thanks the US National Institutes of Health for partial support through grants 1U54CA217378-01A1 for a National Center in Cancer Systems Biology at UC Irvine and P30CA062203 for the Chao Family Comprehensive Cancer Center at UC Irvine.

\end{document}